\RequirePackage{lineno} 
\documentclass[preprint]{elsarticle}

\usepackage{multicol}
\usepackage{color}
\usepackage{url}
\usepackage{xspace}
\usepackage{rotating}

\newcommand{\nubar}[0]{\overline{\nu}}
\newcommand{\numu}[0]{\nu_{\mu}}

\newcommand{\nue}[0]{\nu_{e}}
\newcommand{\nuebar}[0]{\overline{\nu}_{e}}
\newcommand{\numubar}[0]{\overline{\nu}_{\mu}}

\begin{document}

\title{Recent advances and open questions in \\ neutrino-induced quasi-elastic scattering \\ and single photon production}

\author[lanl]{G. T. Garvey} \ead{garvey@lanl.gov}
\author[fnal]{D. A. Harris} \ead{dharris@fnal.gov}
\author[ubc]{H. A. Tanaka} \ead{tanaka@phas.ubc.ca}
\author[indiana]{R. Tayloe} \ead{rtayloe@indiana.edu}
\author[fnal]{G. P. Zeller} \ead{gzeller@fnal.gov}

\address[lanl]{Los Alamos National Laboratory, P.O. Box 1663, Los Alamos, NM 87545, USA.}
\address[fnal]{Fermi National Accelerator Laboratory, P.O. Box 500, Batavia, IL, 60510-5011, USA.}
\address[ubc]{Insitute of Particle Physics and Department of Physics and Astronomy, University of British Columbia, 6224 Agricultural Road, Vancouver, BC V6T 1Z1, Canada.}
\address[indiana]{Department of Physics, Indiana University, 727 E. Third St., Bloomington, IN 47405-7105 , USA.}

\begin{abstract}
The study of neutrino-nucleus interactions has recently seen rapid development with a new 
generation of accelerator-based neutrino experiments employing medium and heavy nuclear targets for the study of neutrino oscillations. A few unexpected results in the study of quasi-elastic 
scattering and single photon production have spurred a revisiting of the underlying nuclear 
physics and connections to electron-nucleus scattering. A thorough understanding and resolution 
of these issues is essential for future progress in the study of neutrino oscillations.

\noindent A recent workshop hosted by the Institute of Nuclear Theory at the University of Washington (INT-13-54W)
examined experimental and theoretical developments in neutrino-nucleus interactions and related
measurements from electron and pion scattering. We summarize the
discussions at the workshop pertaining to the aforementioned issues in quasi-elastic scattering and
single photon production, particularly where there was consensus on the highest priority issues to
be resolved and the path towards resolving them.

\end{abstract}

\maketitle

\section{Introduction}
\label{sec:introduction}
In this report, we focus on two aspects of neutrino-nucleus scattering that have seen significant developments and were discussed at the ``Neutrino-Nucleus Interactions for Current and Next Generation Neutrino Oscillation Experiments'' workshop hosted by the Institute for Nuclear Theory in December 3-13, 2013\footnote{The workshop website with slides presented at the meeting can be found at \url{http://www.int.washington.edu/talks/WorkShops/int_13_54W/} }:

\begin{itemize}
\item charged current quasi-elastic scattering and
\item single photon production.
\end{itemize}
Both topics have seen intriguing experimental observations that have instigated a thorough re-investigation of the fundamental nuclear physics involved in understanding and modeling these processes. Furthermore, both processes are critical in interpreting neutrino oscillations measurements, where the intrinsic properties of neutrinos are studied through interference effects using neutrino-nucleus interactions. As rapid progress is made in neutrino oscillation measurements, commensurate developments in understanding and improving the modeling of the underlying nuclear physics will need to keep pace.

We begin with a brief introduction to neutrino charged current quasi-elastic scattering, followed by a review of recent experimental studies of this process, the intimately related issues in electron-nucleus scattering, and the theoretical developments essential towards understanding both processes. We then discuss the progress in single photon production in neutrino neutral current interactions, before summarizing the next steps towards resolving the remaining issues.

\subsection{The role of Neutrino Charged Current Quasi-Elastic Scattering in Neutrino Oscillation Experiments}
Neutrino charged current quasi-elastic (CCQE) scattering, along with the corresponding anti-neutrino process:
\begin{equation}
\label{eq:ccqe_process}
\begin{array}{lll}
\nu_\ell+n & \to & \ell^-  + p \\
\bar{\nu}_\ell + p & \to & \ell^+ + n
\end{array}
\end{equation}
represent the simplest form of neutrino-nucleon interaction, with the weak charged current inducing a transition of the incoming neutrino into its corresponding charged lepton $\ell$, and an analogous transition in the target nucleon. A detailed discussion of the theoretical framework and history of the study of neutrino CCQE interactions can be found elsewhere (see for instance Reference~\cite{Gallagher:2011zza}); here we review the formalism for CCQE interactions on a free nucleon to highlight the complications that arise once we consider scattering off nuclei and to illustrate the fundamental connections to electron scattering, which provides some of the essential nucleon form factors and an alternative probe of many of the same nuclear physics.
Considerations of Lorentz symmetry allow the differential cross section for the CCQE process to be written as~\cite{Llewellyn Smith:1971zm}:
\begin{equation}
\label{eq:ccqe_dsdqsquared}
\frac{d\sigma}{dQ^2} = \frac{G^2_F M^2|V_{ud}|^2}{8\pi E^2}
\left[ A\mp \frac{s-u}{M^2}B + \frac{(s-u)^2}{M^4}C \right]
\end{equation}
where:
\begin{itemize}
\item $G_F$ is the Fermi weak coupling constant and $V_{ud}$ is the Cabibbo-Kobayashi-Maskawa matrix element characterizing the strength of the weak charged current coupling between the $u$ and $d$ quarks.
\item $E$ is the incident neutrino energy.
\item $M$ and $m$ are the proton and outgoing lepton masses, respectively.
\end{itemize}
and $A$, $B$, and $C$ are defined as the following function of various nucleon form factors that are functions of $Q^2$, the square of the four-momentum transferred from the incoming neutrino to the nucleon:
\begin{equation}
\begin{array}{lll}
A & \equiv & \frac{(m^2+Q^2)}{M^2} \Bigl[(1+\tau)G_A^2 - (1-\tau)F_1^2 +\tau(1-\tau)F_2^2 + 4\tau F_1F_2 - \\ 
  &      & -\frac{m^2}{4M^2}\bigl((F_1+F_2)^2 + (G_A +2G_P)^2 - (\frac{Q^2}{M^2}+4)G_P^2 \bigr)\Bigr]\\
  &        & \\
B & \equiv & \frac{Q^2}{M^2}G_A (F_1+F_2) \\
  &        & \\
C & \equiv & \frac{1}{4} (G_A^2 + F_1^2 + \tau F_2^2) 
\end{array}
\end{equation}
Here,
\begin{itemize}
\item  $\tau=Q^2/4M^2$ and $s$, $u$ are the standard two-body Mandelstam variables with $s-u=4ME-Q^2-m^2$.
\item $F_1$ and $F_2$ are the isovector form factors, related to the electric and magnetic form factors $G_E$ and $G_M$ by $G_E = F_1 - \tau F_2$, $G_M = F_1+ F_2$.
\item $G_A$ and $G_P$ are the axial vector and pseudoscalar form factors, respectively.
\end{itemize}
The conserved 
vector current (CVC) hypothesis equates the vector form factors $F_1$ and $F_2$ in electromagnetic
interactions to the corresponding form factors in the weak interaction. Note that the second
term in Equation \ref{eq:ccqe_dsdqsquared} switches sign depending on whether neutrino or
antineutrino CCQE scattering is under consideration.

Since CVC allows vector form factors measured in electron scattering to be directly applied
to the corresponding neutrino scattering processes, the study of the CCQE process has
historically focused on the axial structure of the nucleon encapsulated in the axial form 
factor. To this end, the axial vector form factor is typically described by the ``dipole''
form:
\begin{equation}
\label{eq:dipole}
G_A(Q^2) = \frac{g_A}{1+Q^2/M_A^2}
\end{equation}
with two empirically determined parameters:
\begin{itemize}
\item $g_A$: a normalization factor that sets the form factor at $Q^2=0$ determined by neutron $\beta$ decay.
\item $M_A$: the so-called ``axial mass'' parameter that characterizes the $Q^2$ dependence of the form factor.
\end{itemize}

Studies of the CCQE process were performed in deuterium bubble chambers
~\cite{Mann:1973pr,Barish:1977qk,Baker:1981su,Kitagaki:1983px,Asratyan}
where the process: 
\begin{equation}
\nu + d \to \ell^- + p + p_s
\end{equation}
can be examined with minimal effects arising from nuclear physics, and all the outgoing particles, including the spectator nucleon $p_s$, can be identified and reconstructed, leading to high efficiency and high purity selections of CCQE events with over-constrained kinematic reconstruction. 
These measurements focused on constraining the axial vector contributions to the 
CCQE cross section using the dipole parametrization (Equation \ref{eq:dipole}), leading to measurements of the axial mass with $M_A=1.026\pm 0.021~\mbox{GeV}$~\cite{Bernard:2001rs}. 

\subsection{Neutrino Oscillations} 
In addition to its role in studying the axial structure of the nucleon, the CCQE process
plays an essential role in many accelerator-based studies of neutrino oscillations. A full
discussion of neutrino oscillations is beyond the scope of this discussion (see, for instance,
the review in Reference \cite{Beringer:1900zz} for more details); here we discuss the most
relevant points in connection to our discussion of the CCQE interaction.

Neutrino oscillations arise from the non-trivial unitary relation between the flavor
eigenstates of the neutrino, which determine their weak interaction properties, 
in particular which charge lepton  results from the interaction of a neutrino, 
and their mass eigenstates. This gives rise to flavor change
as the neutrino propagates in space-time, with the transition probability determined by the
elements of the unitary mixing matrix relating the flavor and mass eigenstates and a sinusoidal
dependence in $L/E$ (distance $L$ traveled by the neutrino divided by its energy $E$) with frequencies set by the
mass eigenvalue differences.

Following their discovery in atmospheric neutrino and solar neutrino data, neutrino 
oscillations have been studied in experiments using accelerators and reactors as neutrino 
sources. Recently, with various oscillation modes well-established, measurements of the 
detailed $L/E$ dependence of the oscillation pattern have been performed with an aim 
towards precise extraction of the oscillation parameters and to search for 
any deviations that may signal non-standard physics in the process. In accelerator-based experiments, 
where neutrinos oscillations are typically studied with a well-defined baseline $L$ between 
the production of the neutrinos and their detection, this requires the determination of the 
neutrino energy based on the outgoing particles in the neutrino interaction. 

CCQE interactions have played an essential role in both past and current neutrino
oscillation experiments for several reasons:
\begin{itemize}
\item At $\sim 1\;\mbox{GeV}$, the CCQE process has the largest cross section of any 
neutrino interaction channel, and thus will represent a large fraction of the interactions 
for experiments using neutrino beams in this energy range.
\item It has a simple topology with a minimal predicted number of outgoing particles, simplifying
identification and reconstruction, including tagging the charged lepton to determine the flavor of the incident neutrino.
\item Due to the quasi-two body kinematics of the interaction, the neutrino energy can be 
inferred from the lepton kinematics alone. Thus, detectors can still determine the neutrino 
energy without fully reconstructing the recoiling nucleons if they can identify the 
interaction as CCQE. For neutrino CCQE scattering, assuming the target neutron to be at 
rest:
\begin{equation}
\label{eq:enuqe}
E^{QE}_\nu=\frac{m_p^2 - (m_n-E_b)^2 -m_\ell^2 + 2(m_n-E_b)E_\ell}{2(m_n-E_b-E_\mu + 
|\bf{p}_\ell|\cos\theta_\ell)}
\end{equation}
where $m_p$, $m_n$, and $m_\ell$ are the mass of the proton, the neutron, and outgoing 
lepton, respectively, $E_b$ is the binding energy of the target neutron, and $E_\ell$, 
$\bf{p}_\ell$, and $\cos\theta_\ell$ are the total energy, (3-)momentum and angle relative 
to the incident neutrino of the outgoing lepton. The role of the neutron and proton are 
reversed in antineutrino CCQE scattering. The $Q^2$ of the interaction can be estimated 
under the same assumptions:
\begin{equation}
\label{eq:q2qe}
Q^{2}_{QE}= 2E_\nu^{QE} \left(E_\ell - |{\bf p}_\ell| \cos\theta_\ell\right)-m_\ell^2
\end{equation}
\end{itemize}

\begin{figure}[t]
\centering
\includegraphics[width=10 cm]{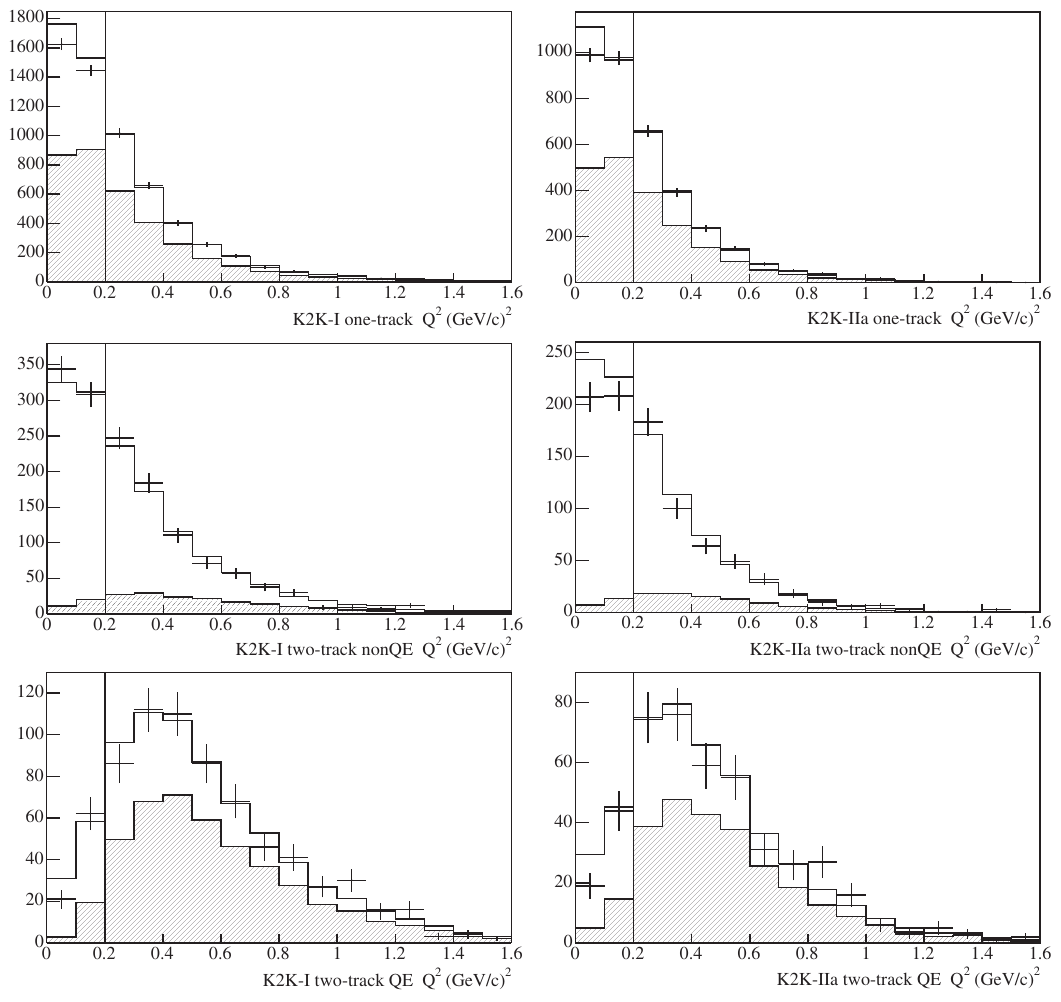}
\caption{\label{fig:k2k_dsdq2}The $d\sigma/dQ_{QE}^2$ distributions of $\nu_\mu$ interactions in water  observed in two data-taking periods (left/right) of the K2K SciFi detector on a water target. The data are from a one-track sample with no recoil proton (top), a two-track non-CCQE control sample (middle), and a two-track CCQE-enhanced sample (bottom). The open histogram represents the best fit to the shape of these distributions, leading to $M_A=1.20\pm 0.12\;\mbox{GeV}$, where the hashed distribution shows the expected CCQE contribution. The data to the left of the vertical line is excluded from the fit \cite{Gran:2006jn}.}
\end{figure}

\begin{figure}[t]
\centering
\includegraphics[height=8 cm]{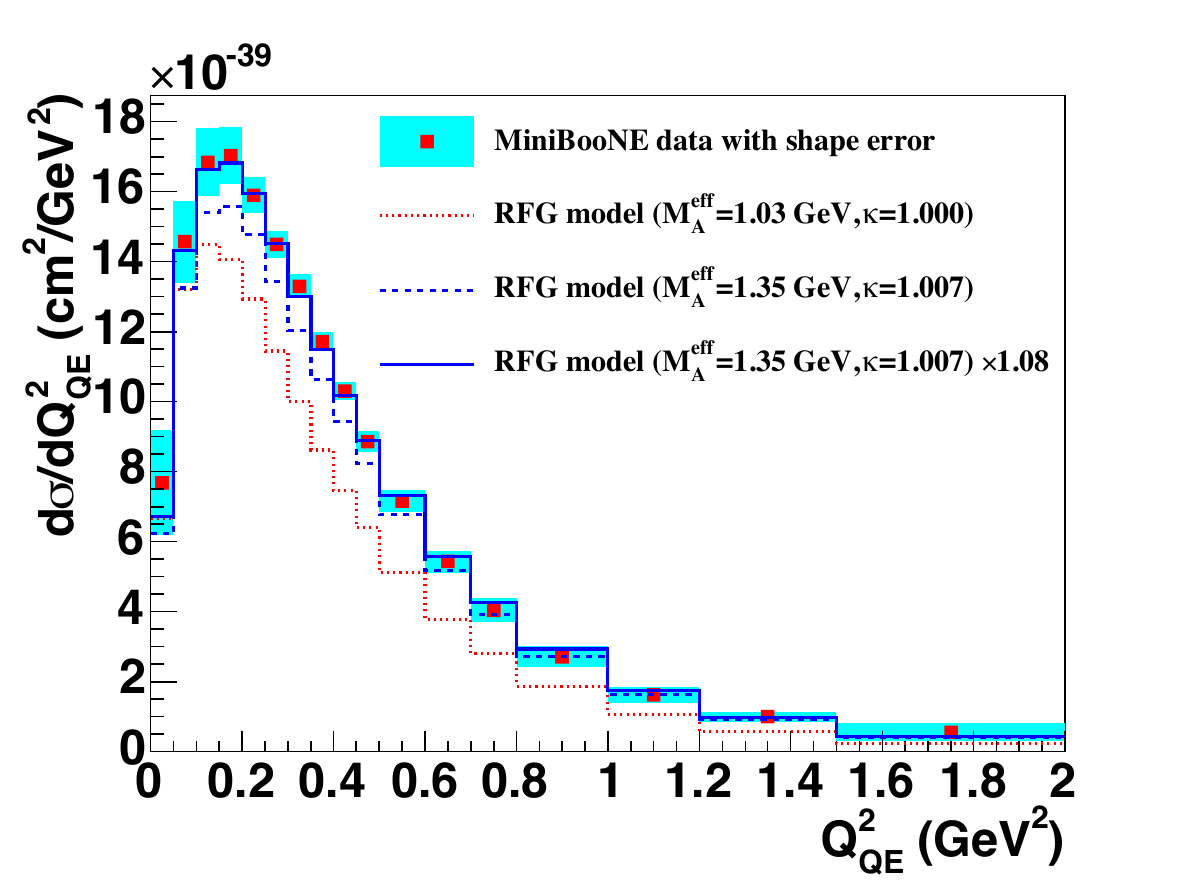}
\caption{\label{fig:mb_dsdq2}
 The $d\sigma/dQ_{QE}^2$ distribution for $\nu_\mu$ CCQE interactions on carbon observed at MiniBooNE. The data points are shown as red points, where the cyan bands represent the systematic uncertainties excluding a 12\% overall normalization uncertainty. The Monte Carlo predictions are shown based on an RFG model with the $M_A=1.35\;\mbox{GeV}$) obtined from a shape fit scaled overall by a factor of 1.08\% (solid), best fit $M_A$ without scaling (dashed), and with $M_A=1.03\;\mbox{GeV}$ and no scaling (dotted)\cite{AguilarArevalo:2007ab}.}
\end{figure}

Due to these properties, CCQE has been the primary channel for neutrino oscillation experiments using accelerator-based beams of $\sim 1\;\mbox{GeV}$. 
Note that the binding of the target neutron is accounted only through an average binding energy and its Fermi motion is ignored, and that Equations \ref{eq:enuqe} and \ref{eq:q2qe} carry an inherent assumption that the interaction can be understood as a two-body process.  
We will see that a number of nuclear effects greatly complicate the implications of employing this simple prescription. 

Current and future experiments will continue to use the CCQE channel for more precise measurements of neutrino oscillation parameters and to look for new phenomena such as $CP$-violating effects and non-standard interactions. Thus, the demands on our understanding of the CCQE process and its modeling will continue to grow as the goals of the accelerator-based studies of neutrino oscillations become more ambitious. 
 
\subsection{Recent studies of CCQE scattering on medium and heavy nuclei}
Over the past decade or so, CCQE interactions have been mainly studied on heavier nuclear 
targets such as those used in detectors studying neutrino oscillations. These targets usually take 
the form of hydrocarbons (plastic or liquid scintillator), water, or iron. The neutrino-nucleus CCQE interaction has typically been modeled using the Relativistic Fermi Gas (RFG) model, which 
models the nucleons as a degenerate Fermi gas with Fermi momentum $k_F$ and a fixed average 
binding/separation energy~\cite{Casper:2002sd,Gallagher:2002sf,Hayato:2009zz,Andreopoulos:2009rq}. The impulse approximation is also usually employed, where the 
yield is treated as the incoherent sum of the scattering off the individual nucleons.  

With the accumulation of higher statistics CCQE samples, recent neutrino experiments first repeated
the historical measurements and examined the role of the axial-vector contribution in these
interactions.
Using the dipole parameterization for the axial form factor, these experiments have measured  
axial masses values ($\sim 1.2-1.4\;\mbox{GeV}$) that are systematically higher than those 
obtained from deuterium targets \cite{Gran:2006jn,AguilarArevalo:2007ab}. Examples of the $d\sigma/dQ^2_{QE}$ distributions from the K2K and MiniBooNE experiments leading to higher $M_A$ values are shown in Figures \ref{fig:k2k_dsdq2} and \ref{fig:mb_dsdq2}.
The interpretation of this discrepancy was complicated by
the large nuclear effects present in these interactions and the perceived difficulties in predicting the neutrino flux. The purities for these data samples are also lower due to the inability to 
reconstruct all the final state particles and topologically irreducible backgrounds arising 
from processes other than CCQE, such as pion production processes where the pion is absorbed 
in the target nucleus and hence is unobservable.

Different values of $M_A$ result in different expected $Q^2$ distributions for the CCQE 
interactions, with larger $M_A$ leading to more events at higher $Q^2$ and therefore a 
``harder'' $Q^2$ distribution. Larger $M_A$ values also result in higher overall  cross 
sections for the CCQE process. Due to the large uncertainties and difficulties in predicting 
neutrino fluxes from accelerator-based beams, particularly in the absolute fluxes necessary for 
absolute cross sections, CCQE analyses have usually used the observed
$Q^2$ distribution rather than the absolute cross section in extracting $M_A$, though  
attempts were made to check the consistency between the two observables~\cite{Barish:1977qk}. We
refer to such measurements as ``shape'' fits or analyses to distinguish them from those which measure absolute cross sections.

It was clear that the CCQE kinematic distributions observed in heavier nuclei did not agree with inferences based on results from deuterium.  These differing kinematic distributions could be accommodated by increasing the value of $M_A$ without regard for the consequences on the resulting cross section. It was noted in the K2K analysis that neither the parameters of the RFG (the Fermi momentum and mean binding energy), nor a more accurate treatment of the vector form factor, could accommodate the observed data without changing $M_A$.


It was not until 2007 that CCQE interactions were revisited with an aim towards determining the absolute cross sections in addition to their differential behavior. Analysis of CCQE interactions at  MiniBooNE with energies around $\sim 1\;\mbox{GeV}$ on carbon indicated that the absolute differential cross sections were consistent not only with the harder $Q^2$ distribution implied by higher values of $M_A\sim 1.35\;\mbox{GeV}$, but also with the absolute rate~\cite{AguilarArevalo:2007ab}. At around the same time, NOMAD reported measurements of CCQE interactions on a carbon target at higher energies ($>3~\mbox{GeV}$) on carbon, which resulted in $M_A$ values closer to those obtained on deuterium $(1.06\pm 0.06 \;\mbox{GeV}/c^2)$~\cite{Lyubushkin:2008pe}. Details on these results, as well as other recent experimental developments in CCQE measurements, are presented in Section \ref{sec:experiments}. The nature of what appears to be tension or disagreement between these two measurements on similar targets, but at different energies and very different detection methods, remains under discussion.


The larger CCQE cross sections measured by MiniBoonE led to the recollection of two past developments that identified potential mechanisms for higher cross sections. First were calculations performed using the Random Phase Approximation (RPA) framework which predicted additional interactions channels corresponding to multi-nucleon knockout (often referred to as ``$np-nh$'')~\cite{Marteau:1999jp}. These would give rise to pionless final states that would be classified as ``CCQE'' in the topological classification used at MiniBooNE, thus leading to an excess of such events that could be interpreted as a higher absolute cross section. The RPA-based calculations were revisited in response to the MiniBooNE CCQE results and found to predict an additional $np-nh$ contribution that could explain the excess of ``CCQE'' events observed at MiniBooNE.

Second were large enhancements in the transverse electromagnetic response function extracted from inclusive electron-nucleus quasi-elastic scattering data, particularly in light nuclei such as $\;^{3}\mbox{He}$, $\;^{4}\mbox{He}$, and $\;^{6}\mbox{Li}$, relative to the longitudinal response~\cite{Carlson:2001mp}. Calculations of the response functions based on the Green's Function Monte Carlo (GFMC) technique reproduced the transverse response enhancement and attributed their source  to nucleon-nucleon interactions that directly produce two-body hadronic weak currents. These interactions are absent in the RFG and other models, leading them to predict equal transverse and longitudinal responses.


Thus, two very different theoretical approaches to studying lepton-nucleus scattering predict nuclear effects in both electron and neutrino scattering that could lead to significant enhancements in the observed cross sections for what is identified as ``CCQE'' (in particular pionless final states). At NOMAD, where a more detailed analysis of the recoiling proton is performed, the implications are less clear. 

This interpretation of the excess of ``CCQE'' events at MiniBooNE has potentially significant implications for neutrino oscillation experiments using the CCQE process. This extra contribution, particularly in the RPA-based $np-nh$ model, would differ kinematically in the relation between the incoming neutrino energy and the outgoing lepton distribution from Equation \ref{eq:ccqe_process}. Thus, the neutrino energy inferred from the observed lepton kinematics assuming Equation \ref{eq:ccqe_process} as the underlying process will be incorrect for these events, and the measured $L/E$ distribution of the observed oscillations will be incorrectly interpreted (see, for example,~\cite{Martini:2012fa, Martini:2012uc, Nieves:2012yz, Lalakulich:2012hs,Mosel:2013fxa}). A detailed understanding of this situation is needed to account for the presence of these events and to constrain any systematic uncertainties that may result from uncertainties in modeling their effects.

In the following sections, we review recent experimental developments in the study of neutrino CCQE scattering (Section \ref{sec:experiments}) with particular attention to analysis methods and results that elucidate potential issues in modeling the CCQE process
and identify experimental signatures of the relevant nuclear effects. This is followed by
experimental and theoretical developments in electron-nucleus scattering (Section \ref{sec:electron}). The status of theoretical developments in CCQE interactions, informed by both neutrino and electron scattering is presented in Section \ref{sec:theory}. Following a brief discussion on the important issue of inferring the neutrino energy in CCQE interactions in light of these developments (Section \ref{sec:neutrinoenergy}) and final state interactions (Section \ref{sec:fsi}), we turn to the topic of single photon production in neutrino interactions (Section \ref{sec:photon}).

\section{ Neutrino-Nucleus Experiments: Summary of Current Measurements}
\label{sec:experiments}
In this section, we review the experimental results on neutrino CCQE scattering presented at the workshop. We start with a brief introduction to the important issue of predicting the neutrino flux before visiting each experiment in turn, and conclude with some discussion and observations that were made at the workshop.

\subsection{Neutrino Flux}

\begin{figure}[t]
\centering
\includegraphics[height=4 cm]{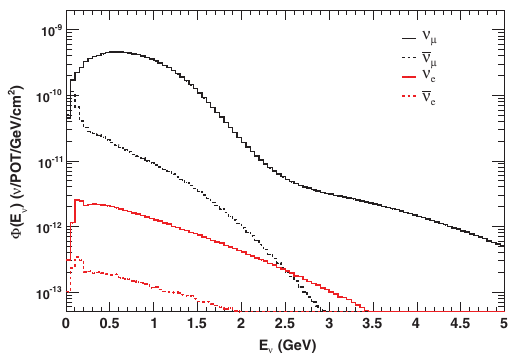}
\includegraphics[height=4.7 cm]{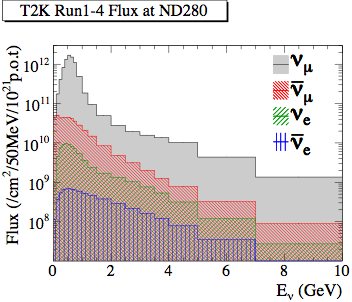}
\caption{\label{fig:nuflux_mbt2k}Predicted neutrino fluxes from the Booster Neutrino Beam line at MiniBooNE~\cite{AguilarArevalo:2008yp} (left) and the T2K beam line at the ND280 near detector (right)\cite{Abe:2012av}. Note that T2K uses an ``off-axis'' beam configuration in which the beam axis is directed $2.5^\circ$ away from ND280.}
\includegraphics[height=5 cm]{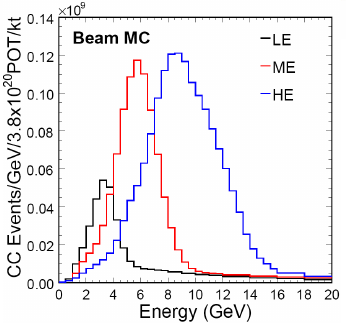}
\includegraphics[height=5 cm]{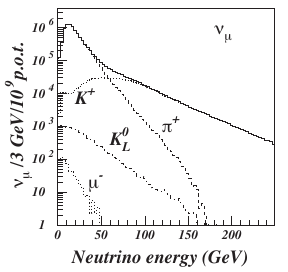}
\caption{\label{fig:nuflux_numinomad}Predicted total neutrino event rates in the NuMI beam line at 1~km (with the target and horns in the original Low (LE), Medium (ME) and High (HE) energy configurations)~\cite{Adamson:2007gu} that supplies beam to MINOS, ArgoNeut, and MINERvA (left) and the WANF beam line at CERN which delivered beam to NOMAD~\cite{Astier:2003rj}(right). Note that the NuMI plot shows the expected $\nu_\mu$ CC event rate versus neutrino energy in the MINOS near detector, and not the neutrino flux.}
\end{figure}

The prediction of the incoming neutrino flux is a critical element in the study of neutrino interactions, but has historically been a challenging endeavor. The overall principle for the production of neutrinos in the experiments discussed here is the same:
\begin{itemize}
\item Primary protons extracted from an accelerator complex impinge on a target, typically consisting of a few interaction lengths of low-$A$ material.  In the experiments discussed at the workshop, the primary proton energies range include 8 GeV (Booster Neutrino Beam at Fermilab), 30~GeV (T2K at J-PARC), 120~GeV (NuMI at Fermilat), and 450~GeV (CERN West Area Neutrino Facility), with beryllium (Booster Neutrino Beam, WANF), carbon (NuMI, T2K), and aluminum (K2K) used as a target material.
\item Secondary particles, primarily pions, produced by the interactions are focused by one or more pulsed electromagnetic devices into a decay region. These devices are usually ``horns'' that produce a toroidal magnetic field. This results in a sign selection that focuses $\pi^+$ and defocuses $\pi^-$, or vice versa.
\item The focused secondary particles enter a decay region to produce neutrinos. The primary decay mode with positive focusing is $\pi^+\to\mu^++\nu_\mu$, resulting in a beam primarily composed of $\nu_\mu$. However, there are also contributions from the decay of kaons produced in the target ($K_{\mu2}$, $K_{\mu3}$, $K_{e3}$) that give rise to higher energy neutrinos and $\nu_e$ in the beam. Muons, resulting from the pion decay, themselves decay ($\mu^+\to e^+ +\bar{\nu}_\mu + \nu_e$) to give an additional source of $\nu_e$ and a ``wrong-sign'' contribution of $\bar{\nu}_\mu$ along with the decay $\pi^-\to\mu^-+\bar\nu_\mu$ arising from $\pi^-$ that reach the decay region. 
Alternatively, by reversing the polarity of the electromagnets, a beam that is enhanced in $\bar\nu_\mu$ can be produced.
\item A beam stop at the end of the decay region absorbs all remaining particles except for muons above a threshold and the neutrinos. Gaps or alcoves within or behind the beam stop region can be instrumented to detect the muons produced in the beam line for monitoring purposes. The muons are eventually absorbed in shielding (or earth) following the beam stop.  
\end{itemize}

The predicted neutrino flux is obtained through detailed Monte Carlo simulations that account for each stage of the process, incorporating precise material and  geometric descriptions of the beam line components, and the various electromagnetic and hadronic interactions of the particles as they propagate through the beam line. The primary sources of uncertainty relate to hadronic processes, starting with the particle production induced by proton interactions in the target. Dedicated experiments \cite{Ambrosini:1999id, Catanesi:2007ab,Abgrall:2011ae} have been carried out to measure particle production at the chosen primary proton energy and target material. These measurements have taken the form of ``thin target'' measurements, where the underlying production cross sections for the proton-nucleus interaction are measured with minimum secondary interaction effects, and ``long'' or ``replica'' target measurements, where particle production from a replica of the target used in the neutrino beam or another target with similar dimensions is measured. In the latter case, the outgoing particles are subject to the additional secondary processes in the target as they are in the actual neutrino beam line.  Additional hadronic uncertainties result from the interaction of these particles on other beam line components, such as  the focusing horns, decay region walls, and beam stop. 

The predicted neutrino fluxes for several recent neutrino beam lines are shown in Figures ~\ref{fig:nuflux_mbt2k} and ~\ref{fig:nuflux_numinomad} respectively. For NuMI, the plot shows not the neutrino flux, but the expected number of $\nu_\mu$ CC interactions in the MINOS near detector, which incorporates the neutrino interaction cross section. In the case of T2K, the experiment has an ``off-axis'' configuration where the beam is directed $2.5^\circ$ away from ND280 and the Super-Kamiokande detector, resulting in a higher yield of low energy neutrinos peaked at 600 MeV, while the NuMI beam line incorporates a variable target-horn geometry system that allows the neutrino spectrum to be varied by focusing different regions of the secondary pion phase space.

Next, we summarize recent studies of CCQE events at various neutrino experiments, including MiniBooNE, SciBooNE, MINERvA, NOMAD, ArgoNeut, T2K, MINOS, and NOvA.

\begin{figure}[t]
\centering
\includegraphics[height=4.1 cm]{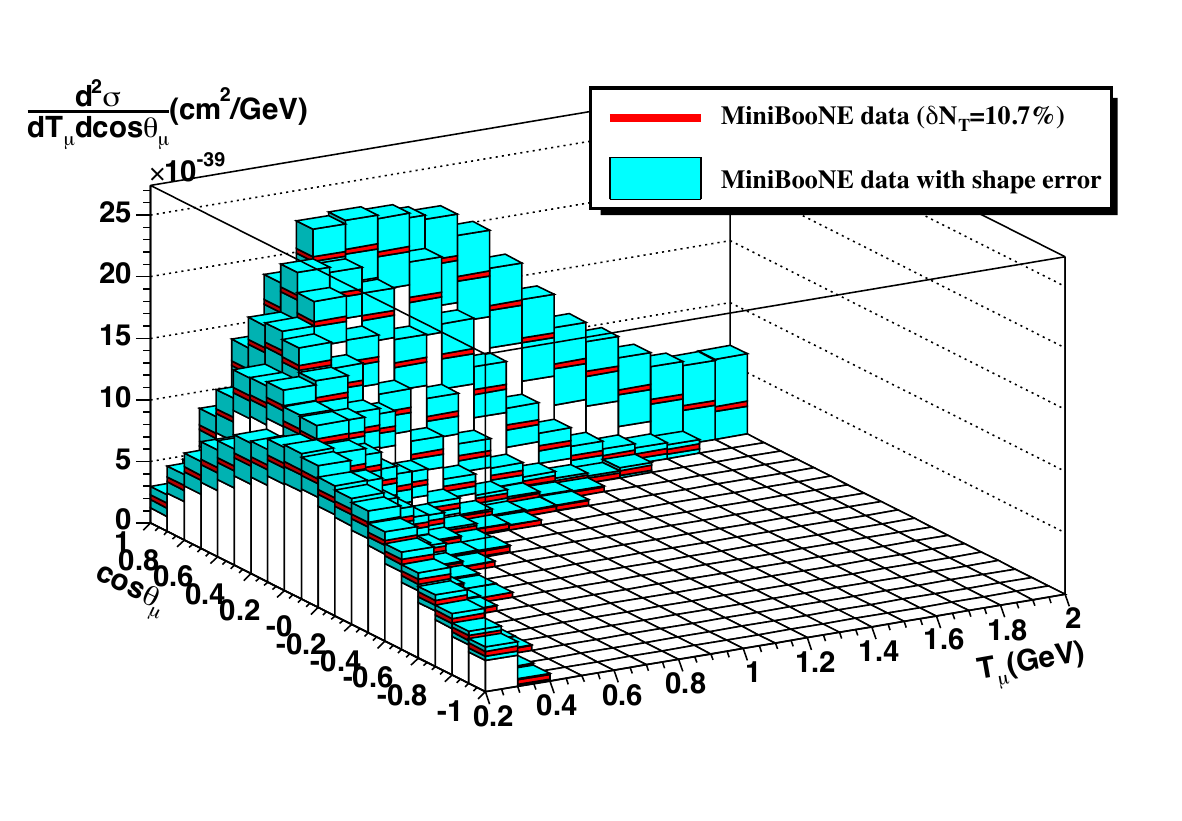}
\includegraphics[height=4.1 cm]{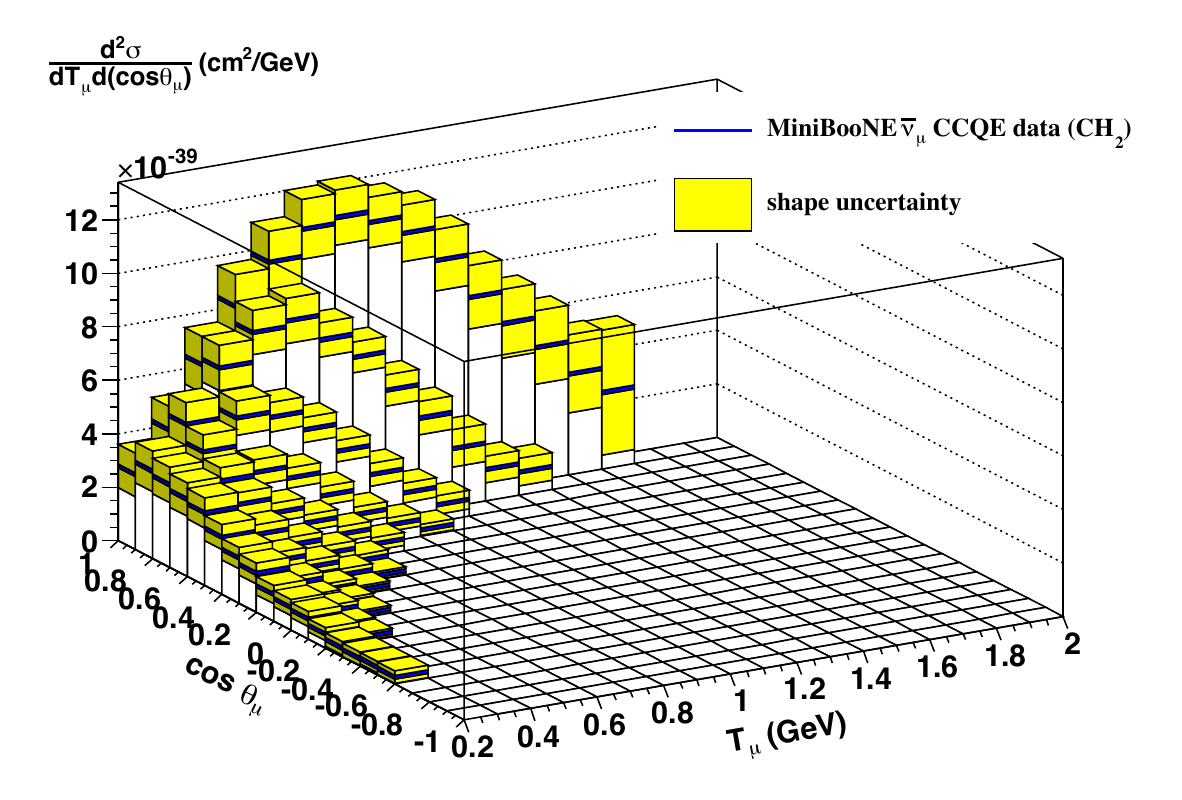}
\caption{\label{fig:dsdtdtheta_mb}Double differential cross sections $d^2\sigma/dT_\mu\;d\cos\theta_\mu$ of the outgoing muon for CCQE events observed in MiniBooNE neutrino (left)~\cite{AguilarArevalo:2010zc} and antineutrino (right)~\cite{AguilarArevalo:2013hm} data, where $T_\mu$ is the kinetic energy of the muon and $\theta_\mu$ is its polar angle relative to the incident neutrino beam. The colored regions indicate the total uncertainty in each bin.}
\end{figure}

\subsection{MiniBooNE}
\label{sec:miniboone}
Geralyn Zeller summarized measurements of the CCQE process from MiniBooNE. MiniBooNE was built to study the ``short baseline'' neutrino oscillations indicated by the LSND experiment\cite{Aguilar:2001ty}. The experiment utilized the Booster Neutrino Beam line with a 800 ton (445 ton fiducial volume) mineral oil-based Cherenkov detector.
After 10 years of data taking, MiniBooNE has published cross sections for neutrino and antineutrino CCQE scattering using the largest samples of CCQE interactions ever reported to date. This includes the first measurement of the double differential cross section, $d^2\sigma/dT_\mu\;d\cos\theta_\mu$, and  measurement of the differential cross section, $d\sigma/dQ^2_{QE}$ for both neutrino and antineutrino CCQE scattering. For historical comparisons, MiniBooNE also reported the CCQE cross section as a function of reconstructed neutrino energy, $\sigma(E^{QE}_\nu)$, and a value of $M_A$, although these measurement are now known to be more model dependent and are hence less useful for comparisons. The $d^2\sigma/dT_\mu\;d\cos\theta_\mu$ distribution obtained as MiniBooNE's main measurement is shown in Figure \ref{fig:dsdtdtheta_mb}.

MiniBooNE is a Cherenkov detector and hence makes use of ring imaging for event reconstruction and particle identification. CCQE interactions are selected in the spherically symmetric detector by relying solely on particle decays and the identification of a contained muon-like Cherenkov ring in the fiducial volume. There is no tracking of protons or pions in the event, lessening some of the model dependence. CCQE events are identified as those containing exactly one muon and one Michel decay electron arising from $\mu\to e$ decay. The requirement of one Michel decay electron eliminates backgrounds from events containing a charged pion in the final state, since the charged pion decay will lead to a second Michel electron in the event. With this selection, CCQE events are defined as CC events with no pions and any number of nucleons in the final state. No attempt is made to reconstruct or count the final state proton(s) in these events. The dominant non-QE background in the neutrino (antineutrino) sample are CC $\pi^+$ (CC $\pi^-$) events where the pion is absorbed in the target nucleus. Such events have the same final state as CCQE interactions in the MiniBooNE detector and are part of the sample often called ``CCQE-like'' events. The size of this background is partially constrained by the measured rate of CC events with an observed pion in the final state in the MiniBooNE detector. The contribution from such pion absorption processes is then subtracted before extracting the CCQE cross sections but is reported in the provided cross section tables so that model builders can adjust and/or add back this contribution.

The $\numu$ CCQE sample contains 146,070 events with an estimated $26\%$ efficiency (which includes a fiducial volume cut) and purity of $77\%$. The $\bar\nu_\mu$ sample is selected in a similar way and includes 71,176 events with an estimated $29\%$ efficiency and purity of $61\%$. For the $\bar\nu_\mu$ CCQE analysis, there is an additional added complication of having to subtract neutrino backgrounds from the sample. Because the MiniBooNE detector is not magnetized, this required the development of several novel techniques for measuring and removing ``wrong sign'' neutrino events from the antineutrino CCQE sample, which compose roughly $20\%$ of the events~\cite{AguilarArevalo:2011sz}.

From these samples, measurements of $d^2\sigma/dT_\mu d\cos\theta_\mu$, $d\sigma/dQ^2_{QE}$, $\sigma(E_\nu)$, and $M_A$ over the kinematic range $0.2 < E_\mu < 2$ GeV, $0 < \theta_\mu < 360^\circ$, $0.4 < E_\nu < 2$ GeV were extracted. Note that the $\numu$ CCQE measurements are on $\mbox{C}$ while the $\numubar$ CCQE measurements are on $\mbox{CH}_2$, {\em i.e.} there is an additional free proton contribution in the antineutrino scattering case; however, the carbon-only contribution is reported separately after subtracting off a hydrogen estimate. The MiniBooNE CCQE cross section measurements are dominated by the uncertainty in the predicted fluxes. The integrated flux uncertainty is $\sim10\%$ in both cases after multiple years of work to incorporate improved hadro-production measurements (for example, data taken using a replica of the MiniBooNE proton target at HARP) and beam modeling using a custom GEANT4-based beam simulation~\cite{AguilarArevalo:2008yp}. The flux uncertainties grow to larger than $10\%$ at both the low and high end of the predicted energy spectrum. Note that no use is made of MiniBooNE neutrino data to determine the neutrino flux. This is in part due to the insufficient statistics in ``standard candle'' processes with {\em a priori} well-known cross sections, such as inverse muon decay (IMD) and deep inelastic scattering (DIS), to measure the flux and the desire to not use the same event samples to both measure a flux and a cross section.

With the broader definition of CCQE scattering and expanded kinematic coverage, MiniBooNE observed a substantially ($\sim 40\%$) larger cross section than impulse approximation-based predictions. The effect is larger for larger muon scattering angles, hence revealing the benefit of having a detector with $4\pi$ coverage. These results are what initiated discussions regarding the possible connections between the enhanced cross sections observed in neutrino and electron nucleus scattering.

\begin{figure}[t]
\centering
\parbox{0.49\textwidth}{
\includegraphics[height=4.0 cm]{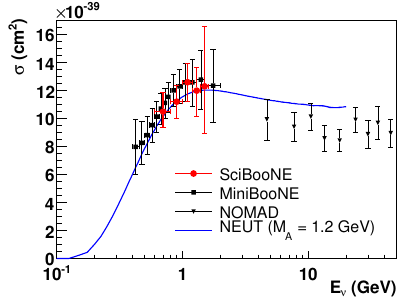}
}
\parbox{0.49\textwidth}{
\includegraphics[height=3.4 cm]{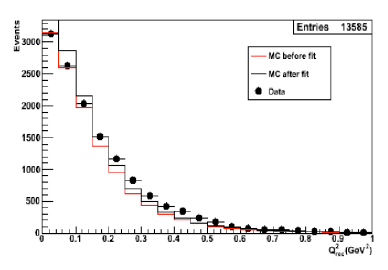}\\
\includegraphics[height=3.4 cm]{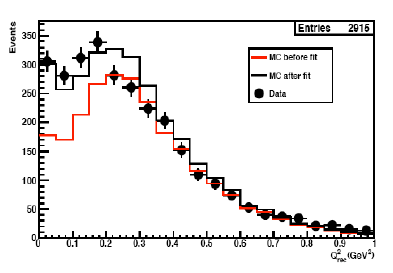}
}
\caption{\label{fig:sciboone}Left: Measured cross section of CCQE events in SciBooNE as a function of $E_\nu^{QE}$ using events with either zero and one proton tracks and no pion tracks ~\cite{Nakajima:2011zz}. Right: Data/MC comparisons of $Q^2_{QE}$ distribution in the 1- (top) and 2-track (top) samples, with the {\texttt neut}-based MC prediction shown before and after a fit to the muon momentum and angule distribution~\cite{AlcarazAunion:2009ku}.}
\end{figure}

\subsection{SciBooNE}
\label{sec:sciboone}
Kendall Mahn presented measurements of $\nu_\mu$ charged current interactions at SciBooNE, where the SciBar scintillating tracking detector was placed in the same Booster Neutrino Beamline (BNB) used by MiniBooNE, affording an opportunity to study neutrino interactions in the BNB with a detector with completely different capabilities. The SciBar detector consists of 32 layers of $1.3 \times 2.5 \times 300~\mbox{cm}^3$ plastic scintillator bars in alternating orientations to enable three dimensional tracking. The scintillator bars are read out with wavelength shifting fibers coupled to 64 channel multianode photomultipliers. In total, there are 14,336 bars with a total mass of 15 tons. A two plane electron ``spaghetti'' calorimeter (EC) consisting of  scintillating fibers and lead foil and a muon range detector (MRD)  with 12 iron plates interspersed with 362 scintillating counters are placed downstream of SciBar detector. The neutrino flux prediction was derived from the same Geant4-based Monte Carlo simulation tuned to HARP and other hadron interaction data used at MiniBooNE.

Event identification at SciBooNE is based on tracks reconstructed in SciBar, with the track range (including the MRD) used to measure momentum. $dE/dx$ measurements based on the deposited charge and the range provide particle identification between muon/pions and protons. Based on this information, single muon track (``1-track $\mu$'') events and two track events consistent with a muon and proton (``2-track $\mu+p$'') are selected to form a CCQE sample and the muon momentum and angle distribution from each sub-sample fit to obtain absolute cross sections as a function of reconstructed neutrino energy.  The measured cross section versus $E^\nu_{QE}$ (Equation \ref{eq:enuqe}) is shown on the left in Figure \ref{fig:sciboone}~\cite{Nakajima:2011zz}, which agrees with the \texttt{neut} prediction using a RFG-based CCQE model with $M_A=1.2\; \mbox{GeV}$. It is also consistent with the MiniBooNE measurements described in Section \ref{sec:miniboone}. A closer look at the data reveals some inconsistencies. The plots on the right of Figure \ref{fig:sciboone} show the $Q^2_{QE}$ distributions observed in the 1-track $\mu$ (top) and 2-track $\mu+p$ samples (bottom)~\cite{AlcarazAunion:2009ku}. The {\texttt neut}-based expectation from the Monte Carlo simulation with the default parameters (red) and following the fit and a retuning of the  the final state interaction model in \texttt{neut} (black) are also shown. While the fit and retuning improve the agreement significantly in some areas, particularly the low $Q^2_{QE}$ region of the 2-track $\mu+p$ sample, the $Q^2_{QE}$ distribution of the 1-track $\mu$ sample exhibits a harder spectrum in data relative to the simulation even after the fit. 
\subsection{MINERvA}
\label{sec:minerva}

The MINERvA experiment measures interactions using a totally active solid scintillator detector, segmented using nestled triangular scintillator extrusions that measure 1.7 cm in height with a 3.3 cm base and are between 2 and 3m in length~\cite{Aliaga:2013uqz}.  The extrusions are oriented in three different views to allow stereoscopic reconstruction of particle tracks.  The MINERvA detector sits just upstream of the MINOS near detector so that muons from MINERvA that enter the magnetized MINOS near detector can be charge- and momentum-analyzed. For MINERvA's first published quasi-elastic cross sections, only events with muons matched to MINOS tracks were used.  The fiducial mass of the MINERvA scintillator is 5.56 tons, although there are also several passive nuclear targets that are in the upstream region of the detector which can be used to study neutrino interactions on different nuclei.  The MINERvA detector sits in the NuMI beam line, which in the low energy configuration produced a broad band $\nu_\mu$ and $\bar\nu_\mu$ beams which peak at 3.5~GeV~\cite{Anderson:1998zza}. MINERvA took data in both $\nu_\mu$ and a $\bar\nu_\mu$ configurations, and with the magnetic field of the MINOS near detector was able to measure both neutrino and antineutrino cross sections with essentially no wrong-sign contamination.  

Gabriel Perdue presented the analysis of neutrino and antineutrino CCQE events at MINERvA. The selection of CCQE events starts by requiring that the muon have the correct sign (negative/positive muons in $\nu_\mu$/$\bar\nu_\mu$ beam data) depending on the horn polarity. The muon momentum is determined based on the distance the muon travels in MINERvA plus the momentum measured 
in the MINOS near detector.  The muon angle at the vertex is determined using the MINERvA tracking.  From these two quantities the neutrino energy ($E_\nu^{QE}$) and momentum transfer squared ($Q^2_{QE}$) can be approximated using the CCQE hypothesis (Equations \ref{eq:enuqe} and \ref{eq:q2qe}), assuming $E_b=34(30)$ MeV for neutrinos (antineutrinos).  

To study any remaining hadronic energy in the event, MINERvA divides the energy deposited in the scintillator into two categories. The {\em recoil energy} includes energy depositions not associated with the muon that occur at least 30~cm (10~cm) away from the muon track vertex in neutrino (antineutrino) mode.  The {\em vertex energy} includes energy depositions that are less than that distance from the muon track vertex.  In order to eliminate potential bias due to mismodeled nuclear effects that could cause extra activity near the vertex, MINERvA makes no requirement on vertex energy, but selects CCQE events by making a $Q^2_{QE}$-dependent cut on the recoil energy designed to maintain $95\%$ efficiency in selecting CCQE event at each $Q^2_{QE}$. In order to subtract the non-CCQE backgrounds from the CCQE sample, a fit to the sideband regions is done in bins of $Q^2$.  The fits reduces the non-CCQE backgrounds relative to the default predictions by 5\% for $Q^2_{QE}<0.8\;\mbox{GeV}^2/c^2$ and 15\% for $Q^2_{QE}=[0.8,\;2.0]\;\mbox{GeV}^2/c^2$.  The dominant background is from resonance production.  The selected sample contains 16,467 (29,620) events in antineutrino (neutrino) mode with an expected purity of CCQE events of 77\%(49\%).

\begin{figure}[t]
\centering
\includegraphics[height=4.2 cm]{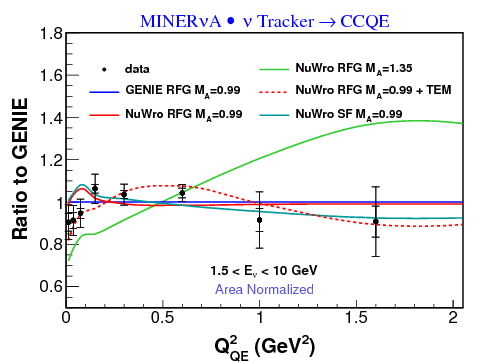}
\includegraphics[height=4.2 cm]{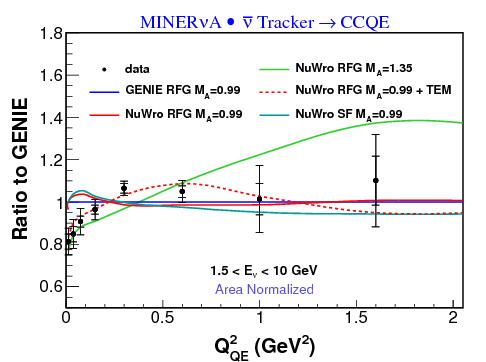}
\caption{\label{fig:minerva_ccqe}Ratio of data and predicted CCQE cross sections as a functin of $Q^2_{QE}$ to a GENIE prediction with a RFG prescription with $M_A=0.99\;\mbox{GeV}$, where all predictions are normalized to the data, for neutrinos (left)~\cite{Fiorentini:2013ezn} and antineutrinos (right)~\cite{Fields:2013zhk}.}
\end{figure}

The resulting sample is unfolded and corrected for detector efficiency, and the extracted cross sections as a function of $Q^2_{QE}$ are compared with several different models for both neutrino~\cite{Fiorentini:2013ezn} and antineutrino~\cite{Fields:2013zhk} CCQE interactions in Fig.~\ref{fig:minerva_ccqe}.  The largest systematic error is from the flux and is largely constant over the different $Q^2_{QE}$ regions. There is more discriminating power in the measurement by comparing the assigned $Q^2$ dependence of different models to the data.  In particular, the MINERvA data are compared to two models which can both describe the enhanced cross section observed in the MiniBooNE CCQE data: one which includes an enhanced transverse cross section constrained by electron-nucleus data (red dashed line in Figure 7) and the RFG model with $M_A=1.35$~GeV (green line in Figure 7) . Both of these models can describe the MiniBooNE CCQE data in shape and rate, but MiniBooNE is unable to distinguish between the two choices given the incident neutrino beam energy and detector technology. This issue is very important to resolve for neutrino oscillation experiments given that the two effects (whether one changes the transverse response or axial form factor) have very different consequences on the energy dependence of the CCQE cross section, the determination of $E_\nu$, and the final state particle composition. With its higher beam energy and added detector capability, MINERvA is able to resolve this issue. The MINERvA data more strongly favors an increase in the transverse response than an increased axial mass. This highlights the power of having data at multiple beam energies: while enhancements in the transverse cross section and axial mass have similar effects at 1 GeV, this is not the case at higher neutrino energies where the two effects can be differentiated. This has been nicely demonstrated in the case of the MINERvA CCQE data and the conclusions seem to hold for both neutrino and antineutrino scattering.

An additional test of the models comes from examining the energy around the vertex for both neutrino and antineutrino CCQE candidates.  The neutrino data show additional energy around the vertex in neutrino mode, consistent with an extra proton of momentum below 200 MeV being emitted, $29\pm9$\% of the time.  In contrast, the antineutrino data show no such excess, and a fit to that distribution would suggest a $-9\pm9$\% deficit of energy around the vertex.  This is in agreement with a signature from correlated neutron-proton pairs in the nucleus:  neutrino CCQE scattering would tend to eject two protons from the nucleus, while antineutrino CCQE scattering would tend to eject two neutrons from the nucleus.  

Since the time of the INT workshop, MINERvA has released an additional CCQE analysis, whose results have been submitted for publication.  
This new analysis measures the hadronic side of the interaction by considering those events that have an outgoing muon and an outgoing tracked proton.  The muon in this case is not required to enter MINOS, since the kinematics of the event are determined by the proton kinetic energy rather than by the muon momentum and angle.  This analysis finds that the model that best describes the hadron kinematics of these CCQE-like events is in fact not the model that best describes the muon kinematics described above~\cite{Walton:2014esl}.  This could be from either a mis-modeling of the inelastic component of the sample (which consists of pions that are absorbed inside the nucleus), or a mis-modeling of the final state interactions of the leading proton.

\begin{figure}[t]
\centering
\parbox[b]{0.44\textwidth}{
\includegraphics[height=5.0 cm]{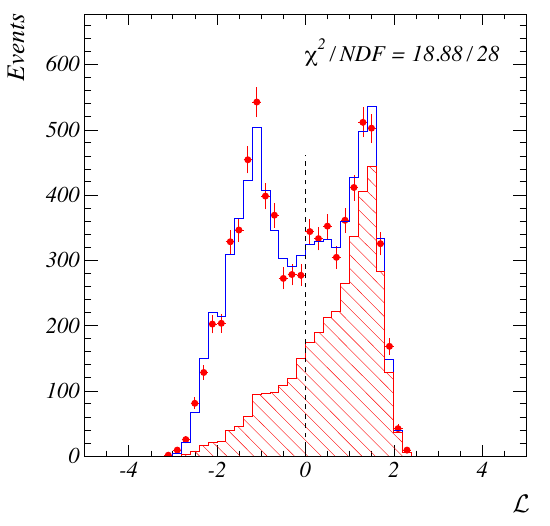}
}
\parbox[b]{0.55\textwidth}{
\includegraphics[width=6.0 cm]{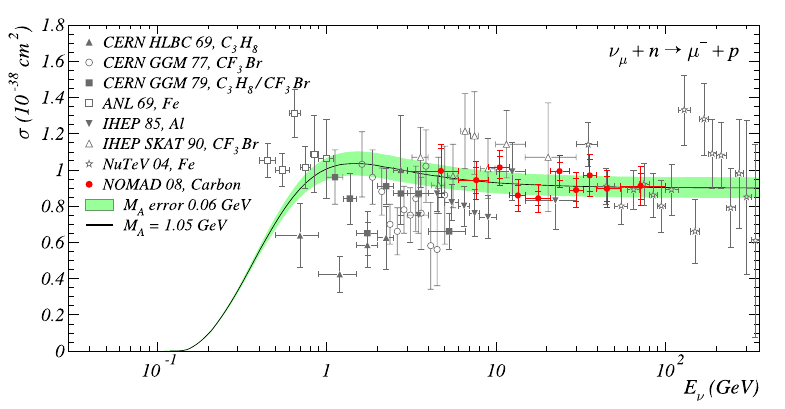}\\
\includegraphics[width=6.0 cm]{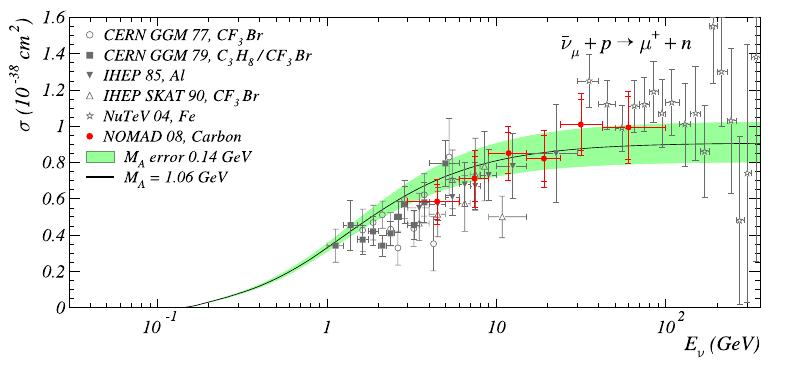}
}
\caption{\label{fig:nomadccqe}Likelihood variable used by NOMAD to select two-track CCQE events based on kinematic consistency of the reconstructed $\mu+p$ system with a two-body CCQE interaction. Right: Measured cross sections as a function of $E^{QE}_{\nu}$ for $\nu_\mu$ (top) and $\bar\nu_\mu$ CCQE interactions\cite{Lyubushkin:2008pe}.}
\end{figure}

\subsection{NOMAD}
The main goal of the NOMAD experiment was to search for $\nu_\mu\to\nu_\tau$ oscillations in the wide band CERN SPS neutrino beam, but despite having concluded its data-taking in 1998, the experiment continues to extract relevant measurements of neutrino-nucleus interactions. These measurements are of particular interest because they include a large span of neutrino energies ranging from 2.5 to 300 GeV~\cite{Astier:2003rj}. The NOMAD detector consisted of an active target containing 44 low-$Z$ drift chambers with a fiducial mass of 2.7 tons located in a 0.4 T magnetic field. Neutrino interactions were predominantly on carbon (the NOMAD detector was $64\%$ carbon, $22\%$ oxygen, $6\%$ nitrogen, $5\%$ hydrogen, $1.7\%$ aluminum). The magnetic field allowed the charge and momentum reconstructin of the final state muons and other charged tracks emerging from an interaction. For cross section analyses, the NOMAD neutrino and antineutrino fluxes were constrained to $\sim2\%$ (integrated) using a combination of hadroproduction data from SPY/NA56 and inverse muon decay and deep inelastic scattering events observed in NOMAD.

Roberto Petti presented NOMAD results on neutrino and antineutrino CCQE scattering published in 2009 ~\cite{Lyubushkin:2008pe}. Both 1- and 2-track CCQE events were selected for analysis. Much like the SciBooNE CCQE analysis, the 1-track CCQE sample includes events with one identified muon and no other reconstructed tracks, while the 2-track CCQE sample are events with one identified muon and one identified proton with a proton detection threshold of $p_p \sim 200\;\mbox{MeV}/c$. Muon tracks could be reconstructed for $0 < |\theta_\mu| < 180^\circ$ and charge-identified for $0 < |\theta_\mu| < 50^\circ$, $E_\mu>2$  GeV. After selection, the $\numu$ 
sample included 14,021 CCQE (1-track $+$ 2-track) events with an estimated efficiency of $34\%$ and purity of $50\%$. Similarly, the $\numubar$ sample included 2,237 CCQE (1-track) events with an estimated efficiency of $64\%$ and purity of $38\%$. From these samples,
NOMAD reported measurements of both the absolute CCQE cross section:

\begin{equation}
\begin{array}{lcl}
\sigma^{\nu}_{CCQE}    & = & [0.92 \pm 0.02\;\mbox{(stat)} \pm 0.06\;\mbox{(syst)}] \times 10^{-38}\;\mbox{cm}^2  \\
\sigma^{\nubar}_{CCQE} & = & [0.81 \pm 0.05\;\mbox{(stat)} \pm 0.08\;\mbox{(syst)}] \times 10^{-38}\;\mbox{cm}^2  \\
\end{array}
\end{equation}
the axial mass ($M_A$):
\begin{equation}
\begin{array}{lcl}
M_A^{\nu}            & = & 1.06 \pm 0.02\;\mbox{(stat)} \pm 0.06\;\mbox{(syst)}\;\mbox{GeV} \\
M_A^{\nubar}         & = & 1.06 \pm 0.07\;\mbox{(stat)} \pm 0.10\;\mbox{(syst)}\; \mbox{GeV}
\end{array}
\end{equation}
\noindent
and measurements of the CCQE QE cross section as a function of neutrino energy over the range $3<E_\nu<100$ GeV. At the INT workshop, NOMAD additionally reported an updated QE analysis using improved selection, a larger fiducial volume, and the complete kinematic range $0 < \phi_\mu < 2\pi$. From a combined 1 and 2-track sample now containing 2.5 times more events than the prior analysis, NOMAD reports:

\begin{equation}
\sigma^{\nu}_{QE} = [0.914 \pm 0.013\;\mbox{(stat)} \pm 0.038\;\mbox{(syst)}] \times 10^{-38} \;\mbox{cm}^2  
\end{equation}

where the uncertainty on the total $\numu$ CCQE cross section has been further reduced from $6.9\%$ to $4.4\%$. In the future, NOMAD plans to measure the CCQE differential cross section, $d\sigma/dQ^2$, and perform model-independent studies of nuclear effects from comparison of the 1- and 2-track samples.

\begin{figure}[t]
\centering
\includegraphics[height=3.75 cm]{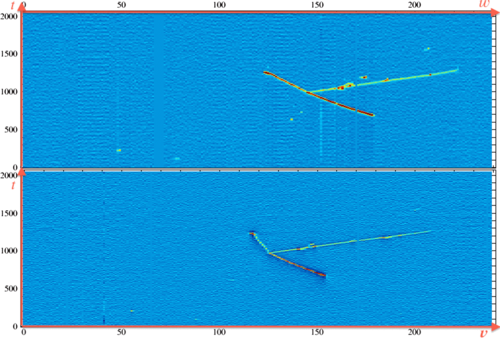}
\includegraphics[height=4 cm]{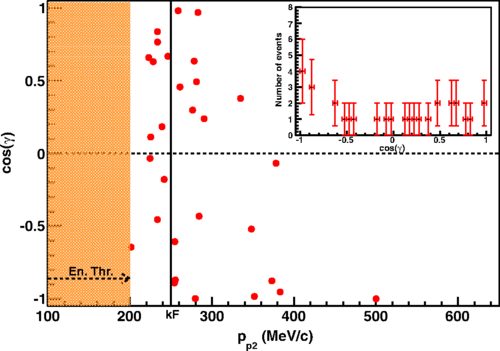}
\caption{\label{fig:argoneut}Left: Event display of a $\nu_\mu$ CC ``hammer'' event in ArgoNeuT with back-to-back protons emerging from the target nucleus. Right: Distribution of $\cos\gamma$ versus $p_{p2}$for the ArgoNeuT CC$0\pi$ sample, where $\gamma$ is the opening-angle of the two protons and $p_{p2}$ is the momentum of the less energetic proton. The inset shows the projection onto $\cos\gamma$~\cite{Acciarri:2014gev}.}
\end{figure}

\subsection{ArgoNeuT}
\label{sec:argoneut}
Ornella Palamara presented results from ArgoNeuT, a 170 liter Liquid Argon Time Projection Chamber (LArTPC) situated in the NuMI beamline at Fermilab just upstream of the MINOS near detector. The bubble chamber-like tracking capabilities of LArTPCs, the low thresholds for proton tracking ($T=21~\mbox{MeV}$), and powerful particle identification capabilities based on $dE/dx$ measurements, allow for a detailed topological classification of interactions according to the outgoing tracks from the interaction. In particular, they allow the further decomposition of the sample of $\nu_\mu$ CC events without outgoing pions (``CC0$\pi$'') into categories based on the number of outgoing protons. With the explicit reconstruction of these outgoing protons, the energy can be estimated by the relation:
\begin{equation}
\label{eq:enuargoneut}
E_\nu = E_\mu + \sum_i T_{p_i} +T_X + E_{miss}
\end{equation}
where $T_{p_i}$ are the kinetic energies of the outgoing protons, $T_X$ is the recoil energy of the residual nucleus estimated from tranverse momentum balance, and $E_{miss}$ is the removal energy of the outgoing nucleons. This definition accounts for the observed energy in the recoil hadronic system and thus relies less on the assumed kinematic configuration of the interaction compared to  $E_\nu^{QE}$ (Equation \ref{eq:enuqe}), which is determined from the muon kinematics alone. Evidence for secondary neutron and/or photon emission can also be observed in the form of electrons and positrons arising from the conversions or scattering of photons emitted either from the residual target nucleus or from secondary interactions of neutrons. ArgoNeuT accumulated approximately 7000 events in short exposures to the NuMI beam in neutrino mode ($8.5\times 10^{18}$ protons-on-target) and antineutrino mode ($1.20\times 10^{20}$ protons-on target), using the MINOS near detector for sign selection and momentum reconstruction of muons exiting downstream of ArgoNeuT.

The detailed event topology information from the ArgoNeuT detector dramatically illustrate evidence for various nuclear effects that were discussed at the workshop. These include $\nu_\mu$ CC interactions in which two protons are seen to emerge back-to-back from the target nucleus that might arise from short-range correlations that bind nucleons at high momenta above the nominal Fermi sea~\cite{Acciarri:2014gev}. Figure \ref{fig:argoneut} shows an event display of such a ``hammer'' event (left) and the opening angle distribution versus momentum of the less energetic proton (right). A cluster of events with wide opening angle ($\cos\gamma\sim-1$) and momentum above the Fermi sea ($k_F=250~\mbox{MeV}/c$) is observed. These distributions can be compared to those observed in the $(e,e'pp)$ interactions shown in Figure \ref{fig:misse_ppcorrelation}.

Detailed model comparisons of the outgoing proton multiplicity with the \texttt{GENIE}~\cite{Andreopoulos:2009rq} and \texttt{GiBUU}~\cite{Buss:2011mx} models were also presented.  While the \texttt{GENIE} model reproduces the low-multiplicity behavior in both neutrino and antineutrino data well, it predicts a significant fraction of events with large ($>5$) proton multiplicity that is not observed in data. \texttt{GiBUU}, on the other hand, seems to model the absence of these large proton multiplicity interactions well. It is noted that the two models predict significantly different compositions for the ``CC0$\pi$'' sample, with \texttt{GiBUU} predicting a contribution of non-CCQE events (50\%) nearly twice that of \texttt{GENIE} (30\%). This represents the first measurement of proton multiplicities in neutrino-nucleus interactions.

\begin{figure}[t]
\centering
\parbox{0.39\textwidth}{
\includegraphics[width=5.0 cm]{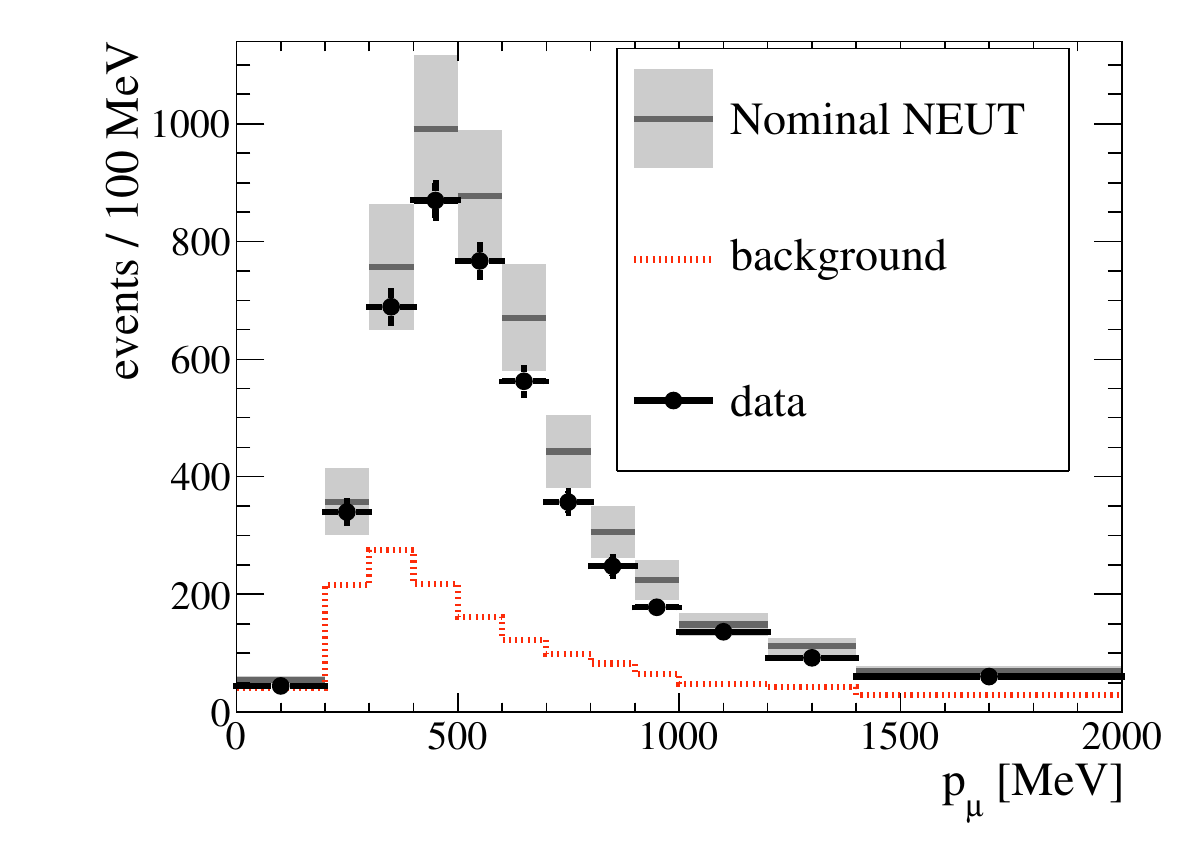}
\includegraphics[width=5.0 cm]{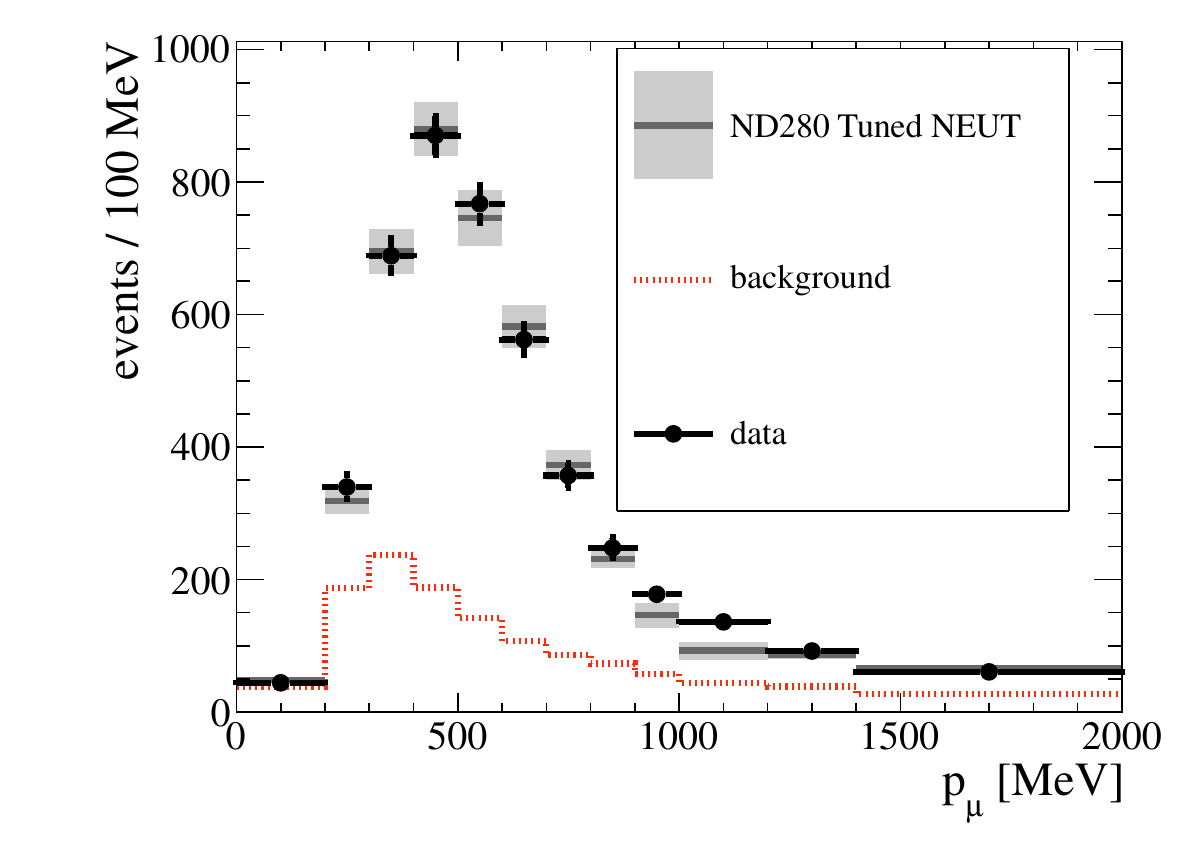}
}
\parbox{0.60\textwidth}{
\includegraphics[width=7.0 cm]{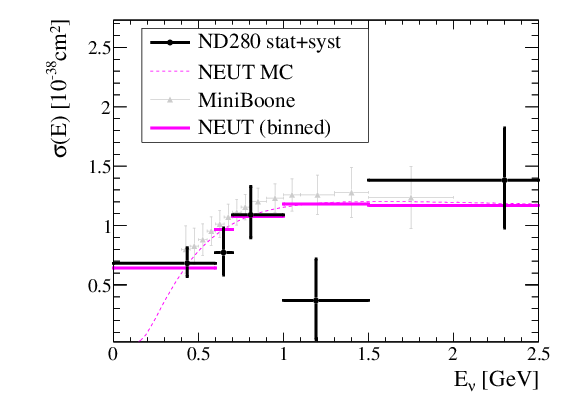}
}
\caption{\label{fig:t2k}Left: The measured muon momentum distribution in CC events with no identified pion in the T2K ND280 tracker compared to the default {\texttt neut} prediction (top) and best fit $M_A$ value (bottom) Right: CCQE cross sections as a function of $E_\nu^{QE}$ compared with the best fit using the RFG-based model in the {\texttt neut} neutrino event generator.} 
\end{figure}

\subsection{T2K}
\label{sec:t2k}
The Tokai-to-Kamioka (T2K) neutrino oscillation experiment~\cite{Abe:2011ks} sends a beam of muon neutrinos produced at the Japan Proton Accelerator Research Center (J-PARC) 295 km across Japan to the Super-Kamiokande (SK) detector~\cite{Fukuda:2002uc} for the study of neutrino oscillations. The experiment is the first to utilize the ``off-axis'' beam configuration~\cite{offaxis}, in which the beam axis is directed away from the detectors. The muon neutrinos result primarily from the decay $\pi\to\mu+\nu_\mu$ of pions produced in the interactions of 30 GeV protons from the J-PARC main ring on a carbon target. The properties of relativistic two-body kinematics result in a beam of $\nu_\mu$ with lower and more narrowly peak energy spectrum as one moves away in angle from the axis of the pion decays. This property is exploited at T2K to give a beam of neutrinos with a narrow spectrum peaked at $\sim 600~\mbox{MeV}$ optimized to maximize neutrino oscillation effects at $L=295~\mbox{km}$. The neutrino flux prediction~\cite{Abe:2012av} is extensively tuned to NA61 particle production meausurements~\cite{Abgrall:2011ae,Abgrall:2011ts} and other hadronic interaction and cross section measurements, as well as {\em in situ} monitors which measure the optics of the incident primary protons and beam direction. The predicted neutrino flux is shown on the left in Figure \ref{fig:t2k}, where uncertainties of $10-15\%$ have been achieved. Continuous improvements in the uncertainty are expected as additional NA61 data is analyzed and incorporated into the prediction.

The T2K near detector complex includes two major systems. The Interactive Neutrino GRid (INGRID) is a set of 16 tracking detectors with scintillator bars planes interspersed with iron plates arranged in a grid spanning the beam axis in horizontal and vertical axes to monitor the beam direction. ND280 is a complex of detectors situated in the off-axis direction towards the SK detector. A tracking detector comprised of a series of fine-grained scintillating detectors (FGDs) and time projection chambers (TPCs) and a dedicated detector for the study of $\pi^0$ production (P0D) are surrounded by a sampling electromagnetic calorimeter (ECAL) consisting of scintillator bars interweaved with lead sheets. These detectors are contained within the UA1 magnet, which provides a 0.2 T magnetic field and is instrumented to detect muons passing through the iron yolk. 

Kendall Mahn reported on studies of $\nu_\mu$ CC events in the tracking detector, where the FGDs provide target mass (primarily $\mbox{CH}$ from plastic scintillator) and tracking of particles emerging from the interaction vertex. For charged tracks entering the TPCs, the magnetic field allows sign selection and momentum reconstruction of charged tracks by measuring their curvature. Measurements of the track ionization yield in the TPC also provide powerful particle identification capabilities. The tracks are matched to hits in the FGDs to form combined FGD-TPC tracks,  while tracks originating from the FGDs that do not enter the TPC are reconstructed from FGD hits only.
Once a negative outgoing muon is identified in the FGD and TPC, the $\nu_\mu$ CC sample is divided into three topological categories based on other particles identified in an event. Events without additional matched tracks, nor with a delayed electron resulting from  the $\pi\to\mu\to e$ decay chain, are classified as ``$\mbox{CC0}\pi$''. Based on additional pion tracks or decay electrons found in the event, the remaining events are categorized as ``$\mbox{CC}1\pi$'' if the topology is consistent with CC interaction with a single $\pi^+$, otherwise it is categorized as ``CC other''. Measurements of the flux-integrated double-differential cross section for the outgoing muon momentum and angle in the inclusive $\nu_\mu$ CC sample have been published~\cite{Abe:2013jth}.

The $\mbox{CC0}\pi$ selection results in a $72\%$ pure CCQE sample with $40\%$ efficiency, with which a CCQE cross section as a function of $E_\nu^{QE}$ is measured and $M_A$ extracted as an effective parameter using the outgoing muon momentum and angle distribution in the RFG-based CCQE model implemented in {\texttt neut} as shown on the right n Figure \ref{fig:t2k}. The cross section as a function of $E^\nu_{QE}$ is found to be in agreement with the nominal {\texttt neut} model. The $M_A$ measurement is performed with and without the absolute normalization, leading to $M_A = 1.43^{+0.28}_{-0.22}\;\mbox{GeV}$ and $1.26^{+0.21}_{-0.18}\;\mbox{GeV}$, respectively. The measurements are consistent with previous reported values of $M_A$ on carbon using similar RFG-based CCQE models with $M_A$ as an effective parameter.
\subsection{MINOS}
The Main Injector Neutrino Oscillation Search (MINOS) studies neutrino oscillations by sending a neutrino beam produced with protons from the FNAL Main Injector (NuMI) 735 km to the Soudan mine in Minnesota~\cite{Michael:2008bc}. 
The MINOS detectors are iron scintillator sampling calorimeter consisting of 282 steel plates that are each 2.5cm thick, interleaved with 1 cm thick planes of scintillator~\cite{Michael:2008bc}. The plates are magnetized so charged particle tracks that traverse enough steel planes can have both momentum and charge measured through curvature.  Momentum measurements through range are also possible for those tracks that end within the active area.  Although the detector mass is 0.98~kton, analyses tend to use a much smaller fiducial mass of 0.03~kton, where the acceptance for events can be easily modeled. MINOS selects charged current events by requiring a reconstructed track, and then the sample is further enriched by using a k-nearest-neighber (kNN) multi-variate technique constructed from several variables that isolates muon tracks from proton or pion tracks.
The analysis achieves a 98\% purity and 95\% efficiency for selecting muons in the final state.   

MINOS determines its incoming neutrino flux with a GEANT-based simulation of the beam line  that has embedded in it a FLUKA-based hadron production model.  That prediction is corrected 
via reweighting based on the inclusive neutrino charged current neutrino energy spectrum measured in several different beam line configurations.  The weights depend on a fit to several different data sets, where the parameters that vary in the fits compensate for imperfect knowledge not only of hadron production, but also of the detector's absolute energy scales (muon and hadronic) and beam line geometry uncertainties.  The neutrino spectrum in the low energy configuration in which the data were taken peaks at 3.5~GeV and extends from roughly 2 to 5 GeV.

Nathan Mayer presented the analysis of CCQE events at MINOS~\cite{Mayer:zqa}. In addition to the muon's momentum, direction, and charge, MINOS reconstruct the total hadronic energy, $E_{had}$.  Event kinematics are reconstructed in two different ways, corresponding to the two kinds of neutrino interactions:  for events identified as CCQE, the neutrino energy and kinematics are reconstructed using a quasi-elastic hypothesis ({\em i.e.} Equations \ref{eq:enuqe} and \ref{eq:q2qe}) with $E_b=34\;\mbox{MeV}$ corresponding to iron.  All other events have the total energy reconstructed $E_\nu=E_\mu+E_{had}$. The hadronic invariant mass, $W$ is reconstructed using the following: 
\begin{equation}
W^2 = m_N^2 + 2m_NE_{had}-Q^2 
\end{equation} 
where $m_N$ is the nucleon mass, and $Q^2$ is determined by the following: 
\begin{equation} 
Q^2 = 2E_\nu(E_\mu - p_\mu \cos (\theta_\mu)) - m_\mu^2
\end{equation}
where the neutrino energy $E_\nu$ is determined either by the quasi-elastic hypothesis ({\em e.g.} $E_\nu^{QE}$) or by summing the hadron and muon energy.  

CCQE events are selected by requiring the hadron energy to be less than 225~MeV, and the muon tracks are required to stop in the detector, which allows a more accurate momentum estimation by range. In addition, three sidebands define a resonance-enhanced (RES) region ($E_{HAD}>250\;\mbox{MeV}$ and $W<1.3\;\mbox{GeV}$), a deep inelastic scattering-enhanced (DIS) region ($W>2\;\mbox{GeV}$), and a RES-DIS transition region ($W=[1.3,\;2.0]\;\mbox{GeV}$).
The models are tuned to reproduce the data in the sidebands so that the backgrounds from RES and DIS can then be subtracted from the signal region.

The predicted $Q^2$ distribution does not match the data in either the RES or the RES-DIS transition regions.  A suppression factor as a function of the true $Q^2$ is applied to simulated resonance events to improve the agreement. The suppression is strongest at low $Q^2$ with a value of 0.46 at $Q^2=0$ and rises quickly to $>0.9$ at $Q^2=0.2$GeV$^2$, and plateaus near unity for $Q^2>0.6\;\mbox{GeV}^2$. An uncertainty on that suppression factor is also determined using the uncertainties in the muon and hadron energy scales, DIS event migration to low $Q^2$ that is not modeled, and uncertainties in final state interactions, axial mass parameters, and the coherent cross section.  
\begin{figure}[t]
\centering
\includegraphics[height=4.0 cm]{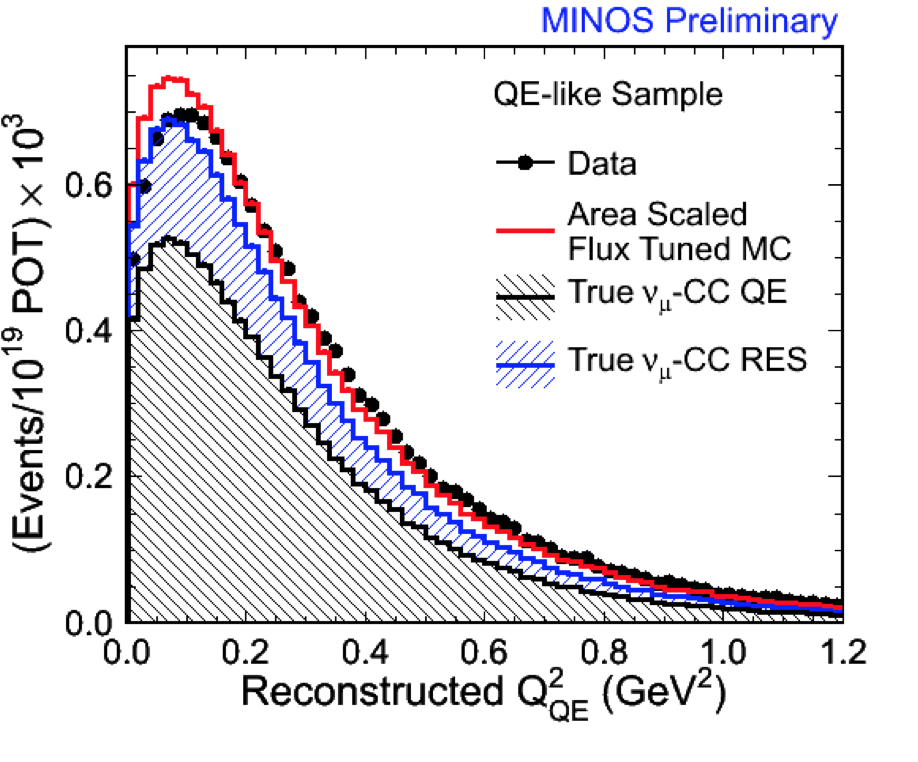}
\includegraphics[height=4.1 cm]{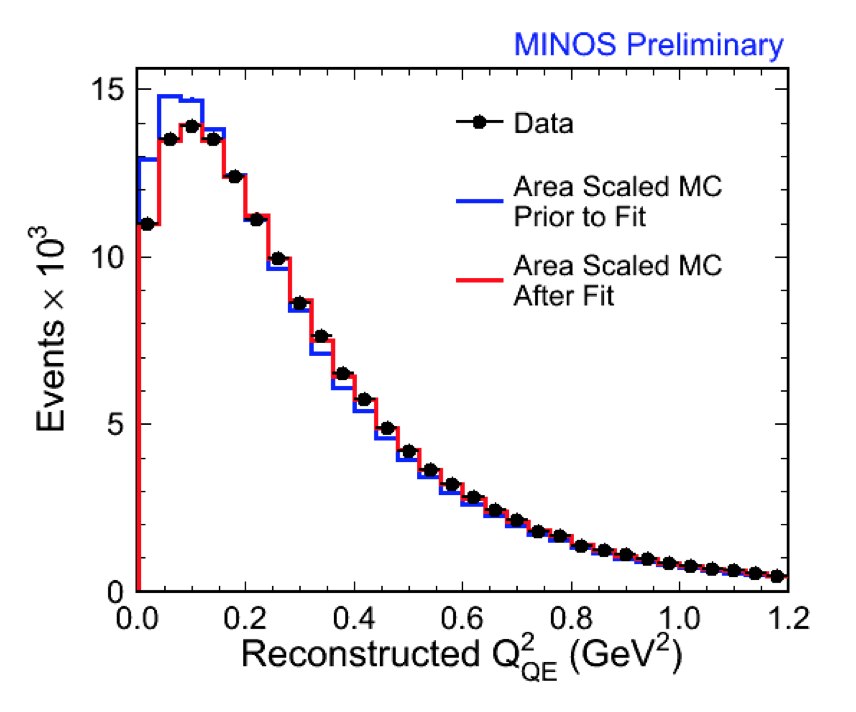}
\caption{\label{fig:minos_ccqe} Left:  $Q^2_{QE}$ distribution for the MINOS CCQE 
candidates before a shape fit with $M_A$ and background tuning.  Right: $Q^2_{QE}$ after a shape fit to a Pauli blocking suppression factor and $M_A$~\cite{Mayer:zqa}.}
\end{figure}

The shape of the predicted $Q^2$ distribution of the CCQE events is compared to the data with the corrected background prediction, as shown in Fig.~\ref{fig:minos_ccqe}.  A fit varying the axial mass parameters for CCQE ($M_A^{QE}$) and resonance production ($M_A^{RES}$), along with the muon and hadron energy scales, and a $k$ factor to modify the Pauli suppression, similar to that used in the initial MiniBooNE CCQE analysis~\cite{AguilarArevalo:2007ab}. Several systematic uncertainties are considered, and the total systematic uncertainty is dominated by the hadron energy cut, the muon angular resolution, and the FSI parameters.  The fit returned an $M_A^{QE}$ value of $1.21^{+0.18}_{-0.10}(\mbox{fit}) ^{+0.13}_{-0.15}(\mbox{syst.})$~GeV when $Q^2$ was fit between 0 and $1.2\;\mbox{GeV}^2$.

\subsection{NOvA}

Nathan Mayer also presented the first look at CCQE events in NOvA's Near Detector On the Surface (NDOS) prototype that was placed 110 mrad off the NuMI beam axis~\cite{Betancourt:2013mba}.  
The detector consists of extruded cells $4~\mbox{cm}\times 6\;\mbox{cm}$ in cross section, stacked in layers alternating in horizontal and vertical directions to provide three dimensional tracking. Each cell contain liquid scintillator read out by a wavelength shifting fiber coupled to an avalanche photodiode.
The very far off-axis configuration resulted in a neutrino energy spectrum with two peaks:  one from off angle pion decays at about 200~MeV, and another from off angle kaon decays, which peaked at about 2~GeV.  The flux prediction comes from the same GEANT-based simulation used by MINOS, but because the flux is so far off the beam line axis, the flux uncertainties are considerably larger.  

NOvA selects charged current events by identifying a muon track originating from withing the fiducial volume. A multivariate analysis using the energy observed around the vertex, the average energy deposited per detector plane, and the number of hit detector planes is used to separate CCQE events from non-CCCQE events.

\begin{figure}[!htpb]
\centering
\includegraphics[height=5.5 cm]{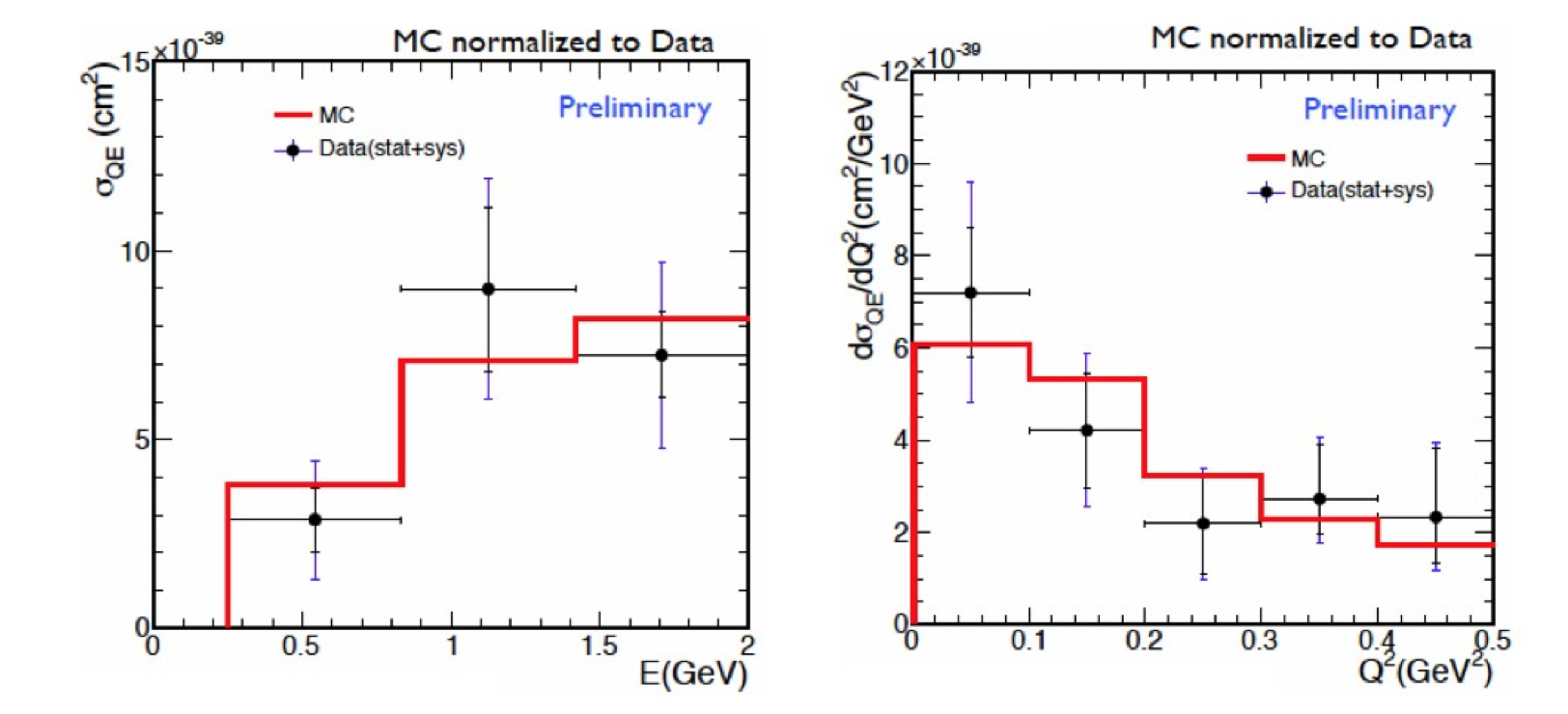}
\caption{\label{fig:nova_ccqe} CCQE cross section measured by NOvA as a function of $E^{QE}_\nu$ (left) and $Q^2_{QE}$ (right).  The inner bars are statistical and the outer bars are the combined systematic and statistical uncertainties~\cite{Betancourt:2013mba}.}
\end{figure}

The analysis reconstructs both the energy and the angle of the outgoing muon, and from that reconstructs $E_\nu^{QE}$ and $Q^2_{QE}$.  The sample has very low statistics, so no sideband tuning techniques were used.  The predicted background is subtracted from the events and then the $E_\nu^{QE}$ and $Q^2_{QE}$ distributions unfolded to yield cross sections as a function of these two variables, as shown in Fig.~\ref{fig:nova_ccqe}.  The dominant systematic uncertainty is the flux at $25\%$, and is evaluated by comparing the flux predictions from FLUKA~\cite{fluka} and GEANT4~\cite{geant4}.  The cross section as a function of neutrino energy from 0.2 to 2.0 GeV is  found to be somewhat lower than the prediction from {\texttt Genie} assuming a RFG model with $M_A=0.99\;\mbox{GeV}$.

\subsection{Comparison and Discussion of Experimental Results on QE Scattering}

Tables~\ref{tab:qe-mega-table-1} and \ref{tab:qe-mega-table-2} summarize how each of the various experiments have defined their CCQE event samples and the kinematic phase space covered by the selected events. Several points can be noted based on the table and the presented results, some of which were discussed at the workshop:

\begin{itemize}
\item The experiments utilize different means to identify CCQE samples that relate primarily to their proton reconstruction capabilities. Detectors with fine-grained tracking capabilities (SciBooNE, NOMAD) use recoil proton information when available to identify CCQE interactions, whereas MiniBooNE and MINOS, due to higher tracking thresholds (arising from the Cherenkov threshold in the former case, and the presence of passive steel plates in the latter), rely on identifying a muon with restrictions on additional activity from hadrons (pions, etc.) in the event without explicitly identifying a recoil proton.

\item Due to their geometrical configurations, the detectors have very different geometric acceptances. In particular, experiments with planar tracking geometries (MINERvA, MINOS, ND280, SciBooNE) have acceptances concentrated in the forward and backwards regions with respect to the neutrino beam, while MiniBooNE has full $4\pi$ acceptance. The importance of the angular acceptance depends on neutrino energy; at lower energies ($<1\;\mbox{GeV}$), the outgoing particles distribute themselves more isotropically, thus a limited acceptance can result in a large fraction of the events being inaccessible for analysis. At higher energies, the particles are boosted into the forward direction, reducing significantly the detrimental impact of a limited wide-angle or backward acceptance. In the ArgoNeut case, what would inherently be $4\pi$ acceptance is limited by the use of the MINOS near detector as a muon spectrometer. The $4\pi$ acceptance is maintained, however, for other particles emerging from the interaction.

\item The NOMAD analysis uses the formation time of the recoil proton in the nucleus as a fit parameter in obtaining a cross section from its one- and two-track CCQE samples. The formation time governs the probability that the proton will interact within the target nucleus, which affects whether a proton track is observed in the detector, and whether the proton will emerge with kinematics consistent with the naive CCQE kinematic hypothesis. The former will dictate whether it ends up in the one-track sample (no identified proton), and the latter impacts the efficiency with which an event with two tracks is accepted into the two-track selection, which utilizes transvere momentum balance, the angle between the two tracks in the transverse plane, and the proton emission angle relative to the incident neutrino direction to select events consistent with a neutrino CCQE scattering off a free nucleon at rest. 

There was considerable discussion about this method and the interpretation of the results, particularly with regard to the apparent discrepancy between the MiniBooNE and NOMAD CCQE cross sections; the former indicates a significant enhancement in the effective cross section from multi-nucleon contributions relative to what would be expected from the free nucleon case, while the latter observes no such enhancements. 

One issue is the energy dependence of the additional processes and whether one would in fact expect multi-nucleon effects to be smaller at the higher neutrino energies probed by NOMAD.  Currently, the RPA-based approaches have difficulty in extrapolating to higher neutrino energy, so further theoretical process is needed to answer this question.

On the experimental side, another issue is whether the topological and kinematic requirements in the NOMAD selections remove events where multi-nucleon effects significantly distort the kinematics or result in higher proton multiplicity. If so, such interactions should reappear as the selection is loosened. The continued study of the impressive NOMAD CCQE data should allow us to better understand this apparent discrepancy and CCQE interactions in general.

\item MINERvA has made the first detailed analysis of its $Q^2_{QE}$ distribution with 
models that go beyond the standard RFG prescription. The results favor a model incorporating the enhancement of transverse response with $M_A=0.99\;\mbox{GeV}$ (consistent with deuterium measurements) over increasing $M_A$ as an effective parameter to $1.35\;\mbox{GeV}$ in a ``basic'' RFG model. The analysis of energy depositions around the interaction vertex  indicate that there is an additional contribution in the neutrino-mode data that may be consistent with more low energy proton emission than predicted by the RFG-based model. No such excess is observed in antineutrino-mode data. The MINERvA data, with its intermediate energies between MiniBooNE and NOMAD, its fine-grained tracking capabilities, and multiple target materials, will play an important role in resolving the above-mentioned apparent discrepancy between the these two measurements, and in further understanding the CCQE interaction.

\item While impressive agreement is achieved in the outgoing lepton kinematics observed in the MiniBooNE CCQE data by the RPA-based calculations, the ability to predict the corresponding hadronic component of the interaction remains a challenge. With availability of data from fine grained detectors (NOMAD, MINERvA, ND280) and LAr TPCs capable of resolving the detailed final state of CCQE interactions (ArgoNeuT, MicroBooNE), including outgoing proton multiplicities and kinematics, it will be important to find a suitable means of consistently assigning a predicted hadronic state with the lepton kinematics in an event.

\item Continued effort to improve {\em ab initio} flux estimates will become increasingly important as we further probe the nature of CCQE interactions. Extensive efforts have already been mounted by MiniBooNE and NOMAD to achieve 5-10\% uncertainties in the flux estimate. The NA61 data is being used to great effect in the T2K flux prediction; further analyses with improved statistics and replica target data under way that should significantly reduce uncertainties. With the effects of secondary and tertiary interactions, as opposed to the primary production measured by SPY, HARP, and NA61, becoming an increasing portion of the systematic uncertainty, the community may consider devoting additional effort to perform the relevant measurements to accurately model these effects. This will be necessary to ensure that  better than $5\%$ overall systematic uncertainty can be achieved in the future.

\item Experiments that use the NuMI beamline (MINOS, MINERvA, ArgoNeuT, NOvA) use a variety of methods to estimate the neutrino flux from the beam line. The differences include the underlying software and hadronic interaction model (FLUKA ~\cite{fluka} vs. GEANT4 FTFP~\cite{geant4}) and data to tune the simulation. The resulting flux predictions differ by as much as $20\%$. In order to fully profit from the measurements in these experiments, it is important to understand how these differences arise and to have a common understanding of the methods and resulting systematic uncertainties. Since the workshop, there has been considerable progress towards this end, with the NuMI-based experiments forming a  collaborative effort to improve the flux estimation.
\end{itemize}

\begin{table}[p]
\centering
{\scriptsize 
\begin{tabular}{|c||c|c|c|c|c|}\hline
            & Neutrino  & Neutrino & $E_\nu$ range   & Neutrino Flux    & $\nu$ Event   \\ 
Experiment  & Target   & Type     & (GeV)            & Determination    & Generator         \\ \hline \hline
ArgoNeuT    & Ar      & $\numu$, $\numubar$ &  $0.5<E_\nu<10$ & MINOS low $\nu$ data  & GENIE  \\  
            &         &&                 & FLUGG                 &         \\ \hline
MINERvA     & CH      & $\numu$, $\numubar$ & $1.5<E_\nu<10$  &  NA49 data                     & GENIE   \\ 
            &         &&                 & GEANT4                &          \\ \hline
MiniBooNE   & CH$_2$  & $\numu$, $\numubar$ & $0.4<E_\nu<2$   & HARP, E910 data   & NUANCE       \\
            &         &&                 & GEANT4 &               \\ \hline
MINOS       & Fe      & $\numu$ & $0.5<E_\nu<6$   & MINOS low $\nu$ data   &  NEUGEN        \\
            &         &&                  & FLUGG, FLUKA      &               \\ \hline
NOMAD       & \tiny{$64\%$ C, $22\%$ O, $6\%$ N}  & $\numu$, $\numubar$ & $2.5<E_\nu<300$ & DIS, IMD, SPY data & NOMAD \\ 
            & \tiny{$5\%$ H, $1.7\%$ Al}          &                  &                 & FLUKA                    & MC \\  \hline
NOvA ND     & CH$_2$       & $\numu$ & $0.5<E_\nu<2$  & NA49 data &  GENIE     \\ 
            &              &&                 & FLUGG &  \\ \hline
SciBooNE    & C$_8$H$_8$  & $\numu$ & $0.6<E_\nu<2$  & HARP, E910 data   & NEUT \\ 
            &              &&                & GEANT4 & NUANCE     \\ \hline
T2K         & C$_8$H$_8$   & $\numu$ & $0.2<E_\nu<30$ & NA61/SHINE data   & NEUT \\ 
            &              &&                 & GEANT4 &      \\ \hline
\end{tabular}
}
\caption{Summary of the experiments that have measured $\nu_\mu$ CCQE scattering: the neutrino target, neutrino type used in CCQE analyses, energy range of the CCQE data, how the neutrino flux is determined including the brand of beam simulation used, and the default neutrino event generator used to extract CCQE results from the experiment. The corresponding CCQE selection and kinematics are provided in Table~\ref{tab:qe-mega-table-2}.}
\label{tab:qe-mega-table-1}
\end{table}

\begin{table}[p]
{\scriptsize 
\centering
\begin{tabular}{|c||c|c|c|c|c|c|}\hline
            & CCQE Event  & Sign & Muon Angle & Muon KE & Proton   & CCQE   \\ 
Experiment  & Selection   &  ID?         & ($^\circ$) & (GeV)   & KE (MeV) & Results \\ \hline \hline
ArgoNeuT    &  $1\mu$ + $0\pi$  & Yes & $0<\theta_\mu<40$ & $T_\mu>0.4$     & $T_p>21$ & $\#$ protons \\ 
            &                    &     & \tiny{(MINOS-match)}   &
            &   &  \\ \hline
MINERvA     &  $1\mu$ + recoil  & Yes & $0<\theta_\mu<20$  & $1.5<T_\mu<10$  & $T_p>80$   & $d\sigma/dQ^2$ \\ 
            & consistent        &     & \tiny{(MINOS-match)}    &                  & \tiny{(tracking)} & \\ 
            & with CCQE $Q^2$     &     &                    &                  &            &  \\ \hline
MiniBooNE &  $1\mu$ + $0\pi$    & No  & $0<\theta_\mu<360$ & $0.2<T_\mu<2$   & --- & $d^2\sigma/dT_\mu d\theta_\mu$ \\ 
          & \tiny{(Michel $e^-$ ID)}         &     &                    &                 &     & $d\sigma/dQ^2$ \\ 
          &                           &     &                    &                 &     & $\sigma(E_\nu)$, $M_A$  \\ \hline
MINOS     &  $1\mu$ + \tiny{($E_{had}<225$ MeV)} & Yes & $0<\theta_\mu<180$ & $0<T_\mu<5$     & --- & $M_A$ \\ \hline
NOMAD & ($1\mu$) or ($1\mu$ + 1p) & Yes & $0<\theta_\mu<100$   & $T_\mu>2$ & $T_p>21$ & $\sigma(E_\nu)$ \\ 
      & \tiny{(accepted)}                 &     &                      &           &          & $M_A$ \\  \hline
NOvA ND   & $1\mu$ + multi-variate  & No & $0<\theta_\mu<45$ & $0<T_\mu<1.4$   & --- & $\sigma(E_\nu)$     \\ \hline
SciBooNE  & ($1\mu$) or ($1\mu$ + 1p) & No  & $0<\theta_\mu<60$ & $0.1<T_\mu<1.1$ & $T_p>100$ & $\sigma(E_\nu)$ \\ \hline
T2K       &  $1\mu$ + $0\pi$          & Yes & $0<\theta_\mu<80$ & $0<T_\mu<30$    & ---  & $\sigma(E_\nu)$     \\ \hline
\end{tabular}
}
\caption{Summary of the atttributes of experimental measurements of $\nu_\mu$ CCQE scattering: how the CCQE events are selected, whether or not final state muons can be charge-identified, muon angle with respect to the incoming neutrino direction, muon kinetic energy, proton detection threshold (where applicable), and list of reported CCQE cross section measurements. Companion to Table~\ref{tab:qe-mega-table-1}.}
\label{tab:qe-mega-table-2}
\end{table}

\section{Lessons from Nuclear Physics and Inclusive and Exclusive Electron Scattering, Scaling, Relativistic Mean Fields}
\label{sec:electron}
\subsection{Background and Contribution of Electron Quasi-elastic Scattering}         
\begin{figure}[t]
\centering
\includegraphics[width=10 cm]{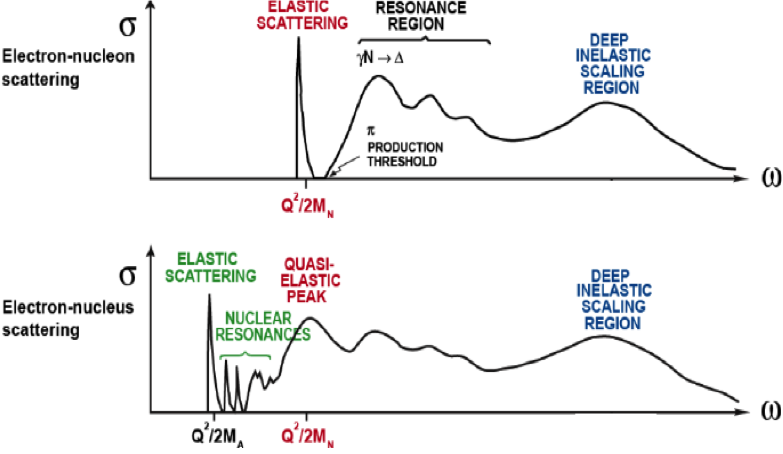}
\caption{\label{fig:escat_omega}Idealized spectra of high-energy electron scattering off a nucleon (top) and off a nucleus (bottom). At momentum transfers appropriate to quasi-elastic scattering, the yield from nuclear excitations and elastic scattering are greatly suppresssed due to form factors.}
\end{figure}

 Most of our knowledge regarding the dynamics of lepton-nucleus interactions for $0.3 < E_{\ell}< 3~\mbox{GeV}$ comes from electron-nucleus scattering.  Several talks on electron quasi-elastic (QE) scattering at the workshop provided valuable background and input on what has been learned from inclusive and exclusive quasi-elastic electron scattering. The flux and energy of the GeV beams from electron accelerators are precisely known and the momentum of the scattered electrons precisely measured in magnetic spectrometers. Figure \ref{fig:escat_omega} shows an idealization of such a scattered electron spectra.   

The following focuses on what is termed quasi-elastic electron scattering. The process is readily measured as a function of the transferred three-momentum (${\bf q}$) and energy ($\omega$) from the electron to the nucleus by measuring the scattered electron without any reference to the hadronic final state. The variable ${\bf q}$ ($\omega$) is reconstructed by taking the vector (scalar) difference between the initial and final electron momentum (energy).  
                                               
\begin{figure}[t]
\centering
\includegraphics[width=10 cm]{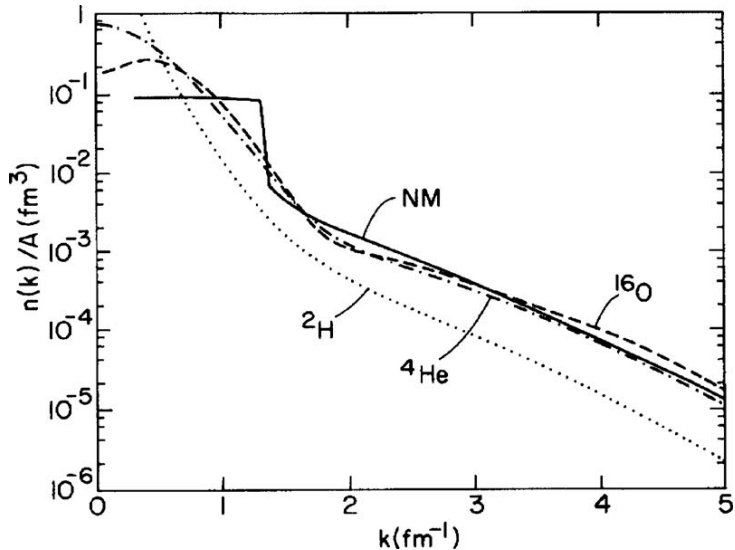}
\caption{\label{fig:nucleonmomentum}Momentum distribution of nucleons in light nuclei. The number of nucleons in a given $dk$ is $4\pi\;A\;n(k)\;k^2\;dk$. The small binding energy (2.2 MeV) of the deuteron and its resulting diffuseness accounts for its difference from the more tightly bound nuclei. ``NM'' refers to the distribution in infinite nuclear matter~\cite{Benhar:2006wy}.}
\end{figure}

The events from quasi-elastic scattering of the electron off a single nucleon in the target nucleus form a peak centered at $\omega = Q^2/2M_N$ with a width approximately $\sqrt{2p_F |{\bf q}|}$ where $p_F$ is the Fermi momentum of the nucleons in the target ground state. QE scattering experiments have been carried out for many years on a variety of targets and incident energies. Reference~\cite{Benhar:2006wy} provides an excellent review of the results as of 2008. The intent of such measurements was often to measure the momentum distribution of nucleons in the target nucleus.  As shown in Figure \ref{fig:nucleonmomentum}, it was very early noted that in every nucleus there appeared nucleons with momenta well beyond the Fermi momentum that are incompatible with scattering off a "free" nucleon moving in a mean field. Theorists soon identified the high momentum nucleons as arising from correlated pairs of nucleons whose high momenta originate in the strong short-range nucleon-nucleon interaction.

Direct observation of these correlated pairs has only become possible in the last decade and there now exist spectacular demonstrations of their existence in nuclei. Ingo Sick's presentation addressed the overall importance of these correlations. While there is approximately a 20\% probability for a nucleon to be involved in such a correlation, in carbon, for example, they are responsible for 37\% of the energy required to remove a nucleon from the nucleus. The existence of these correlated pairs also accounts for approximately 20\% of the depleted occupancy of the nuclear shell model orbits.

Figure \ref{fig:nucleonmomentum} shows a calculation of the normalized nucleon momentum distribution in selected light nuclei as shown by Ingo Sick. The shape of the high momentum distribution ($k>1.5~\mbox{fm}^{-1}$) is very similar in all four cases and almost identical for $\!\;^{4}\mbox{He}$ and $\!\;^{16}\mbox{O}$. The fraction of correlated pairs is smaller in $\!\;^{2}\mbox{H}$ because of the dilute structure of this system. The high momentum components are due to the correlated pairs of nucleons (mostly $n-p$ pairs with $S=1$, $T=0$) in the nucleus.  The nucleon-nucleon interaction inside 1.5 fm is strongly attractive, but becomes sharply repulsive when the two nucleons  overlap. As we shall see, it is crucial to take account of these nucleon-nucleon correlations when calculating lepton-nucleus quasi-elastic scattering.

\begin{figure}[t]
\centering
\includegraphics[height=4 cm]{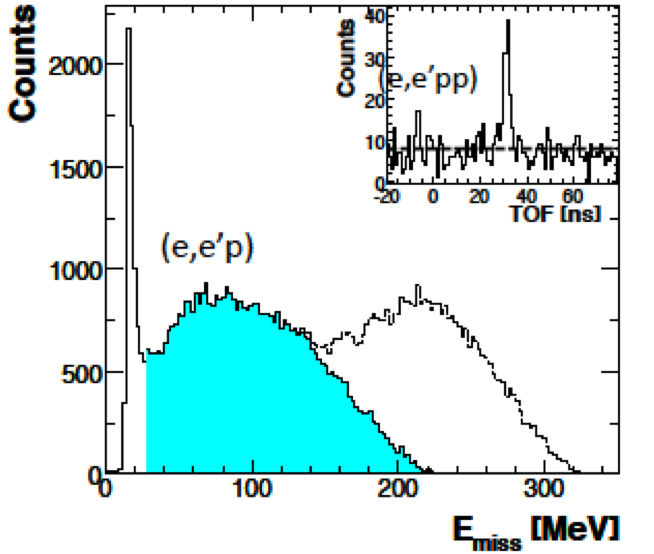}
\includegraphics[height=4.2 cm]{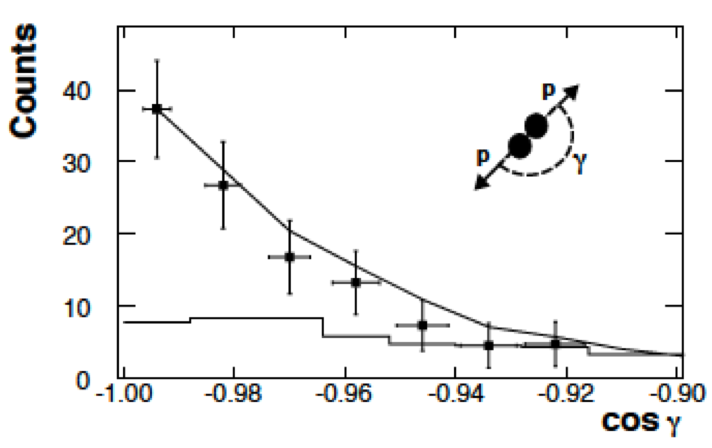}
\caption{\label{fig:misse_ppcorrelation}Left: Missing energy in $e-\!\;^{12}\mbox{C}$ scattering events (see text for definition). Right: back-to-back angular correlation between the initial momentum of the struck proton and the detected momentum of its spectator partner in the pair correlation. From reference~\cite{Shneor:2007tu}.}
\end{figure}

Doug Higginbotham's talk provided dramatic demonstration of the existence of highly correlated pairs as shown in Figure \ref{fig:misse_ppcorrelation} showing the missing energy in an $(e,e'pp)$ measurement on $\!\;^{12}\mbox{C}$\cite{Shneor:2007tu}. The missing energy is defined as $E_{miss}= \omega -T_p-T_{A-1}$, where 
$T_p$ is the kinetic energy of the struck proton and $T_{A-1}$ the inferred kinetic energy $\frac{{\bf p}_i^2}{2(A-1)M_N}$ of the recoiling nucleus assuming it has an initial state momentum (${\bf p}_i$) opposite to that of the struck proton. If one simply assigned the minimal separation energy to the struck proton, the sharp peak at $E_{miss}\sim 20 \mbox{MeV}$ is the expected value. However, the measured energy loss, $\omega$ (shaded in blue) extends for some 200 MeV beyond that minimal value, because the struck proton is involved in a short range correlation, thus requiring its correlated partner to carry off its equal but opposite initial state momentum, and  resulting in a much higher missing energy. It is extremely important to understand the implications of this plot as the incident energies in neutrino experiments are unknown and must usually be inferred from the momentum of the scattered charged lepton.

This measurement is just one of a series of measurements carried out at Jefferson Laboratory~\cite{Shneor:2007tu,Subedi:2008zz,Fomin:2011ng} that have directly observed the correlated high momentum states that theorists have predicted for many years. These measurements have clearly established that approximately 20\% of the nucleons in nuclei with $A\ge 12$ are involved in pair-wise correlations. Proton-neutron correlations dominate over proton-proton correlations by a factor about nine. These $n-p$ pairs are dominantly $T=0$, $S=1$ pairs as a result of the tensor force. 

Another important discovery uncovered by electron scattering measurements is the observation of scaling~\cite{Donnelly:1998xg,Donnelly:1999sw} in inclusive electron QE scattering. The observed scaling is reassuring in that it confirms essential elements of the depiction of QE scattering off individual nucleons, but its real value is found in what is learned from the observed violations of scaling. Maria Barbaro and Juan Antonio Caballero presented the background and current status of what has been learned from the application of scaling to QE scattering. The approach is similar to the deep inelastic scattering off partons in the nucleon where the scaling variable $x=\frac{Q^2}{2M_N\omega}$ allows a coherent account of a large body of data. Appropriate scaling variables have been idenified in electron-nucleus scattering without the benefit of asymptotic freedom.

\begin{figure}[t]
\centering
\includegraphics[height=4.2 cm]{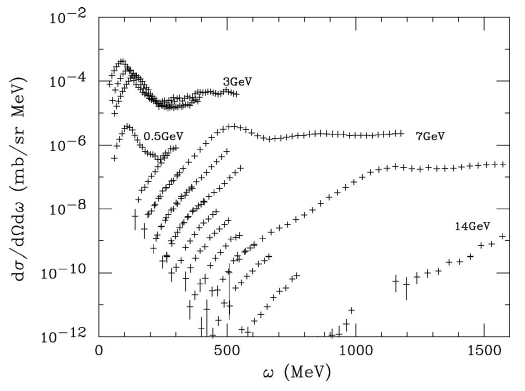}
\includegraphics[height=4.2 cm]{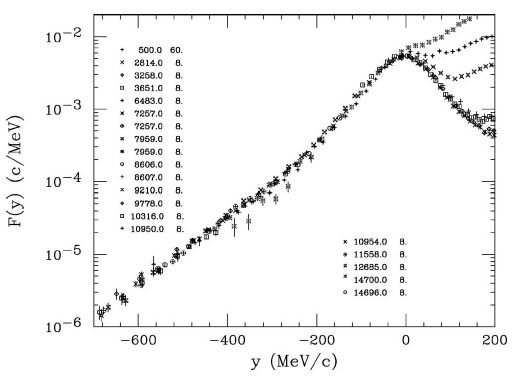}
\caption{\label{fig:yscaling}An example of $y$ scaling. Left: Raw cross section data at a vareity of incident energies and momentum as a function of $\omega$. Right: The same data as a function of the scaling variable $y$. From Reference~\cite{Benhar:2006wy}.}
\end{figure}

Three kinds of scaling have been uncovered in electron-nucleus scattering and each provides valuable lessons. It is straightforward to extract the values for ${\bf q}$ and $\omega$ in electron scattering experiments. Scaling of the first kind employs the conservation of energy and momentum and assumes the observed QE cross section is simply an incoherent sum of the scattering off the individual nucleons in a nucleus.  This simplification reduces the dependence of the cross section to that of single variable $y=\sqrt{\omega^2 + 2M_N \omega} - q$\footnote{In what follows, $q\equiv |{\bf q}|$, the magnitude of the three-momentum transfer, and should not be confused with $q^2\equiv-Q^2\equiv t$.} the   and yields a scaling function $F(y,q)$ given as
\begin{equation}
F(y,q) = \left( \frac{d^2\sigma}{d\Omega d\omega}\right)_{EXP} \left( \frac{1}{Z \sigma_{ep}(q) + N\sigma_{en}(q)}\right) \frac{d\omega}{dy}
\end{equation}
where $\sigma_{ep}$ and $\sigma_{en}$ are the elastic scattering cross sections at three momentum ${\bf q}$ for scattering off a single proton and neutron, respectively. 
Figure \ref{fig:yscaling} shows scaling of this first kind in $\!\;^{3}\mbox{He}$ at a variety of incident energies. Note that scaling of the first kind works very well for $y \le 0$ $(x > 1)$. The value $y=0$ $(x=1)$ corresponds to the scattering off a nucleon at rest. For $y > 0$ $(x<1)$ there must be contributions from some other source than simple scattering off an independent nucleon moving in a mean field.

\begin{figure}[t]
\centering
\includegraphics[height=6 cm]{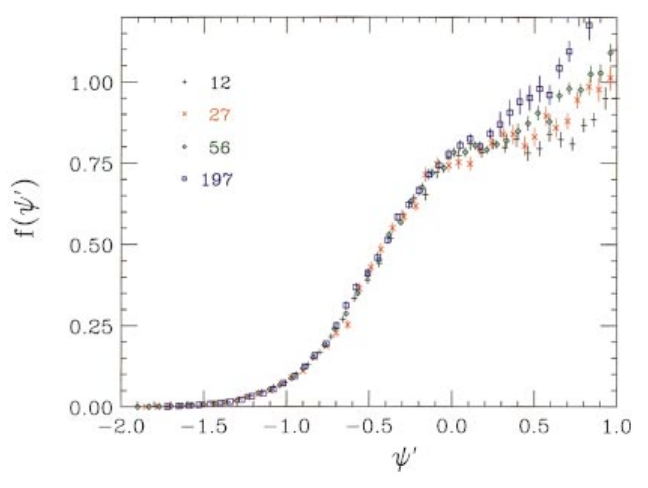}
\caption{\label{fig:scaling_2ndkind}An example of scaling of the second kind with electron scattering data on C, Al, Fe, and Au. Scaling is observed for $\psi'<0$ but again breaks down for $\psi'>0$~\cite{Donnelly:1999sw}.}
\end{figure}

Another form of scaling results from scaling across different nuclei. To allow for the slow changes that occur with increasing $A$, for $A\ge 12$ ,the scaling variable $\psi' \simeq \frac{y}{k_F}$ is employed. Figure \ref{fig:scaling_2ndkind} shows the scaling observed for widely differing nuclei from C to Au at a
fixed electron energy and scattering angle. This kind of scaling is termed scaling of the second kind and illustrates the nearly universal character of nuclear matter for $A > 12$. 

Moving on to what is termed zeroth order scaling, electron scattering can be separated into its longitudinal and transverse components where the terms refer to the polarization direction of the virtual photon relative to the transferred momentum ${\bf q}$. The different kinematic dependence of the longitudinal and transverse couplings allows their separation into the individual responses, $R_L$ and $R_T$:
\begin{equation}
\frac{d\sigma}{d\Omega d\omega} = \sigma_{Mott} 
\left[   \left| \frac{Q^2}{{\bf q}^2}\right|^2 R_L  + \left( \frac{1}{2} \left| \frac{Q^2}{{\bf q}^2}\right| + \tan^2 \frac{\theta_e}{2}\right) R_T\right]
\end{equation} 
                          
The longitudinal response is typically associated with scattering off charge while the transverse response is due to scattering off magnetization. Accounting for the longitudinal and transverse responses of the free neutron and proton, the relativistic Fermi gas (RFG) model for the nucleus predicts that the residual responses, $f_L(\psi',q)$ and $f_T(\psi'âq)$ are equal:
\begin{equation}
(f_L)_{RFG} = (f_T)_{RFG} =  f_{RFT}
\end{equation}

\begin{figure}[t]
\centering
\includegraphics[height=4.5 cm]{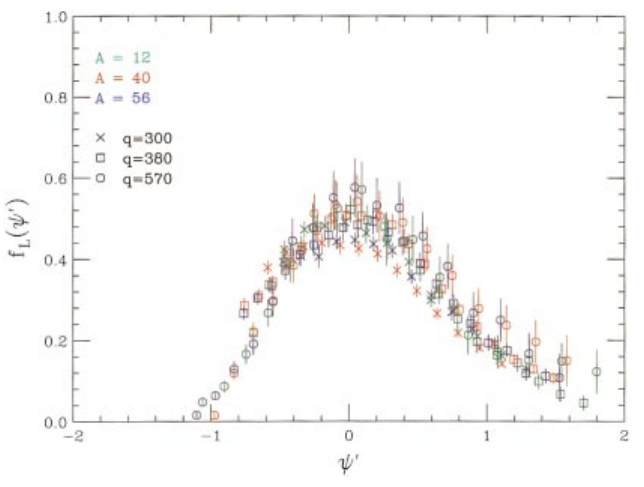}
\includegraphics[height=4.5 cm]{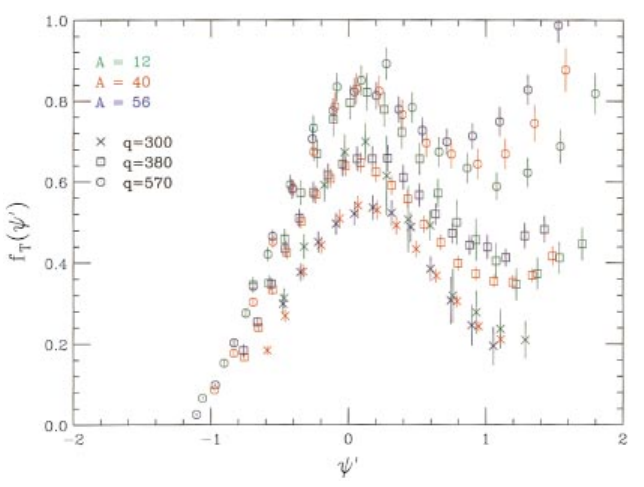}
\caption{\label{fig:scale_longtrans}The longitudinal (left) and transverse (right) response of inclusive electron QE scattering at different three-momentum transfers as observed in three different nuclei~\cite{Donnelly:1999sw}. The longitudinal response scales while the transverse clearly does not and appears to increase with 3-momentum transfer over the range shown.}
\end{figure}

Figure \ref{fig:scale_longtrans}~\cite{Donnelly:1999sw} shows the result of applying this prescription to QE scattering for several different nuclei at different three-momentum transfers.  Some very important facts emerge from such an analysis:  the longitudinal response shows scaling both with momentum transfer $q$ and $A$ and is therefore said to "super-scale" while the transverse response scales with $A$ (scaling of second kind) but not with $q$ (scaling of the first kind). The transverse response is observed to increase with three-momentum transfer at least up to $q = 570~\mbox{MeV}/c$ in contrast to the scaling predicted in the RFG impulse approximation. The RFG also predicts that both nuclear responses should be identical when the longitudinal and transverse responses of free nucleons are taken into account. Clearly, the transverse response receives an appreciable enhancement in the nuclear environment and further this enhancement appears similar in all nuclei. 
 
\begin{figure}[t]
\centering
\includegraphics[height=4.5 cm]{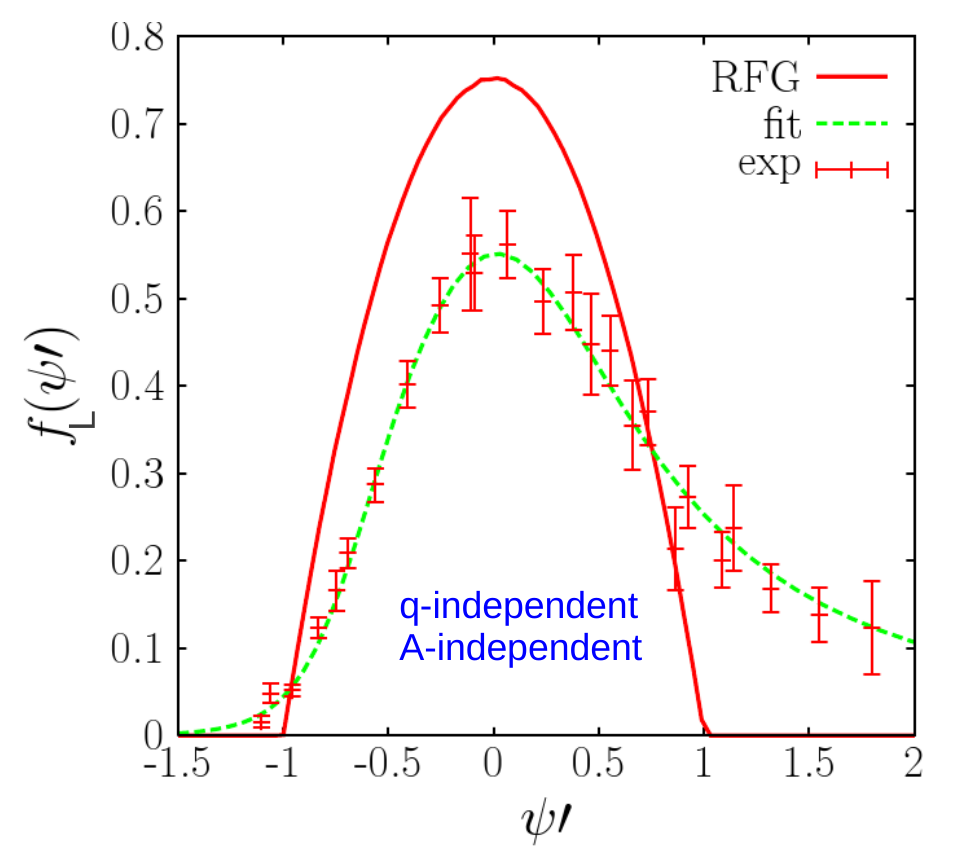}
\caption{\label{fig:superscale_longfit}A fit to the world data of the longitudinal response function shown by M. Barbaro and J. Caballero. The RFG model predicts the response to be symmetric about $\psi'=0$. The asymmetry indicates a violation of the impulse approximation. Data are from Reference~\cite{Jourdan:1996np}. 
}
\end{figure}

While the RFG correctly predicted superscaling of the longitudinal response, it does not predict the magnitude and shape of the longitudinal response nearly as well, as shown in Figure \ref{fig:superscale_longfit}, which compares RFG to a fit of the world's data. Two observations are to be made. The longitudinal response is asymmetric about $\psi'=0$, extending to larger values of $\psi'$ than the RFG predicts, but it appears that the integral over $\psi'$ of the measured response and the RFG prediction are nearly the same. Recalling that $\psi'\simeq y/k_F$, in order that $\psi'>1$, $\omega$ must exceed $(k_F+q)^2/2m_N$. Going beyond a mean field impulse approximation and taking account of the high momentum nucleon pairs produced by the nucleon-nucleon interactions account for both  failures (the asymmetric $f_L(\psi')$ and $f_L(\psi')<f_T(\psi')$ ) of the RFG~\cite{Carlson:2001mp}. This issue is of great importance in treating neutrino-nucleus scattering, as described in Section \ref{sec:theory}.

While RFG and a variety of nuclear mean fields produced with Woods-Saxon optical model potentials all produce super-scaling for longitudinal electron QE scattering, they all fail to generate an asymmetric longitudinal response function $f_L(\psi')$. Given that it is an established fact that approximately 20\% of the nucleons are involved in short range correlations, it is not surprising that such scaling violations are observed. Indeed, both the asymmetry in the longitudinal response and the fact that $f_T(\psi') > f_L(\psi')$ can be accounted for by including the effects of correlated pairs~\cite{Carlson:2001mp}. 

\begin{figure}[t]
\centering
\includegraphics[height=6 cm]{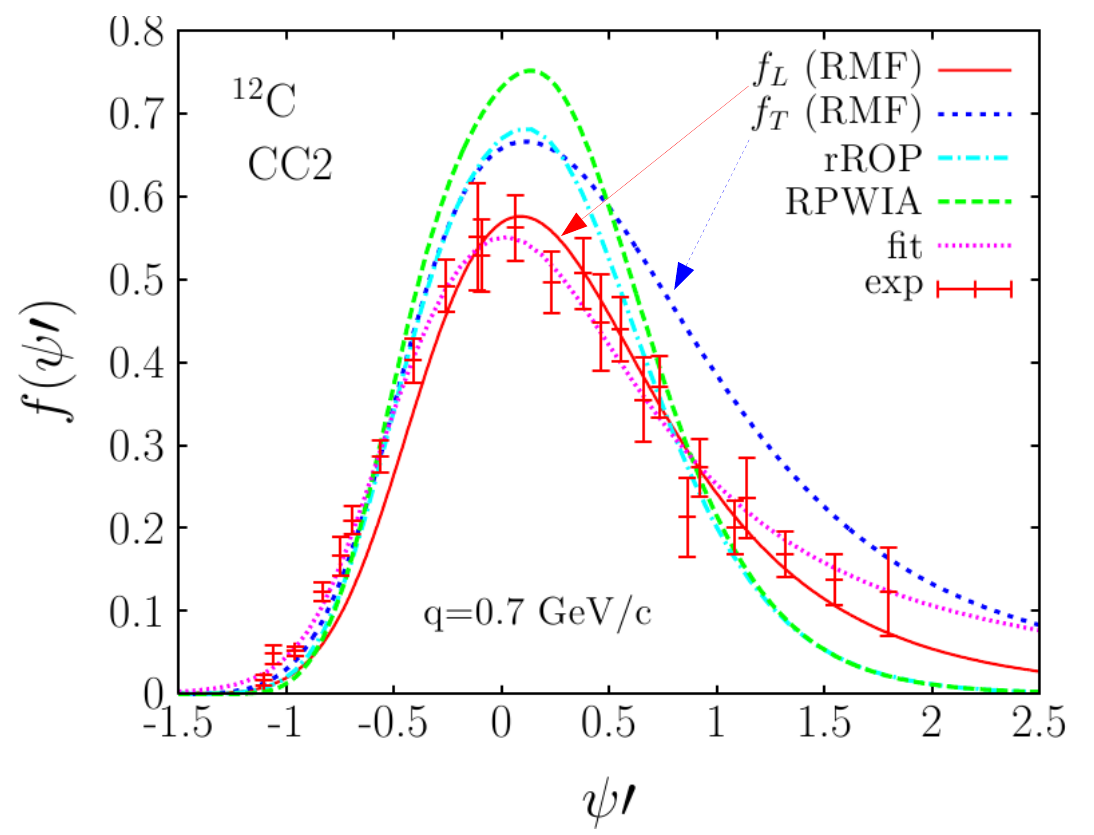}
\caption{\label{fig:psil_rmf}Various theoretical response functions and the experimentally determined longitudinal response for $\!\;^{12}\mbox{C}$ as a function of the scaling variable $\psi'$ shown by J. Caballero. Note especially the asymmetry of $\psi'_L$ achieved with RMF. ``CC2'' refers to the specific form of the one body nucleonic vector current.
}
\end{figure}

In contrast, the use of a relativistic mean field (RMF) rather than RFG can produce the requisite asymmetry in $f_L(\psi')$ and produce $f_T(\psi') > f_L(\psi')$, both in line with experimental observation. The Pavia and Madrid-Seville groups have extensively carried out the use of various forms of relativistic mean fields to deal with electroweak nuclear scattering. The presentation of their work on electron-nucleus scattering was covered in Juan Antonio Caballero's presentation and also appeared in the talks by Maria Barbaro and Carlotta Giusti.  The result identified as RMF in Figure \ref{fig:psil_rmf} employs the same energy independent relativistic mean field both to create the bound nucleon initial state as well as the nucleon final state in the continuum. 
The desired physics is produced by the final state interaction of the struck nucleon with the relativistic mean field.
The relativistic mean field is made up of a scalar and a vector potential~\cite{waleckarmf}
and with the nucleons treated as Dirac particles with upper and lower components. The initial and final states are solutions of a Dirac equation employing the relativistic optical potential. A particular form for the nucleon vector weak current, denoted as CC2 is given by: 
\begin{equation}
j^\mu = F_1(Q^2) \gamma^\mu + i \frac{\kappa}{2M} F_2(Q^2) \sigma^{\mu\nu} Q_\nu
\end{equation}
                                          
where $\kappa$ is the anomalous piece of the isovector nucleon magnetic moment. Note, this current is obviously just a one body current. In this approach, the impulse approximation is employed, the nucleons move in a mean field with the desirable results produced from the interaction of the initial and final state of the struck nucleon with the relativistic mean field. One appears to obtain significant benefits "for free" with such a relativistic-Dirac approach. However, the underlying physics is often obscure. More physically transparent approaches, for example~\cite{Carlson:2001mp}, achieve similar results but employ nucleon-nucleon interactions, require two-body currents, and hence are far more computationally intense.

\begin{figure}[t]
\centering
\includegraphics[height=5 cm]{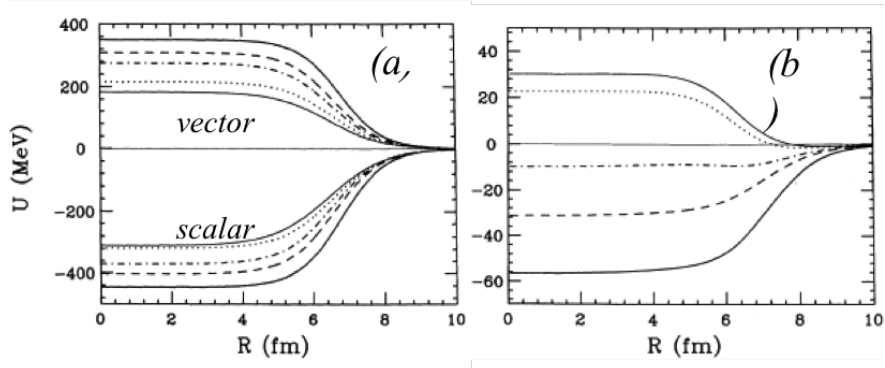}
\caption{\label{fig:svecpotential}a.) The vector and scalar central optical potentials for a nucleon on $\!\;^{208}\mbox{Pb}$ as a function of $R$ with incident proton kinetic energy of 20 MeV (heavy solid line), 100 MeV (dashed line), 200 MeV (dot dashed line), 500 MeV (dotted line), and 800 MeV (light solid line). b.) Net effect of the scalar and vector central potentials~\cite{Cooper:1993nx}.}
\end{figure}

These relativistic potentials have enjoyed considerable success in describing many aspects of   
 nucleon--nucleus scattering~\cite{Cooper:1993nx}. As phenomenologically determined from low energy
 nucleon--nucleus  scattering, the central scalar potential is strongly attractive ($\sim -400~\mbox{MeV}$) while the central vector potential is repulsive  ($\sim 350~\mbox{MeV}$) at low energy. Figure \ref{fig:svecpotential}a shows the central vector and scalar potentials as a function of nuclear radius, while Figure \ref{fig:svecpotential}b shows the net effect with the net central potential becoming repulsive at about 250 MeV. However, recalling what is stated above, to obtain the best fit to longitudinal and transverse response functions, the {\underline same} optical potential is used to generate the initial bound state and the spectrum of final states. While the RMF impulse approximation is most convenient, it is puzzling that for non-relativistic approaches to produce the same physics, they must employ nucleon-nucleon interactions and two-body currents~\cite{Carlson:2001mp}. Apparently, much of the benefit of the RMF comes from the strong modification of the lower component wave function in the presence of the strong vector and scalar fields in the nucleus. 
                                            
Reference~\cite{Antonov:2011bi} is a study investigating the underlying physics accounting for the successful features of using RMFs. We shall return to this when treating 
neutrino–-nucleus CCQE. 

Clearly a great deal relevant to neutrino--nucleus scattering has been learned via the scaling analyses of {\em inclusive} electron QE scattering from nuclei. 
Such measurements are feasible with electrons where the critical variables ${\bf q}$ and $\omega$ are readily obtained from measurement of the momentum of the scattered electron. Unfortunately, for most neutrino experiments these quantities are out of reach. Thus, scaling cannot at the moment be applied to neutrino experiments due to a lack of knowledge of the incident neutrino energy which limits how well ${\bf q}$ and $\omega$ can be established. 
The appearance of short-range (high momentum) pairwise correlations among 20\% of the nucleons in a nucleus is especially important. These correlations are known to be the cause of the enhancement appearing in the nuclear transverse response. These correlations will be shown to be important in neutrino-nucleus CCQE scattering. Also of importance and a cause for optimism is the fact that these correlations appear universally in all nuclei. 

\section{Theoretical developments in neutrino-nucleus CCQE scattering}
\label{sec:theory}
\subsection{Introduction}
  We now move to the analysis of neutrino-nucleus scattering where less data exists and experiments are much harder to compare because of differences in beams, detection techniques, and target composition (see, for example, Table~\ref{tab:qe-mega-table-1}). Unlike in electron scattering, there is further ambiguity in defining exactly what constitutes quasi-elastic scattering. T.W. Donnelly was asked to provide a set of definitions to be used at the workshop, which appear in \ref{sec:definitions}. 

\begin{figure}[t]
\centering
\includegraphics[height=7.5 cm]{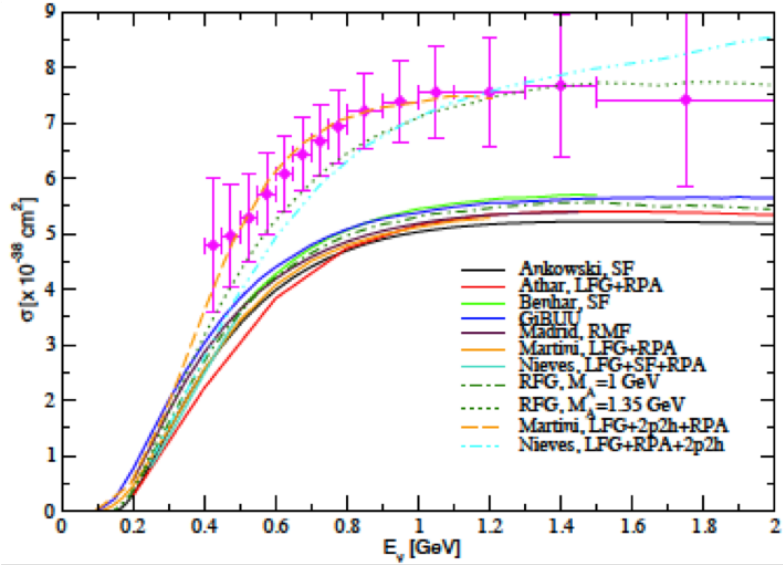}
\caption{\label{fig:mbxsec_vs_e} Early predictions for the CCQE cross section \cite{Alvarez-Ruso:2014bla} for $\nu-\!\;^{12}\mbox{C}$ scattering. Most were based on mean-field impulse approximation models of the process. The MiniBooNE data lies well above such approaches, but the two approaches allowing $2p-2h$ excitations fit the data rather well.}
\end{figure}
   
For reasons that are unclear, many of the lessons learned from electron scattering were not carried over to the early predictions for neutrino CCQE scattering. The simplest view of neutrino CCQE scattering on a nucleus is that it is simply an incoherent sum of the CCQE scattering on the individual nucleons in the nucleus. Nucleon vector form factors are readily available from electron scattering, and the neutron lifetime and the $Q^2$ dependence of neutrino scattering on deuterium fixes the axial form factors. The publication of the MiniBooNE CCQE cross section data in 2009\cite{Katori:2009du,AguilarArevalo:2010zc}  dramatically called attention to the fact that the observed data were some 45\% above the consensus value  predicted for the cross section \cite{Boyd:2009zz} as shown in Figure \ref{fig:mbxsec_vs_e}. Much of the difference is ascribable to the different definition of CCQE used by the theorists and MiniBooNE. While theorists typically defined ``CCQE'' as a process giving rise to a muon and proton final state, MinBooNE termed events to be ``CCQE'' when there was a muon in the final state and no pion, with corrections applied to remove the contribution of events in which a pion was created and subsequently absorbed.  Thus, the experimental signature sought by MiniBooNE was more inclusive than that employed by theorists.

To fit their data under the constraint that the nucleon isovector vector form factors were fixed, MiniBooNE increased the value of the axial mass ($M_A$) in the nucleon's dipole axial form factor from the accepted value of approximately 1 GeV to 1.35 GeV. While this increased axial mass allowed MiniBooNE to fit their data, this procedure is considered to be non-physical and is now understood to be a consequence of the inadequate nuclear model employed. Since then, there has been a large body of theoretical work to investigate this issue, with the approaches presented at the workshop summarized here.

\subsection{The Random Phase Approximation approach to $\nu$-nucleus scattering}
Marco Martini presented the extensive results on neutrino-nucleus scattering achieved by the collaboration of which he is a member. This collaboration very quickly came forward\footnote{Some ten years earlier, this group published results that presaged what was later observed in the MiniBooNE measurements\cite{Marteau:1999kt}.} with an RPA calculation that was in very good agreement with the MiniBooNE measurements~\cite{Martini:2009uj} as shown in Figure \ref{fig:mbxsec_vs_e}. They were able to fit the MiniBooNE cross section with $M_A=1.03\;\mbox{GeV}$ by the inclusion of a variety of many-particle, many-hole ($np-nh$) configurations in the initial and final states. These processes are not included in a Fermi gas/impulse approximation-based approach and might not be considered ``CCQE'' by many authors. In their analysis, Martini {\em et al.} excluded any final states that contain  pions. They followed their initial publication with a series of five others, all dealing with significant aspects of the problem including: antineutrino CCQE scattering from $\!\;^{12}\mbox{C}$\cite{Martini:2010ex}, relativistic extensions \cite{Martini:2011wp}, remarkable fits to details of MiniBooNE data \cite{Martini:2011wp}, and the very difficult issue \cite{Martini:2012fa,Martini:2012uc} of a better reconstruction of the incident neutrino energy; this is further discussed in Section \ref{sec:neutrinoenergy}.

\begin{figure}[t]
\centering
\includegraphics[height=7.5 cm]{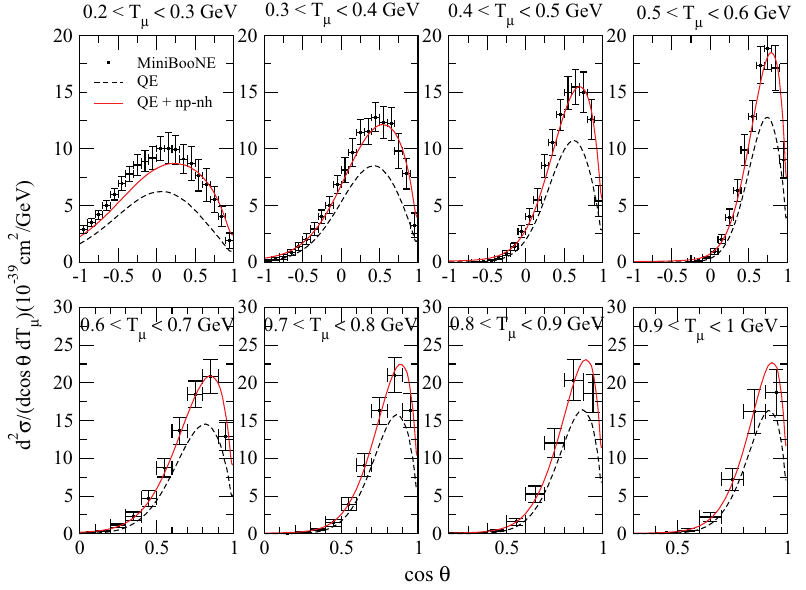}
\caption{\label{fig:mbxsec_vs_martini}The flux integrated double differential cross section measured at MinibooNE as a function of $\cos\theta$ for values of the scattered muon kinetic energy. The curves show a straight CCQE calculation (dashed) and the large improvement obtained with the inclusion of $n$-particle, $n$-hole contributions (solid) \cite{Martini:2011wp}.} 
\end{figure}

In order to report their data without a model-dependent inference on the incident neutrino energy, MiniBooNE provided the double differential cross sections of the scattered $\mu^-(\mu^+)$ in momentum and angle, along with the flux of incident $\nu_\mu(\bar{\nu}_\mu)$. The agreement of the Martini {\em et al.} calculations with the MiniBooNE data shown in Reference \cite{Martini:2010ex} are remarkable; an example is shown in Figure \ref{fig:mbxsec_vs_martini}. 

The formalism employed by Martini {\em et al.} proceeds as follows, starting with the weak Lagrangian:

\begin{equation}
\begin{array}{lll}
L_W & = &  \frac{G_F}{\sqrt{2}}\cos\theta_C l_\mu h^\mu \\
& & \\
\langle k', s' | l^\mu | k,s \rangle & =  & e^{-i q \cdot x} \bar{u}(k',s')  \left[ \gamma^\mu (1-\gamma_5) \right ] u(k,s) \\
& & \\
\langle p', s' | h^\mu | p,s \rangle & =  & e^{+i q \cdot x} \bar{u}(p',s') \times \\
&  & \left[F_1(t)  \gamma^\mu  + F_2(t) \sigma^{\mu\nu} \frac{iq_\nu}{2M_N} +G_A(t) \gamma^\mu\gamma_5 +G_P(t) \gamma_5 \frac{q^\mu}{2M_N}\right]\tau^+ u(p,s)
\end{array}
\end{equation}
where $u(k,s),\;\bar{u}(k',s')$ and $u(p,s),\;\bar{u}(p',s')$ are spinors corresponding to the initial/final lepton and hadron states, respectively.

\begin{figure}[t]
\centering
\includegraphics[height=7.5 cm]{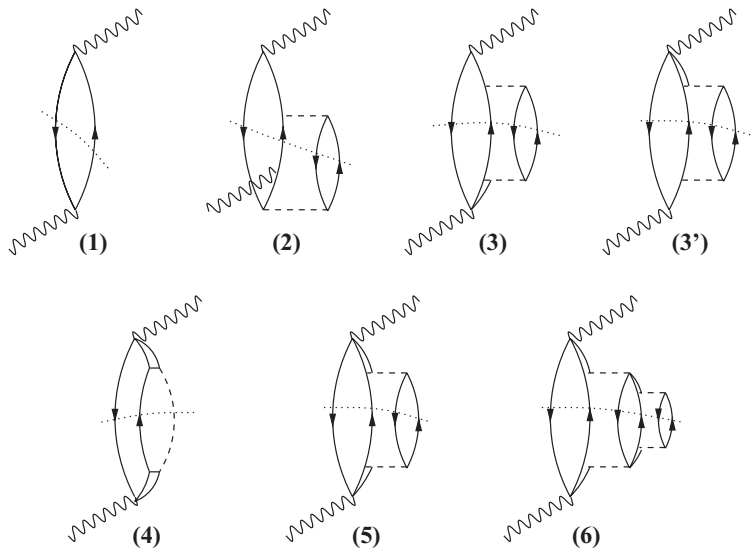}
\caption{\label{fig:rpa_diagrams} Feynman graphs of the RPA polarization propagators: (1) $NN$ quasi-elastic, (2) $NN(2p-2h)$, (3) $N\Delta(2p-2h)$, (3') $\Delta N(2p-2h)$, (4) $\Delta\Delta (\pi N)$, (5) $\Delta\Delta (2p-2h)$, (6)$\Delta\Delta(3p-3h)$~~\cite{Martini:2009uj}. The wiggly lines represent the external probe, the full lines the propagation of a nucleon, the double lines a $\Delta$, the dashed lines an interaction between the baryons.  The dotted lines show the particles that are placed on shell.}
\end{figure}

The expression for the resulting cross section is written as:
\begin{equation}
\label{eq:diffxsc_rpa}
\begin{array}{ll}
\frac{d^2\sigma}{d\Omega dk'} = & \frac{G_F^2 \cos^2\theta_c (k')^2}{2\pi^2} \cos^2\frac{\theta}{2} \times\\
 & [ G^2_E \left(\frac{Q^2}{{\bf q}^2}\right)^2 R_C + G^2_A \frac{(M_\Delta-M_N)^2}{2 {\bf q}^2} R_L' + \\
 &+ (G^2_M \frac{\omega^2}{{\bf q}^2} +G_A^2)\left(\frac{Q^2}{{\bf q}^2} + 2 \tan^2 \frac{\theta}{2} \right)R_T \pm 2G_A G_M \frac{k+k'}{2M_N} \tan^2\frac{\theta}{2} R_T]\\
\end{array}
\end{equation}

where the $t$-dependence ($t\equiv q^2 = -Q^2)$ of the Sachs nucleon form factors ($G_E, G_M$, and $G_A$) has been suppressed. The various nuclear response functions depend on $Q^2$ and $\omega$ ; $R_C$, the charge response; $R_L'$,the isospin-spin longitudinal response; and $R_T$, the isospin-spin transverse response. The $R_L'$ response arises from the effect of soft pion exchange on the axial current \cite{Kubodera:1978wr,Towner:1987ns} and is not a large contribution. On the third line, the enhancement known to exist for the magnetic response is also attributed to the axial response. This enhancement of both the axial and magnetic contributions is further amplified in the interference term for neutrino scattering (last term in Equation~\ref{eq:diffxsc_rpa}). These transverse enhancements are attributed to the effects of the tensor force arising from meson exchanges and are included diagrammatically with examples shown in Figure \ref{fig:rpa_diagrams}~\cite{Martini:2009uj}. This formalism reproduces the MiniBooNE data without any need to increase the axial mass in the nucleon form factors because of the enhancement of both the axial and magnetic responses. 

Equation \ref{eq:diffxsc_rpa} shows that Martini {\em et al.} assumed the enhancement of the transverse magnetic and axial responses are identical. There is no obvious justification for this apart from the similarity of a non-relativistic reduction of their associated operators ($(\vec{\sigma}\times {\bf q}/|{\bf q}|)\tau^\pm$). The effects of pion exchange on vector and axial vector currents are known~\cite{Kubodera:1978wr,Towner:1987ns} to be quite different
    
While very successful in reproducing the MiniBooNE data, it is not clear that this approach is robust. The weak current employed by Martini {\em et al.} is one-body so it is not clear to what degree conservation of the vector current is respected. Furthermore, all hadronic states are represented as plane-waves, which is certainly not correct and requires taming some two body integrals that would otherwise be infinite \cite{Amaro:2010iu}.  

\begin{figure}[t]
\centering
\includegraphics[height=5 cm]{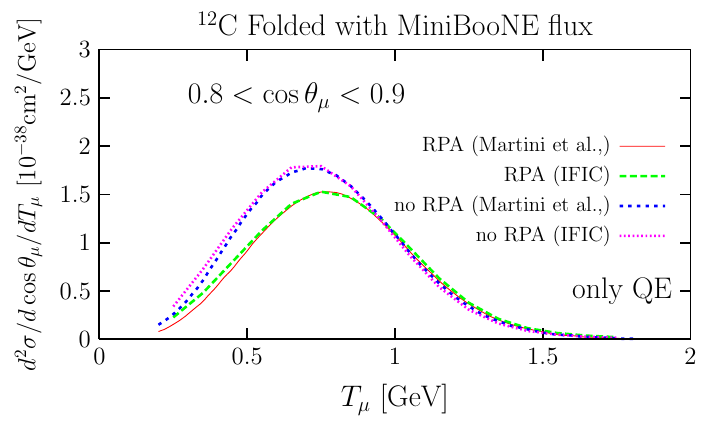}\\
\includegraphics[height=5 cm]{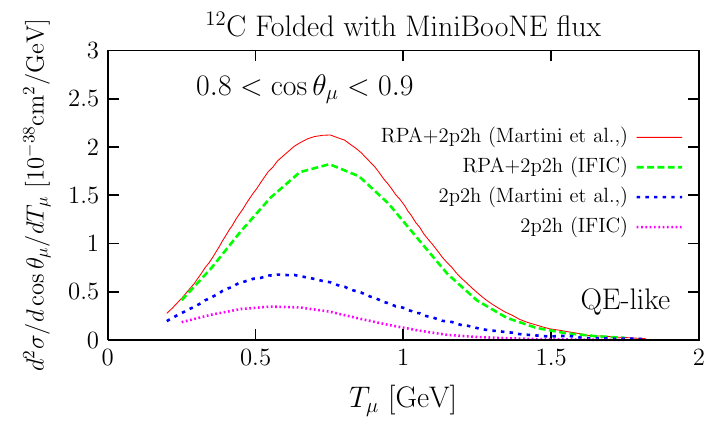}
\caption{\label{fig:nieves_mb}Comparison shown by J. Nieves of the calculated muon spectra at scattering angles $0.8<\cos\theta<0.9$
for $\nu_\mu$ CCQE scattering off carbon integrated over the MiniBooNE flux by Martini {\em et al.} to that of the IFIC collaboration \cite{Nieves:2011pp}. The top figure shows the spectra with and without the inclusion of the long range RPA correlations while the bottom shows the effect of including short range $2p-2h$ correlations.
}
\end{figure}

Another extensive body of work employing RPA was presented by Juan Nieves. This work is chiefly carried out with collaborators at IFIC and CSIC at the University of Valencia. They have published a series of papers \cite{Nieves:2011pp,Nieves:2011yp,Nieves:2012yz,Nieves:2012yz,Nieves:2013fr,Gran:2013kda} that cover all essential aspects of neutrino-nucleus CCQE scattering, particularly with regard to the MiniBooNE data. 
Many aspects of their work are similar to those of Martini {\em et al.} as both employ a Fermi gas model with  long and short-range interactions implemented via the RPA. Figure \ref{fig:nieves_mb} shows their results for the CCQE cross section on $\!\;^{12}\mbox{C}$ calculated with the MiniBooNE neutrino flux. Of note is the large difference between their calculation and that from Martini {\em et al.} in the contribution from $2p-2h$ configurations. We speculate that this may be the result of IFIC having no enhancement of the axial response in their calculation. Such enhancement is still an issue under investigation.  In the presentation of the RPA formalism and results, terms such as ``$n$-particle, $n$-hole'' final states are often used, creating some confusion with $n$-body currents employed in other formalisms. 

\begin{figure}[t]
\centering
\includegraphics[height=6 cm]{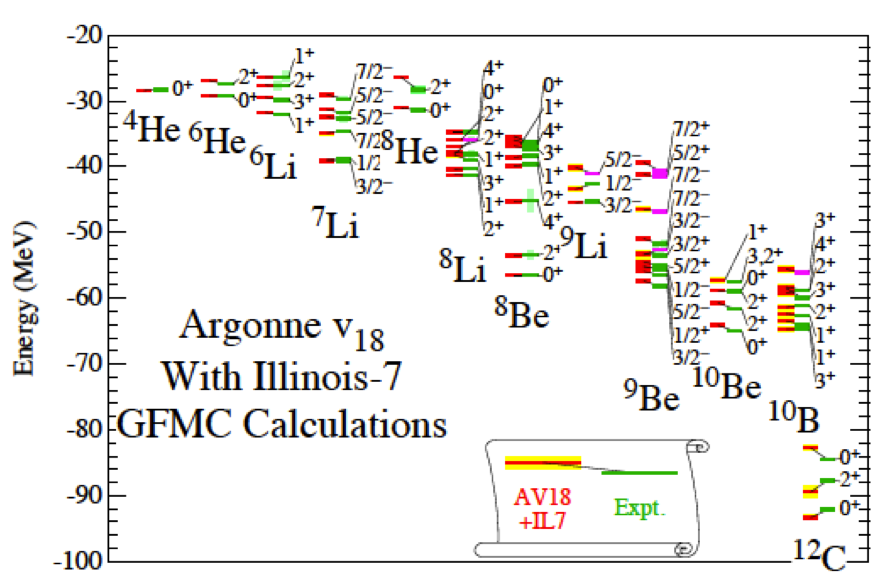}
\caption{\label{fig:gfmc_binding}The calculated binding energies of the ground and excited states of the nuclei from $\!\;^{4}\mbox{He}$ to $\!\;^{12}\mbox{C}$ using the Argonne V18 nucleon-nucleon interaction and the Illinois-7 3-body interaction shown by J. Carlson~\cite{pieper}. }
\end{figure}

\subsection{Green's Function Monte Carlo Approach}
Joe Carlson and Rocco Schiavilla reported on the work done by a group referring to itself as the Nuclear Computational Low Energy Initiative (NUCLEI). The group has a background in electron scattering \cite{Carlson:2001mp} and only recently turned its attention to neutrino CCQE scattering from nuclei. NUCLEI's approach is very different from any other presented at the workshop. They employ a parameterized version of the nucleon-nucleon interaction (VA18), including 3-body forces (TNI), and solve for the low-lying states of the nucleus using quantum Green's Function Monte Carlo (GFMC) techniques. The computational complexity of such an approach is currently limited to light nuclei ($A \le 12$). This very fundamental approach has proven very successful in reproducing the spectra and transitions of the low lying states of light nuclei as shown in Figure \ref{fig:gfmc_binding}. The energy of the nuclear states from $\!\;^{4}\mbox{He}$ to $\!\;^{12}\mbox{C}$ are predicted to better than 0.5 MeV over a scale of $\sim\! 70~\mbox{MeV}$. Of course, this success is no measure of how well quasi-elastic scattering will be predicted. 

\begin{table}[tb]
\begin{tabular}{ccccccccc}\hline\hline
                & \multicolumn{4}{c}{$S_L$}
                & \multicolumn{4}{c}{$S_T$} \\
                & \multicolumn{2}{c}{$\!\;^{3}\mbox{He}$}
                & \multicolumn{2}{c}{$\!\;^{4}\mbox{He}$}
                & \multicolumn{2}{c}{$\!\;^{3}\mbox{He}$}
                & \multicolumn{2}{c}{$\!\;^{4}\mbox{He}$} \\
                & \multicolumn{2}{c}{-----------------}
                & \multicolumn{2}{c}{-----------------}
                & \multicolumn{2}{c}{-----------------}
                & \multicolumn{2}{c}{-----------------} \\
$q\;(\mbox{MeV}/c)$ & 1 & 1+2& 1 & 1+2 & 1 & 1+2 & 1 & 1+2 \\ \hline
300 & 0.787 & 0.763 & 0.670 & 0.649 & 0.929 & 1.31 & 0.893 & 1.67 \\
400 & 0.921 & 0.875 & 0.859 & 0.815 & 0.987 & 1.30 & 0.970 & 1.62 \\
500 & 0.964 & 0.901 & 0.941 & 0.881 & 1.01  & 1.28 & 1.00  & 1.55 \\
600 & 0.982 & 0.908 & 0.973 & 0.910 & 1.01  & 1.25 & 1.01  & 1.49 \\
700 & 0.994 & 0.914 & 0.994 & 0.942 & 1.01  & 1.23 & 1.01  & 1.44 \\ \hline
\end{tabular}
\caption{\label{tab:gfmc_sumrules}The calculated sum rules from Reference \cite{Carlson:2001mp} for the vector longitudinal (L) and transverse (T) responses in $\!\;^{3}\mbox{He}$ and $\!\;^{4}\mbox{He}$ as a function of the transferred 3-momentum $q$. The values are shown for both a one body current (1) and including 2-body currents (1+2).}
\end{table}

\begin{figure}[t]
\centering
\includegraphics[height=4 cm]{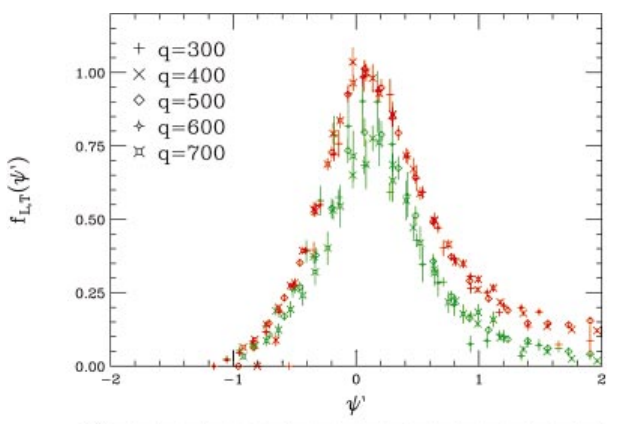}
\includegraphics[height=4 cm]{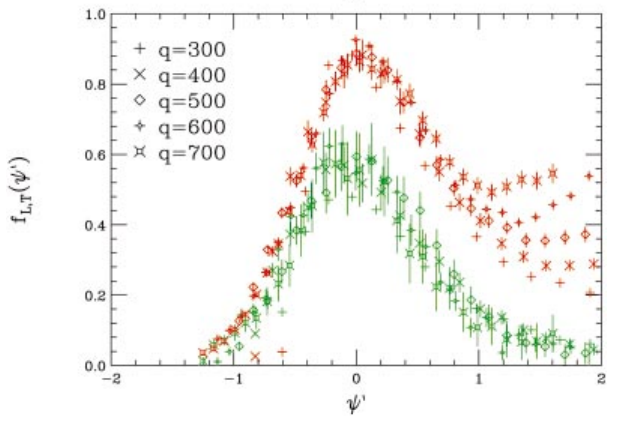}
\caption{\label{fig:gfmc_flt}The experimentally extracted longitudinal (green, lower band) and transverse (red, uppper band) responses for $\!\;^{3}\mbox{He}$ (left) and  $\!\;^{4}\mbox{He}$ (right)~\cite{Carlson:2001mp}. Recall in a relativistic Fermi gas (RFG) impulse approximation description of quasi-elastic scattering, the longitudinal and transverse responses are identical.}
\end{figure}

The most convincing demonstration of the utility of this approach is from quasi-elastic electron scattering from $\!\;^{3}\mbox{He}$ and $\!\;^{4}\mbox{He}$. Figure \ref{fig:gfmc_flt} shows the experimentally extracted longitudinal and transverse response functions in terms of $\psi'$, the scaling variable introduced in Section \ref{sec:electron}~\cite{Carlson:2001mp}. In each case, the contributions of charge, magnetic moment, and form factor of the individual nucleon have been removed from the response functions. The large change between $\!\;^{3}\mbox{He}$ and $\!\;^{4}\mbox{He}$ is presumably due to the increased binding energy and resulting higher density of the $\!\;^{4}\mbox{He}$ nucleus. The longitudinal response is broadened in $\!\;^{4}\mbox{He}$ though the integral over $\psi'$ remains the same, while the transverse response is both broadened and increased in integral. The change in the longitudinal response shows that the nucleon momentum in $\!\;^{4}\mbox{He}$ is higher and that there are more collisions in which the energy transfer is greater than would be inferred from the impulse approximation. The large increase in the relative transverse to longitudinal response shows that effects beyond individual nucleon physics must be at play. Table \ref{tab:gfmc_sumrules} shows the calculated sum rule for the longitudinal and transverse responses for $\!\;^{3}\mbox{He}$ and $\!\;^{4}\mbox{He}$ at different 3-momentum transfers. The longitudinal and transverse sum rules are integrals over the transferred energy at fixed three-momentum transfer as follows: 
\begin{equation}
\begin{array}{ll}
S_{L/T}(q) & =   C_{L/T} \int_{\omega_{th}}^{\infty} d\omega\; R_{L/T}(q, \omega)  \\
                 &  = C_{L/T} \left[\langle 0 | O^*_{L/T} (\vec{q}) O_{L/T}(\vec{q}) |0\rangle - |\langle 0 | O^*_{L/T}(\vec{q}) | 0\rangle|^2 \right] 
\end{array}
\end{equation}
 with $C_L=1/Z$ and $C_T = \frac{2M_N^2}{Z \mu^2_p + N \mu^2_n}$. The respective nuclear ground states are represented as $|0\rangle$  and $\langle 0|$ while the rightmost squared term is the contribution of elastic scattering off the target nucleus. The values of the scale factors $C_{\alpha\beta}$ are chosen such that the value of $S_{\alpha\beta}\to 1$ as $q\to\infty$. For example:
\begin{equation}
C^{-1}_{XY} = -\frac{q}{M} G_A (Q^2_{qe}) \left[Z \bar{G}^p_M(Q^2_{qe} - N\bar{G}^n_M(Q^2_{qe})   \right]
\end{equation}
where $\bar{G}^{n/p}_M(Q^2_{QE})$ are the neutral weak magnetic form factor of the proton.

Table \ref{tab:gfmc_sumrules} shows that the calculated integral of the longitudinal response remains nearly unaffected by the inclusion of two-body currents. On the other hand, the calculated integral of the transverse response shows sizable enhancement consistent with data from electron scattering when two-body currents are included. This is presumably the same effect observed in RPA calculations where much of the same physics is included via inclusion of various strong interactions between hadrons, but in this case current conservation is assured.

The NUCLEI collaboration generates their two-body currents using CVC and PCAC. For the vector current, CVC requires the divergence of the current to equal the commutator of the charge with the Hamiltonian:
\begin{equation}
\vec{q}\cdot \left[ \vec{j}^{(1)} + \vec{j}^{(2)} + \vec{j}^{(3)}(V^{2\pi})\right] = [H, \rho] = \left[T + V + V^{2\pi}, \rho\right]
\end{equation}
A complete presentation of the one and two-body vector and axial currents they employ is found in Reference \cite{Shen:2012xz}. 

\begin{figure}[t]
\centering
\includegraphics[height=11 cm]{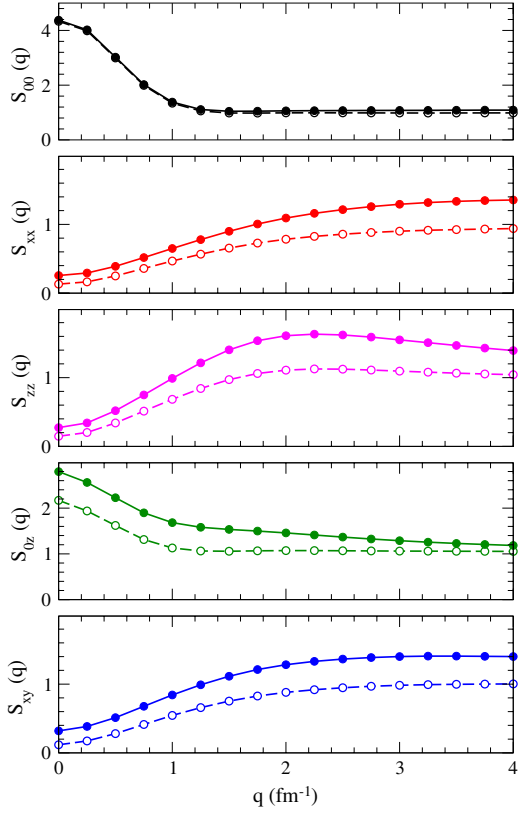}
\caption{\label{fig:gfmc_s}The calculated sum rule value for each of the five response functions as a function of the three- momentum transfer $|{\bf q}|$, where the one-body contribution is shown by the open circles, and the sum of the one- and two-body contributions is shown with solid circles~\cite{Lovato:2014eva}. $S_{00}$, $S_{0Z}$, $S_{ZZ}$ are longitudinal while $S_{XX}$ and $S_{XY}$ are transverse.}
\end{figure}

To date, the NUCLEI collaboration has published two papers dealing directly with neutrino interactions on nuclei. The first \cite{Shen:2012xz} deals with neutrino-deuteron CCQE and NCQE and introduces the formalism. Because the deuteron is so simple and the short range effects so small, the calculation of NCQE and CCQE is sufficiently reliable that it could serve to normalize the neutrino flux via the $\nu_l + d \to l^- + 2 p$  reaction to better than 5\%. A further benefit to the use of a deuterium target arises if measurement of the momentum of the protons in the final state is possible. This fixes the transferred ${\bf q}$ and $\omega$ and the determination of the incident neutrino energy is limited only by the precision of the measured charged lepton energy.

A calculation more directly relevant to neutrino–-nucleus scattering can be found in their recent work~\cite{Lovato:2014eva} on the inclusive neutral current scattering of neutrinos and anti-neutrinos from $\!\;^{12}\mbox{C}$. Figure \ref{fig:gfmc_s} shows the calculated sum rule values for the individual five response functions that characterize the inclusive cross section. The five response functions are defined as follows, where the direction of the three momentum transfer ${\bf q}$ is defined to be along the $z$-axis: 
 \begin{equation}
\begin{array}{ll}
\label{eq:response}
\frac{d^2\sigma}{d\epsilon'd\Omega} = & \frac{G_F^2}{2\pi^2} k' \epsilon' \cos^2\frac{\theta}{2} \times\biggl[ R_{00} + \frac{\omega}{|{\bf q}|^2} R_{zz} - \frac{\omega}{|{\bf q}|} R_{0z} \\
 & \\
 & + \left(\tan^2\frac{\theta}{2} + \frac{Q^2}{2|{\bf q}|^2}\right)R_{xx} \mp \tan\frac{\theta}{2}\sqrt{\tan^2\frac{\theta}{2} + \frac{Q^2}{|{\bf q}|^2}} R_{xy}\biggr]
\end{array}
\end{equation}
where
\begin{equation}
R_{\alpha\beta}({\bf q},\omega) \sim \sum_f \sum_i \delta(\omega+M_A -E_f)\;\langle f | j_\alpha({\bf q},\omega)|i\rangle \langle f | j_\beta({\bf q},\omega)|i\rangle^*
\end{equation}

Figure~\ref{fig:gfmc_s} shows the values for the sum rule of each of the response functions shown in Equation~\ref{eq:response}. The open circles are the values for just the one-body currents while the closed circles are the values when two-body currents are included. As expected the charge response ($S_{00}$) shows no enhancement with the inclusion of two-body currents, while the important transverse summed responses ($S_{xx}$ and $S_{xy}$) show appreciable enhancement (40\%). The transverse responses, $S_{xx}$ and $S_{xy}$, include both the vector magnetic and the axial vector couplings.

\begin{figure}[t]
\centering
\includegraphics[height=6 cm]{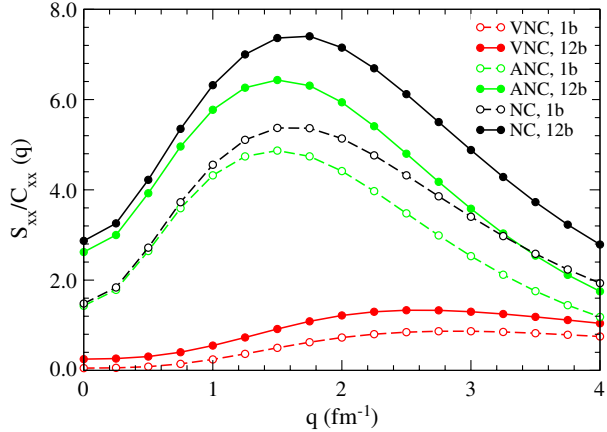}
\caption{\label{fig:gfmc_s_vec_axial} The sum rules for neutral current (NC) transverse response for $\!\;^{12}\mbox{C}$ showing the vector (VNC) and axial vector (ANC) response separately, where the open circles show the one-body response only, and the solid circles show the sum of the one- and two-body response~\cite{Lovato:2014eva}.}
\end{figure}

More surprising is the enhanced axial response, not explicit in the
$S_{xx}$ and $S_{xy}$ of Figure~\ref{fig:gfmc_s}, but clearly present in Figure~\ref{fig:gfmc_s_vec_axial}. The enhancement
of the axial response is sizable and present at all three-momentum transfers and
given its role in the coherent transverse interference response ($S_{xy}$), it is
clear that its neglect leads to seriously under predicting the NCQE and  CCQE cross sections on carbon. This axial enhancement results when two-body currents are included as is  the case for the transverse vector response. The origin of this two-body current  enhancement in the case of the axial response has not yet been identified by the authors. The result, if correct, supports the conjecture of Martini {\em et al.} and presents difficulty to all who would enhance the CCQE neutrino-nucleus  inclusive cross section only via the transverse vector response. This may be the most significant result presented at the workshop.

\subsection{Relativistic Extensions}
The approach of the NUCLEI collaboration is necessarily non-relativistic and serious questions were raised at the workshop as to how to extend this work to the momentum and energy transfers encountered with neutrino beams with energies of ${\mathcal O}(\mbox{GeV})$ and above. It is useful to recall that even at very high neutrino energy, much of the cross section involves energy and momentum transfers in the non-relativistic regime. Bill Donnelly presented specific suggestions on how to proceed.
He pointed out there are at least three places where relativistic effects come into play: 
\begin{enumerate}
\item Kinematics
\item Boost effects on the nuclear current matrix elements
\item Dynamical effects in the wave functions
\end{enumerate} 
The kinematic effects are straightforward and are often employed. The final state ejected nucleon (nucleons) should obey relativistic kinematics, $E=\sqrt{{\bf p}^2+m^2}$. Replacing $\omega$  by $\omega(1+\omega/2m)$ will place the center of the QE peak at the right position, namely $|Q^2|/2m$ rather than $|{\bf q}|^2/2m$. Boost factors typically take the form of factors of $\gamma$ and $\gamma^{-1}$, to first order, with the charge (longitudinal) response enhanced by $\gamma$ while the transverse response is suppressed by $\gamma^{-1}$. The modification of the wave functions can be treated when employing a Dirac-Hartree approach which provides the modification between the upper and lower components of the wave function relative to the free particle case.

Omar Benhar also addressed how one might extend CCQE calculations into the relativistic regime with a consistent treatment of one and two-body currents within a spectral function formalism~\cite{Benhar:2013bba}. He noted that while the initial nuclear state is non-relativistic, the leptonic current and final hadronic state usually are not. Typical CCQE and NCQE neutrino measurements involve a tangled mix of small and large three-momentum transfers. The target response is written as:
 \begin{equation}
 W_A^{\mu\nu} = \sum_N\langle 0|J_A^{\mu\dagger}|N\rangle \langle N | J_A^\nu |0 \rangle \; \delta^{(4)}(P_0 +k -P_N - k')
 \end{equation}
There are no two-nucleon currents in the Fermi gas/impulse approximation approach to quasi-elastic scattering. Thus, the current 
is simply:
 \begin{equation}
J_A^\mu(q) = \sum_{i=1}^A j_i^\mu(q)
 \end{equation}
The resulting final state is then a product: 
\begin{equation}
|N\rangle \to |{\bf p}\rangle \otimes |n_{A-1},{\bf p}_n\rangle
\end{equation}
where $|{\bf p}\rangle$  is a free nucleon with momentum ${\bf p}$   and $ |N_{A-1},{\bf p}_n\rangle$  is the recoiling residual nucleus in the state $n$ with momentum ${\bf p}_n$. Thus, the matrix element of the one nucleon current becomes: 
 \begin{equation}
 \langle N | j_1^\mu | 0\rangle = \int d^3{\bf k} \; M_n({\bf k}) \; \langle {\bf p} \; | j_1^\mu | \; {\bf k}\rangle
 \end{equation}                                                                                                    
with
 \begin{equation}
 M_n({\bf k}) = \left\{ \langle n_{A-1}, {\bf p}_n| \otimes \langle {\bf k} | \right\} |0\rangle
 \end{equation}                          
                     
Employing a nuclear spectral function  $P({\bf k}, E)$ to provide the probability of removing a nucleon with momentum  ${\bf k}$ leaving the residual system with energy $E$,  the differential for scattering from a nucleus $A$ is:
\begin{equation}
d \sigma_A = \int d^3{\bf k} \; dE \; d\sigma_N \; P({\bf k}, E)
\end{equation}                                          
where $d\sigma_N$ is the single nucleon cross section and
\begin{equation}
P({\bf k}, E) = \sum_n |M_n({\bf k}) |^2 \delta(E_0 + E - E_n)
\end{equation}

This approach is incomplete as two-body currents must be included. These can be accurately computed using a non-relativistic GFMC approach as used by NUCLEI.  Following the {\em ansatz} developed above one now defines a final state: 

\begin{equation}
|N\rangle \to |{\bf p}\;{\bf p'} \rangle \otimes |m_{A-2}, {\bf p}_m \rangle
\end{equation}    	 
along with a two-nucleon spectral function $P({\bf k}, {\bf k}', E)$. Benhar {\em et al.} employ an earlier two-nucleon spectral function for uniform isospin symmetric infinite nuclear matter. Note the two-body amplitude:
\begin{equation}
M_n({\bf k}, {\bf k}') = \left\{ \langle n_{A-2}| \langle {\bf k}, {\bf k}' | \right\} \otimes |0\rangle
\end{equation}                      
is independent of ${\bf q}$ and is obtained from non-relativistic theory while the relativistic collision interaction is isolated in: 
\begin{equation}
\langle N | j_{ij}^\mu| 0\rangle = \int d^3{\bf k}\; d^3{\bf k}' \; M_n({\bf k}, {\bf k}') \langle {\bf p},{\bf p}' | j^\mu_{ij} | {\bf k}, {\bf k}'\rangle
\end{equation}

\begin{figure}[t]
\centering
\includegraphics[height=7.5 cm]{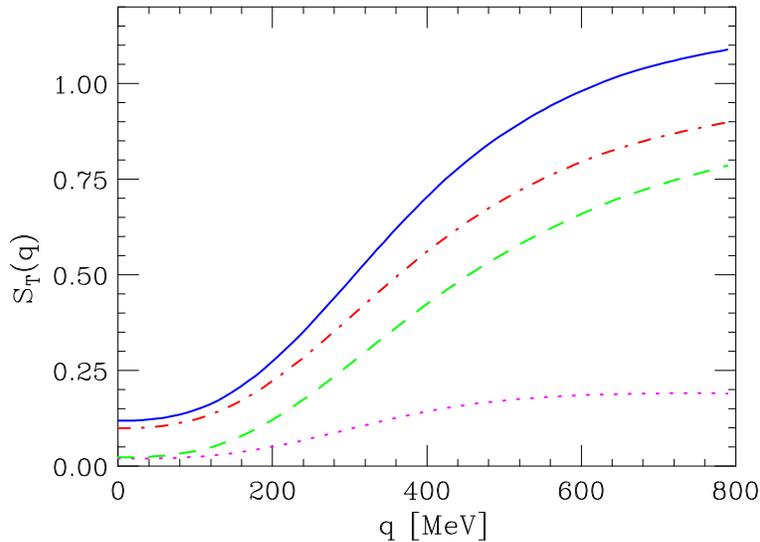}
\caption{\label{fig:sumrule_transvector}
Sum rule of the transverse vector response for $\!\;^{12}\mbox{C}$ the dashed line is for one-body currents~\cite{Benhar:2013bba}. The dot-dashed line is sum of the contributions of one- and two-body currents, neglecting their interference. The solid line is the result including the interference. The dotted line at the bottom shows the size of the one-body/two-body interference term.}
\end{figure}

Figure \ref{fig:sumrule_transvector} shows the calculated enhancement with the inclusion of these two-body currents for the sum rule of the vector transverse response in $\!\;^{12}\mbox{C}$. The large enhancement seen in NUCLEI's non-relativistic calculation is preserved with presumably far less computation. A further point made in Figure \ref{fig:sumrule_transvector} is the significance of the interference between one- and two-body currents in the inclusive CCQE cross section. In a system of {\em interacting nucleons}, one- and two-body currents can readily lead to the same final states, a fact often obscured when using terminology such as one particle-one hole and two particle-two hole final states.

\begin{table}[t]
\begin{tabular}{lcl} \\ \hline\hline

{\bf Present SuSA}                          & & {\bf SuSAv2} \\ \hline
{\footnotesize Based on the superscaling function}& & {\footnotesize The Relativistic Mean Field (RMF)}  \\
{\footnotesize extracted from QE electron-nucleus}& $\Rightarrow$ & {\footnotesize model is employed to improve the data} \\
{\footnotesize scattering data}                   & & {\footnotesize analysis, where RMF accounts for FSI} \\ \hline
                                      & & \\
                                      & & \\ 
{\bf Longitudinal}                          & & {\bf Longitudinal + Transverse} \\ \hline
{\footnotesize Description of nuclear response built} & & {\footnotesize Differences between transverse and} \\
{\footnotesize only on the longitudinal scaling}      & $\Rightarrow$ & {\footnotesize longitudinal scaling functions are}  \\
{\footnotesize function. Assumption of $f_L(\psi)\approx f_T(\psi)$,} & & {\footnotesize introduced in order to describe} \\
{\footnotesize scaling of the 0th kind.}             & & {\footnotesize properly the nuclear response.} \\ \hline
                                                     & &  \\
                                                     & &  \\ 
{\bf Isoscalar + Isovector Structure}                & & {\bf Isovector Structure} \\ \hline
{\footnotesize The scaling function based on QE}     & & {\footnotesize We separate the scaling function into} \\
{\footnotesize electron scattering data takes into}  & & {\footnotesize isovector and isoscalar structure so as}\\
{\footnotesize account isovector and isoscalar}      & $\Rightarrow$ & {\footnotesize to employ a purely isovector scaling} \\
{\footnotesize currents to describe the interaction} & & {\footnotesize function for CCQE neutrino-nucleus} \\
{\footnotesize between the electron and nucleus.}    & & {\footnotesize processes where isospin changes.} \\ \hline
\end{tabular}
\caption{\label{tab:susa_v1v2}Table contrasting the differences between SuSA and SuSAv2 shown by J. Caballero.}
\end{table}

\begin{figure}[t]
\centering
\includegraphics[height=3.8 cm]{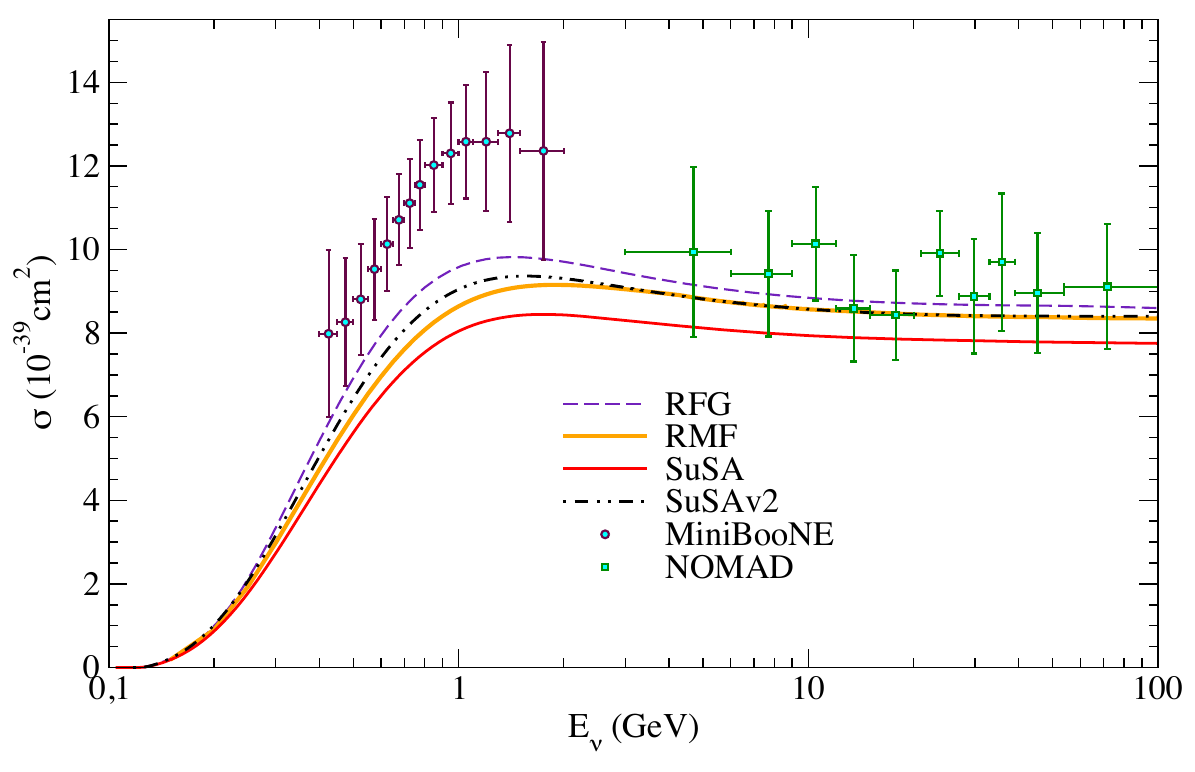}
\includegraphics[height=3.8 cm]{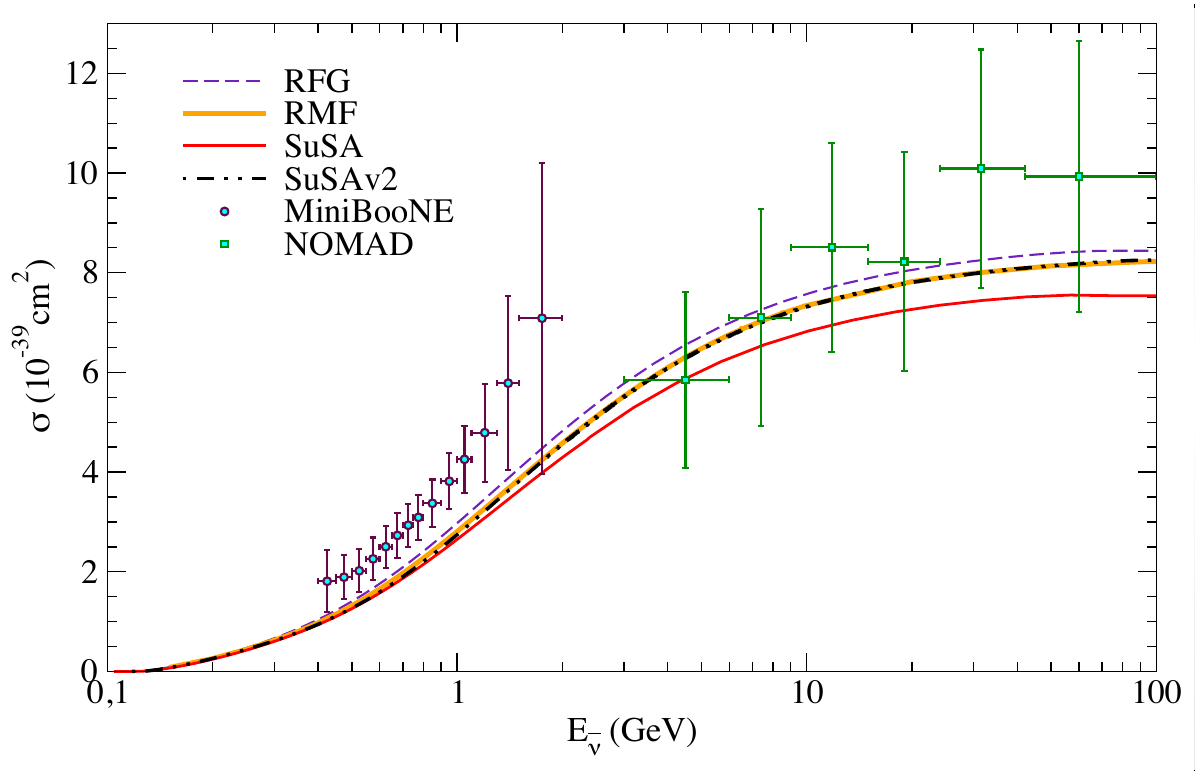}
\caption{\label{fig:susa_nunubar} Comparison to MiniBooNE and NOMAD $\nu_\mu$ (left) and $\bar\nu_\mu$ (right) CCQE measurements as a function of neutrino energy with various SUSA~\cite{Amaro:2013yna} and RMF prescriptions shown by J. Caballero.}
\end{figure}

\subsection{SuSA and Relativistic Mean Field Approaches}
A very different approach to neutrino-nucleus scattering is taken by theorists who employ scaling from electron scattering and various formulations of relativistic mean field theory. Some aspects of this approach were covered earlier in the discussion of scaling in electron-nucleus quasielastic scattering. Barbaro presented an early approach termed the super scaling approach (SuSA) to neutrino scattering \cite{Amaro:2004bs} 
 employed by a Basel, INFN, MIT and Turino collaboration. The approach is attractive as it supplements a conventional view of CCQE with  explicit inclusion of meson exchange currents. However, in assuming  $f_L(\psi')\simeq f_T(\psi')$, it had difficulty reproducing the MiniBooNE data.  Extensions of this work were reported by J. Caballero in an approach designated SuSAv2; Table \ref{tab:susa_v1v2} summarizes some of the major differences between SuSA and SuSAv2. SuSAv2 incorporates the relativistic mean field, which was noted to reproduce the observed shape of the longitudinal vector response (Figure~\ref{fig:psil_rmf}). It also was noted that this formulation also produces a transverse vector response greater than the longitudinal response. Both results are consistent with observation in quasielastic electron scattering. However, application of this procedure underpredicts the MiniBooNE measurements
but is in agreement with NOMAD and MINERvA. The data from the latter two experiments is at higher incident neutrino energy than MiniBooNE. Figure \ref{fig:susa_nunubar} shows the comparison of several models to neutrino CCQE data. It is not clear that the models result in a consistent definition for the CCQE process as selected by MiniBooNE or NOMAD and, possibly more relevant, the RMF shows no axial enhancement. 

\begin{figure}[t]
\centering
\includegraphics[height=4 cm]{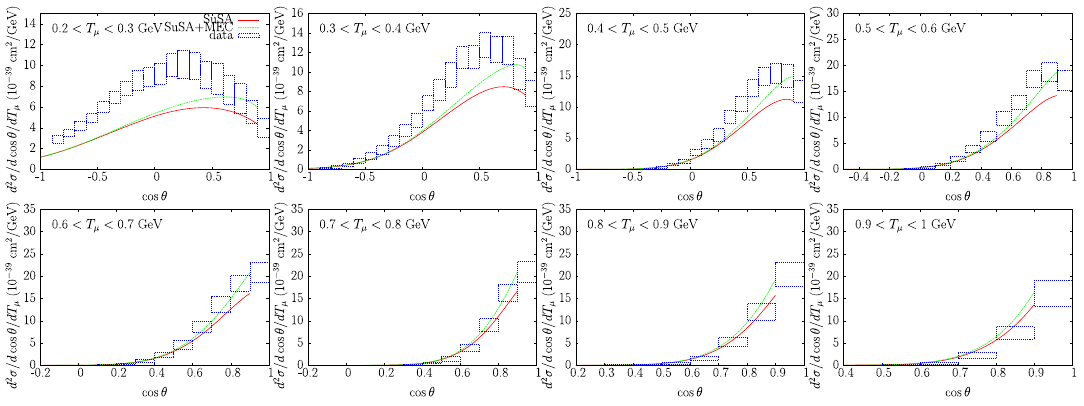}
\caption{\label{fig:susa_mec_mb}SUSA predictions with the inclusion of MEC compared MiniBooNE CCQE differential cross section measurements~\cite{Amaro:2010sd}.}
\end{figure}

The inclusion of $2p-2h$ excitations and the effects of meson exchange channels that are omitted in SuSA and SuSAv2 were addressed in Amaro's presentation. There are two papers in the literature addressing this shortcoming in CCQE electron scattering \cite{De Pace:2003xu,Amaro:2010iu} and presenting the formalism. Applying these corrections to the MiniBooNE data did not appear to produce a sufficient increase in the yield. Figure \ref{fig:susa_mec_mb} shows the increased yield produced by meson exchange currents (MEC) in a more stringent test where it is compared to the MiniBooNE differential cross section measurements. There appear to be several theoretical and computational issues that need to be addressed before this phase of the work is complete. In particular, the contribution to the transverse axial response needs to be better understood. 

\begin{figure}[t]
\centering
\includegraphics[height=8 cm]{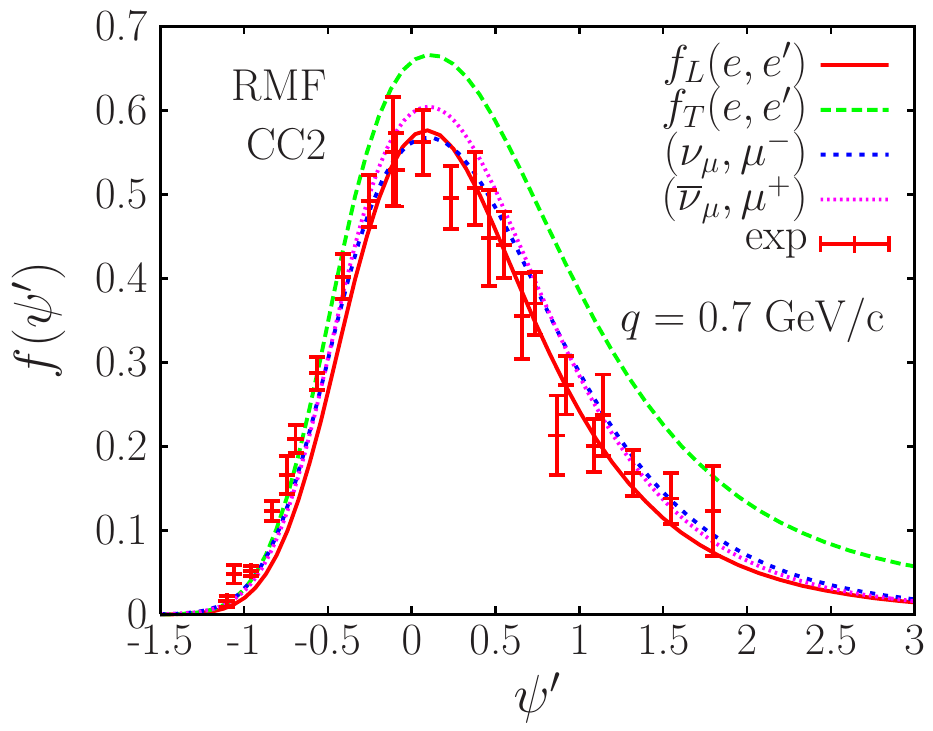}
\caption{\label{fig:susa_rmf_fl}Calculated response functions using RMF compared to the experimentally measured longitudinal response in electron QE scattering.}
\end{figure}

\begin{figure}[t]
\centering
\includegraphics[height=8 cm]{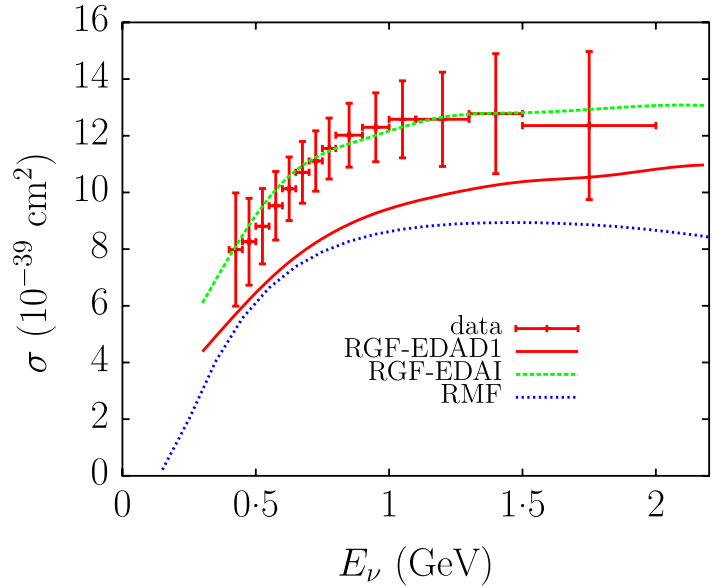}
\caption{\label{fig:rgf_mb}Fit to the MiniBooNE data with RGF employing different (EDAD1, EDAI) relativistic optical potentials for the final state proton~\cite{Meucci:2011vd}.}
\end{figure}

A likely source of the short comings of the approach of this group is displayed in Figure \ref{fig:susa_rmf_fl} that was shown by Caballero.  Recall that the RMF is used to account for the effects of strong interaction dynamics on lepton-nucleus scattering. Quite good results were obtained for electron-nucleus scattering with an asymmetric longitudinal response function and an enhanced transverse response. However Figure \ref{fig:susa_rmf_fl} shows that the total CCQE response lies closest to the $(e,e')$ longitudinal response. This implies that the RMF produces no enhancement of the axial response. In that case, it is impossible to reproduce the MiniBooNE CCQE measurements as not only is the enhancement of the axial response missing, but the further enhancement it provides to the interference term is also absent.
 
Carlotta Giusti presented the work of the Pavia group and its collaborators on employing a Green's function model to a RMF calculation of neutrino-nucleus CCQE scattering. There is a large body of their work starting with QE electron scattering in the early 90s \cite{Capuzzi:1991qd} through their most recent publication \cite{Meucci:2014bva} discussing the MINERvA CCQE data. There is extensive experience in the application of RMF to various inclusive and exclusive QE processes. While the particular recipe being applied may not be obvious to a non-expert, the eventual fit to data using a relativistic Green's function (RGF) is remarkably good \cite{Meucci:2011vd} as shown in Figure \ref{fig:rgf_mb}. The EDA1 and EDAD1 are two different optical model potentials. The former is energy dependent but $A$-independent while the latter is energy independent but $A$-dependent and in this case is chosen to best fit elastic scattering of protons from $\!\;^{12}\mbox{C}$. From Figure \ref{fig:susa_rmf_fl}, it is evident that crucial effects must be obtained from the final state interaction of the outgoing nucleon with the potential generated by the $A-1$ residual nucleons. In this model, it is hoped that that the complex physics is all captured in the final state interaction with a relativistic optical potential.

It is certainly surprising that so much apparently complex physics contained in the GFMC approach can be captured in the final state interactions of the outgoing nucleon with an appropriate RMF.  While there is presently appreciable latitude in determining the appropriate RMF to apply in a particular case, it clearly is of interest and importance to better understand the RMF approach to remove this ambiguity and establish contact with the relevant underlying physics. 

In addition to the presentations covered above, presentations by Bodek, Sobczyk, Garvey, and Mosel, dealing with neutrino-nucleus inclusive CCQE and NCQE scattering obtained the requisite enhancement of the observed cross sections by including the effects of correlated pairs at various levels of sophistication. There seemed to be no question that two-body correlations and extension of the impulse approximation be added to any Fermi gas or non-relativistic mean field model of the nuclear ground state. Some place all of the enhancement of the cross section into the transverse {\em vector} response as this had clearly been observed in electron scattering. A major issue to be settled is the enhancement of the transverse {\em axial} response relative to that predicted in the impulse approximation. Its role in neutrino CCQE is very large due to constructive interference with the transverse vector response. 
 
The experimental community, particularly those associated with LAr-TPCs, is certainly interested in extensions beyond predicting the inclusive CCQE cross section and desire a more exclusive description that requires dealing with final state interactions at a minimum in order to predict the particle emission, As the inclusive CCQE cross section allows integration over all final CCQE channels, a reliable more exclusive treatment requires far more information and computation.

\section{Establishing the Incident Neutrino Energy}
\label{sec:neutrinoenergy}
Establishing the incident neutrino energy in CCQE reactions is critical for neutrino oscillation measurements. It directly affects the determination of $\Delta m^2$ and indirectly the mixing angle. While the increased yield due to more complex nuclear weak currents increases the effective statistics of these experiments, the simple Fermi Gas-based recipes relating the incident neutrino energy to the momentum of the final charged lepton are no longer valid. Several of the presentations at the workshop addressed this issue but none at great length. This subject is still a work in progress with the most relevant work to date from Martini {\em et al.}~\cite{Martini:2012fa,Martini:2012uc}, Nieves {\em et al.}~\cite{Nieves:2012yz}, Lalukulich, {\em et al.}~\cite{Lalakulich:2012hs,Mosel:2013fxa}, and Shen {\em et al.}~\cite{Shen:2012xz}. 

From this body of work, it is clear that the extracted distribution of incident neutrino energies for a given measured charged lepton momentum greatly depends on the energy distribution of the incident neutrino flux, which typically has significant uncertainties. If it were possible to measure the energy of all the final state particles, in addition to the outgoing lepton, it would be possible in principle to unambiguously assign a neutrino energy by exploiting energy conservation.  Practically, all detectors have finite energy thresholds for particle detection and other limitations ({\em e.g.} detection of final state neutrons) that will induce errors that depend on the accuracy of the underlying physics models.

\begin{figure}[t]
\centering
\includegraphics[height=6 cm]{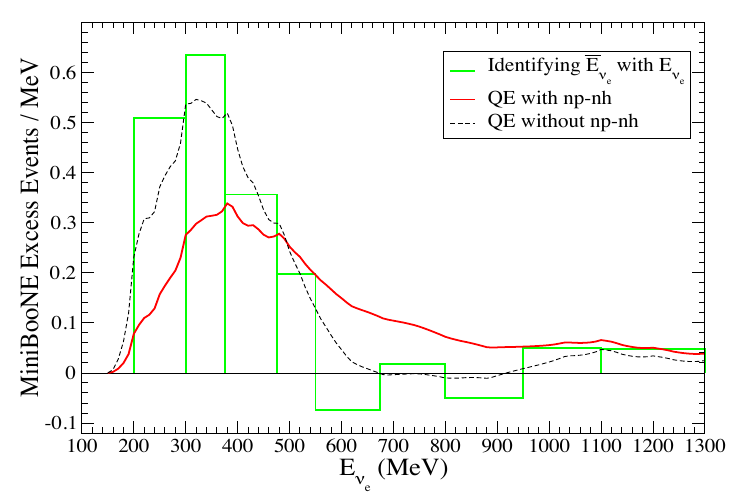}
\caption{\label{fig:martini_mb_excess}The MiniBooNE measured distribution of excess electron neutrino candidate events (histogram) where the neutrino energy was assigned via Equation~\ref{eq:qekine}. The black curve is the best fit to that distribution by Martini {\em et al.}~\cite{Martini:2012fa} without the inclusion of $np-nh$ correlations, while the red curve includes such correlations.}
\end{figure}

As mentioned in Section \ref{sec:introduction}, it is possible in CCQE scattering to assign a neutrino energy $E_\nu$ and $Q^2$ based on the outgoing lepton kinematics and assuming that that the (bound) target nucleon is at rest. We reproduce the relevant formulas here for the case of $\nu_\mu$ CCQE scattering:
\begin{equation}
\begin{array}{lll}
\label{eq:qekine}
E^{QE}_{\nu_\mu} &= & \frac{2(M_n-E_B)E_\mu - \left((M_n-E_B)^2+m_\mu^2 - M_p^2 \right)}
{2\left[M_n-E_B-E_\mu+\sqrt{E_\mu^2-m_\mu^2}\cos\theta_\mu \right]} \\
 & & \\
Q^2_{QE} &= &-m_\mu^2 + 2E_{\nu_\mu}^{QE}\left(E_\mu-\sqrt{E_\mu^2-m_\mu^2 \cos\theta_\mu}\right)
\end{array}
\end{equation}
In the above, $E_\mu=\sqrt{{\bf p}_\mu^2+m_\mu^2}$, $E_B$ is an average binding or separation energy for the neutron in the target nucleus, and $\theta_\mu$ is the muon scattering angle. The assigned neutrino energy for a given measured muon momentum is of course smeared by the Fermi momentum of the nucleon. Referring back to Figure~\ref{fig:misse_ppcorrelation} in Doug Higinbotham's presentation on $(e,e'pp)$, one can directly see the size of errors that can be made in assigning an incident neutrino energy if one can only observe the final state charged lepton energy. Using the momentum of the final state muon to infer the incident neutrino energy can greatly underestimate that energy if the muon was produced by CCQE scattering off a neutron in a correlated pair. The correlated pair can be viewed as being far off-shell and the scattering interaction must put both nucleons back on-shell. As an example of how such considerations may effect the assignment of neutrino energy, Figure~\ref{fig:martini_mb_excess} from Martini {\em et al.}~\cite{Martini:2012fa} shows the altered neutrino energy assignment in their model after accounting for the effect of correlated pairs, thus illustrating the model-dependence of this procedure.

\begin{figure}[t]
\centering
\includegraphics[height=5 cm]{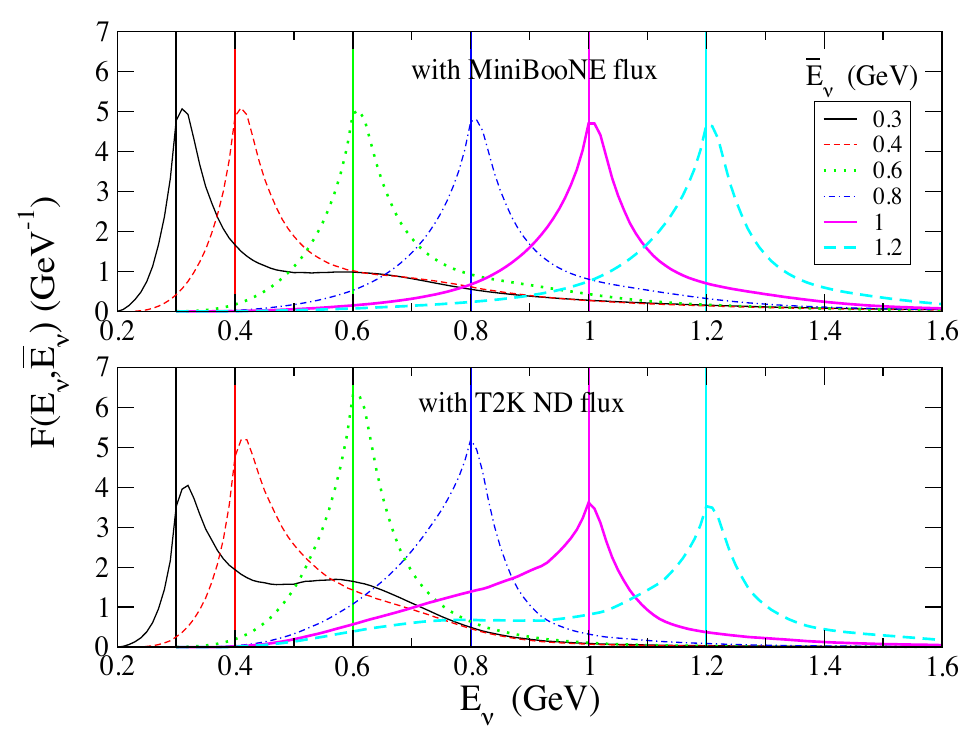}
\includegraphics[height=5 cm]{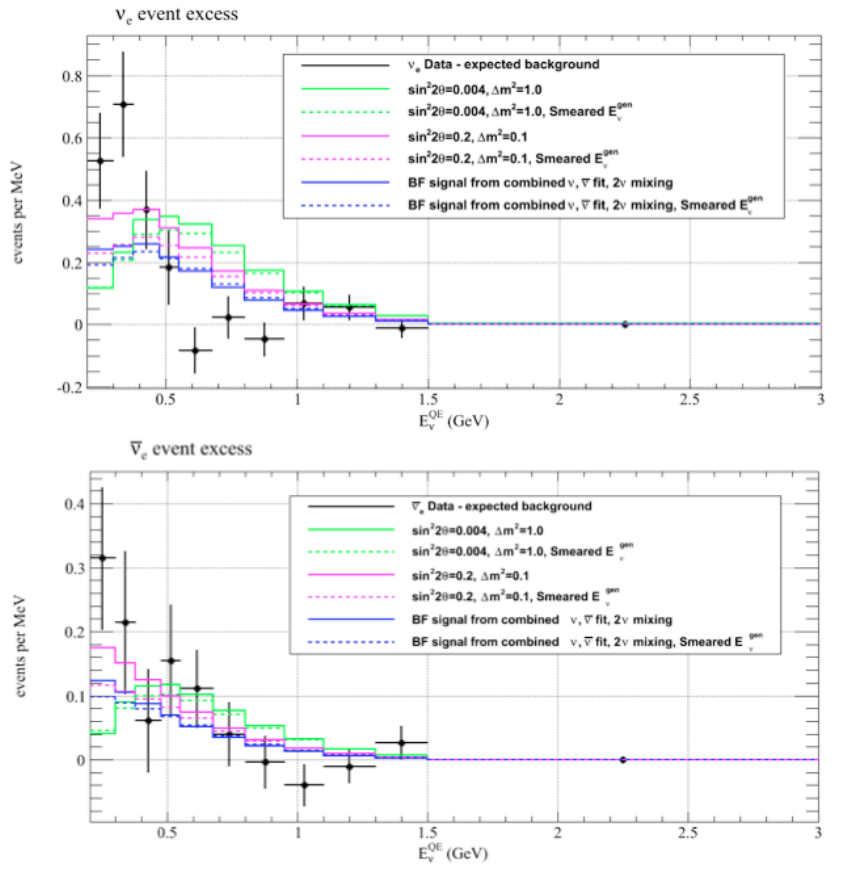}
\caption{\label{fig:martinimb}Left: Expected distribution of true neutrino energies ($E_\nu$) contributing to events reconstructed at a given reconstructed neutrino energy ($\bar{E}_\nu$) according to the model of Martini {\em et al.}~\cite{Martini:2012fa} with the neutrino flux at MiniBooNE (top) and T2K (bottom). Right: Observed excess of electron (anti)-neutrino candidates in neutrino mode (top) and antineutrino mode (bottom) at MiniBooNE with a comparison of predicted distributions for various oscillation parameters with (dashed) and without (solid) smearing to accommodate multi-nucleon contributions as predicted by Martini {\em et al.}}
\end{figure}

In addition to cross section measurements of the CCQE process, the MiniBooNE and T2K collaborations have explored the potential impact of  mismodeling of the neutrino energy on their respective neutrino oscillation results. At MiniBooNE, the predicted reconstructed energy distribution energy distribution assuming CCQE kinematics ({\em i.e.} Equation~\ref{eq:qekine}) from the RPA-based model of Martini {\em et al.}~\cite{Martini:2012fa} was incorporated into the Monte Carlo simulation by reassigning the true neutrino energy of some fraction of events at a given reconstructed neutrino energy to a higher neutrino energy. This simulates the larger expected ``feed-down'' of events from a given true neutrino energy to a lower reconstructed energy expected from this model due to the contribution of multi-nucleon processes as show on the left in Figure~\ref{fig:martinimb}. The reassignment of these true neutrino energies also changes the $\nu_\mu\to\nu_e$ oscillation probability, since this depends on the true neutrino energy.  Based on this redefined distribution between true and reconstructed neutrino energy (and hence oscillation probabilities), the expected reconstructed energy distribution as a function of the oscillation parameters is re-evaluated while incorporating the multi-nucleon effects. 

The results of the study are shown on the right in Figure~\ref{fig:martinimb}, where the expected reconstructed neutrino energy distributions are compared with and without including the additional smearing expected from multi-nucleon effects for several oscillation parameters in both neutrino~\cite{AguilarArevalo:2008rc} and anti-neutrino mode~\cite{Aguilar-Arevalo:2013pmq}. These are also compared to the observed data. While the impact of the multi-nucleon contribution is visible, it does not appear to match the observed excess at lower energies, as evidenced by a worse best-fit $\chi^2$ when the additional smearing is introduced (320.0) relative to the default analysis where no smearing is introduced (317.6). 

\begin{figure}[t]
\centering
\includegraphics[height=5 cm]{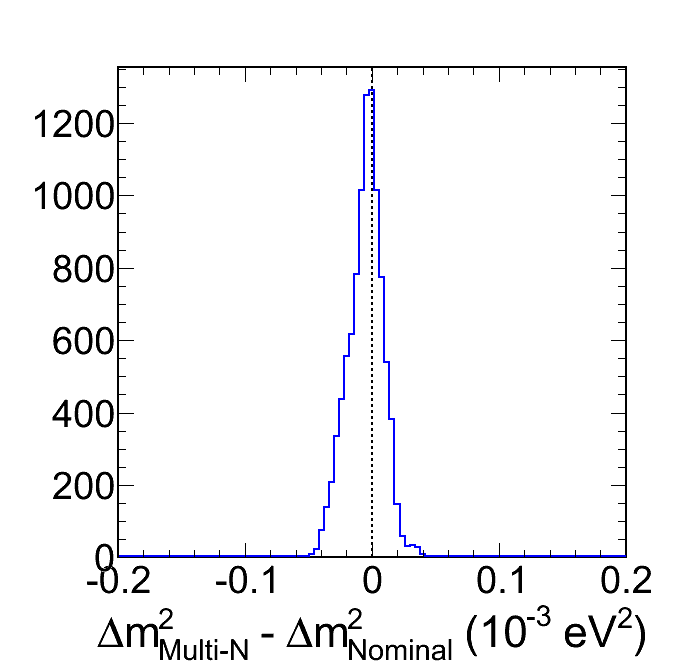}
\includegraphics[height=5 cm]{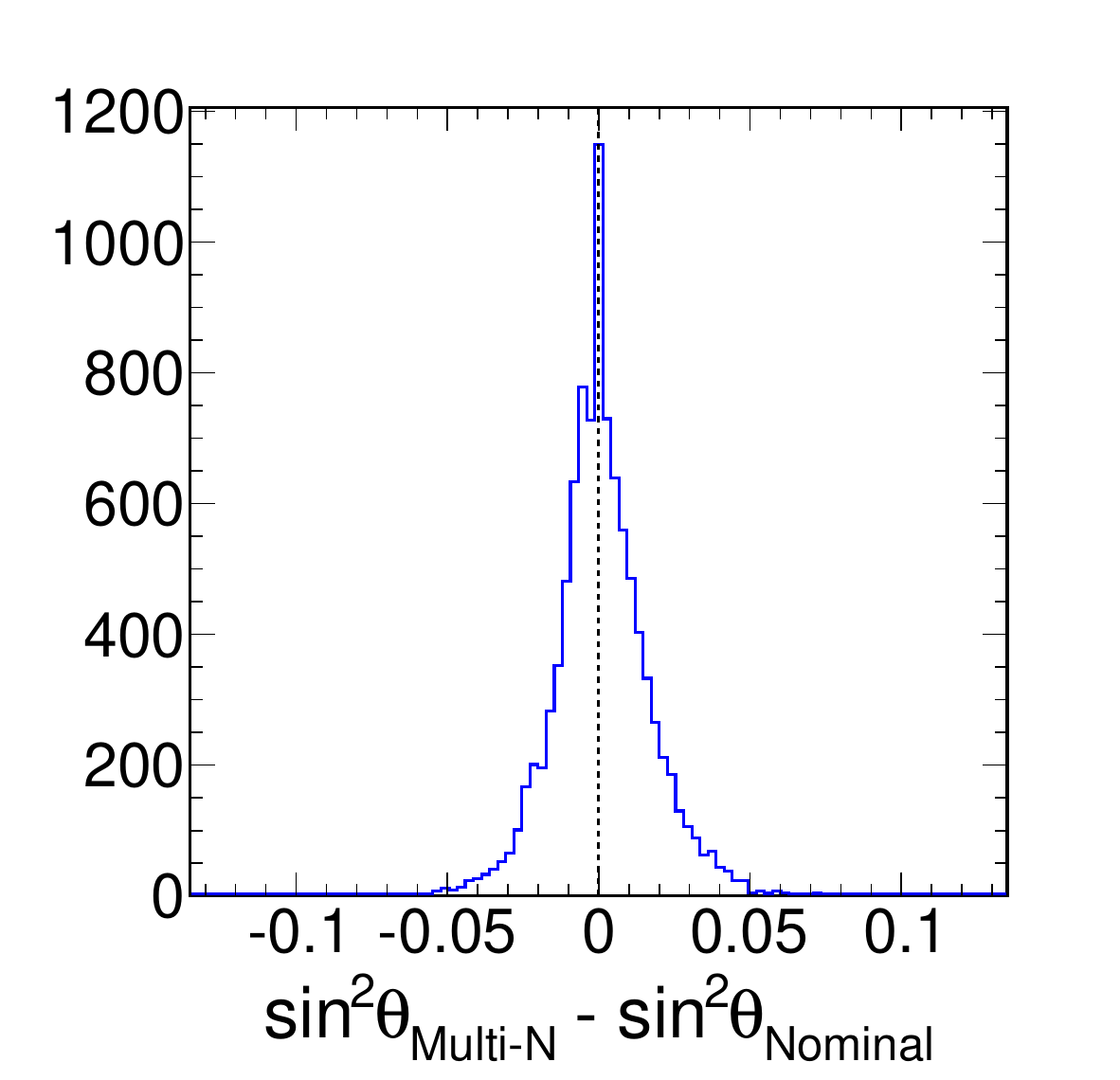}
\caption{\label{fig:t2k_nieves_bias}Disribution of the difference in extracted oscillation parameters $\Delta m^2_{32}$ (left) and $\sin^2\theta_{23}$ (right) where multi-nucleon processes are included in the simulated data 
sets that are fit to two hypotheses that do and do not account for the multi-nucleon contributions. This illustrates the potential additional uncertainty and bias in the T2K Run 1-3 $\nu_\mu$ disappearance analysis~\cite{Abe:2013fuq}, which did not account for multi-nucleon effects. The multi-nucleon processes are simulated according to the model of 
Nieves {\em et al.}~\cite{Nieves:2004wx}.}
\end{figure}

At T2K, the potential impact of multi-nucleon effects have been studied by incorporating the predictions of Nieves {\em et al.}~\cite{Nieves:2004wx} into the Monte Carlo simulations. The details of this implementation were presented by Asmita Redij and Jackie Schwehr. The predictions of the Nieves {\em et al.} model are implemented as tables which provide the double differential cross section in momentum and polar angle for the outgoing lepton emerging from the multi-nucleon interactions. Since this model extends only up to 1.5 GeV in incident neutrino energy due to possible additional channels contributing at higher energies, the extension to higher energies proposed by Gran {\em et al.} \cite{Gran:2013kda} was used to produce a model valid up to 10 GeV in neutrino energy. The prescription of Sobczyk \cite{Sobczyk:2012ms} is used to assign momenta to the ejected nucleons. 

Mark Hartz presented the application of this model to the study of multi-nucleon effects in the T2K $\nu_\mu$ disappearance analysis. Simulated data sets reflecting potentially observed distributions of $\nu_\mu$ CC events at ND280 and at Super-Kamiokande (see Section \ref{sec:t2k} for more details on T2K) were generated with and without the multi-nucleon contributions described above. By fitting these samples with alternative fit models which include or do not include multi-nucleon contributions, the bias from the default fit model (where the multi-nucleon contributions are not included) due to the unaccounted multi-nucleon contribution can be estimated.

The results are shown in Figure \ref{fig:t2k_nieves_bias}, where the left plot shows the expected bias in the extracted $\Delta m^2_{32}$ and the right plot shows the bias in $\sin^2\theta_{23}$ for the Run 1-3 T2K $\nu_\mu$ disappearance results with $3.01\times 10^{20}$ protons-on-target\cite{Abe:2013fuq}. The mean bias and RMS in $\Delta m^2_{32}$ of $-0.05\times 10^{-3}\;\mbox{eV}^2$ and $0.014\times 10^{-3}\;\mbox{eV}^2$, respectively, can be compared to the errors obtained in the actual analysis of $\pm 0.082\;\mbox{eV}^2$. Likewise, the bias and RMS  in $\sin^2\theta_{23}$ is 0.0012 and 0.016, respectively, which can be compared to the uncertainty obtained in the fitted data of $\pm 0.082$. Thus, in the context of the model described in Nieves {\em et al.}, the multi-nucleon contributions do not significantly impact the $\nu_\mu$ disappearance analysis at T2K at this point; the parameter bias and additional uncertainty are small compared to the uncertainties arising from statistics and other systematic errors. However, a wider range of models need to be explored, and as the precision of the measurement increases as more data is accumulated at T2K, a more detailed analysis will be necessary.

\section{Final State Interactions}
\label{sec:fsi}
\begin{figure}[t]
\centering
\includegraphics[height=4.0 cm]{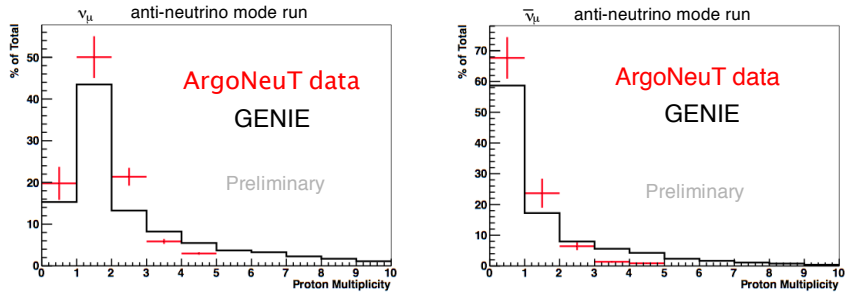}
\caption{\label{fig:argoneut_proton}Measured proton multiplicity in $\nu_\mu$ (left) and $\bar\nu_\mu$ (right) CC events with no final state pions as observed in the exposure of the ArgoNeuT deteector in the NuMI LE beam in antineutrino-mode compared with {\texttt Genie} predictions.}
\end{figure}

Final state interactions (FSI), the interactions of the hadrons created at the electroweak vertex as they traverse the nuclear medium, must be accounted for if one is to understand the exclusive final states that arise from the CCQE process. We note that this definition of FSI is linked to the impulse approximation, and precludes the possibility of FSI affecting the inclusive CCQE cross section. This differs from the same term used in the context of the relativistic mean field, where ``FSI'' can modify the inclusive cross section. 

Formally, a full treatment of FSI requires, among other things, the introduction of a very large number of new and unknown form factors; 
this has not happened to date, and is not likely to happen in the future. However, the possibility of employing large fine-grained tracking detectors and LAr-TPCs in neutrino experiments has greatly increased the interest in improving the FSI treatment in neutrino event generators. Ideally these new detectors could reliably specify the incident neutrino energy, allow better specification of the interaction at the weak vertex, and allow the assignment of ${\bf q}$ and $\omega$ to each event. Indeed, data from SciBooNE (Section~\ref{sec:sciboone}) and MINERvA 
(Section~\ref{sec:minerva}) are already providing evidence of the shortcomings of currently implemented FSI models.
Detailed information on proton emission, such as shown in Figure~\ref{fig:argoneut_proton} that was presented 
by Ornella Palamara at the workshop gives us a preview of the detailed topological and kinematic information we will obtain on CCQE final states.


There is not a great body of work dealing with FSI in CCQE neutrino interactions. The Giessen Boltzmann-Uehling-Uhlenbeck Project (GiBUU) is the most systematic and extensive effort~\cite{Buss:2011mx}. Ulrich Mosel presented the model and its application to neutrino-nucleus interactions. As the names Boltzmann, Uehling and Uhlenbeck imply, GiBUU employs transport theory and provides a transport equation for each baryon and meson in a nuclear medium. The model has considerable generality and has been applied in a wide variety of reactions with considerable success. At present, GiBUU's principal role in CCQE is treating pion absorption following resonance formation at the electroweak vertex. These absorptive processes create a background to CCQE scattering as nucleons are the only hadrons present in the final state. Mosel made a strong case for upgrading present event generators to improve their treatment of FSI by incorporating GiBUU.  This seems an entirely reasonable suggestion, given the huge cost and importance of many proposed neutrino experiments (LBNF, LBNO) that have the possibility of observing many of the final state products. 

A critical element in furthering our understanding of FSI in neutrino reactions are the corresponding studies in electron scattering. 
Pandharipande and Pieper~\cite{Pandharipande:1992zz} investigated FSI in a limited fashion by applying an optical model to describe the propagation of a quasi-elastically struck proton in C, Al, Ni and Ta. Good agreement with $(e,e'p)$ data~\cite{Garino:1992ca} was found but, of course, the fate of the protons absorbed by the imaginary part of the optical
potential was unaccounted for.


It is generally recognized that FSI present a very serious challenge to theory if a useful and reliable procedure is to be developed and incorporated into neutrino event generators. It appears progress on this front will require substantial input from electron scattering and that  GiBUU (as one example) will be a useful tool to digest such data and to connect it to neutrino scattering. A generic concern raised at the workshop is the consistency of any final state interaction model with the initial state calculation.
This is necessary to have a correct accounting
of all the physical processes and to avoid ``double-counting'' or introducing other inconsistencies between the two (or more) parts of the model.

\section{Energetic Photons from Weak Neutral Current Interaction}
\label{sec:photon}
The observation of an excess of low energy electron-like events in MiniBooNE~\cite{AguilarArevalo:2008rc,Aguilar-Arevalo:2013pmq} has recently been the subject of signficant interest and scrutiny. One explanation is that the excess is due to to energetic photons (200-475 MeV) produced in neutral current (NC $\gamma$) reactions  \cite{Harvey:2007rd, Hill:2009ek, Hill:2010zy}. Since the MiniBooNE detector is not capable of distinguishing such photons from electrons emerging from $\nu_e$ CCQE interactions, the estimate of this background relies heavily on theoretical estimates. It is natural to assume on general grounds that the yield of such a process would be of order $G_F^2\times \alpha$. The issue is of considerable importance since sterile neutrinos are required to explain the excess of the electron-like events in MiniBooNE if it is due to $\numu \rightarrow \nue$ ($\numubar \rightarrow \nuebar$)  oscillations. Also, if the observed excess at MiniBooNE is due to an unmodeled source of photon production, this is an important background source to understand for current and future long baseline experiments studying $\nu_\mu\to\nu_e$ oscillation in the standard three neutrino flavor framework such as T2K, NOvA, LBNF~ \cite{Adams:2013qkq}, and Hyper-Kamiokande~\cite{Abe:2011ts}.


\begin{figure}
\centering
\includegraphics[width=10 cm]{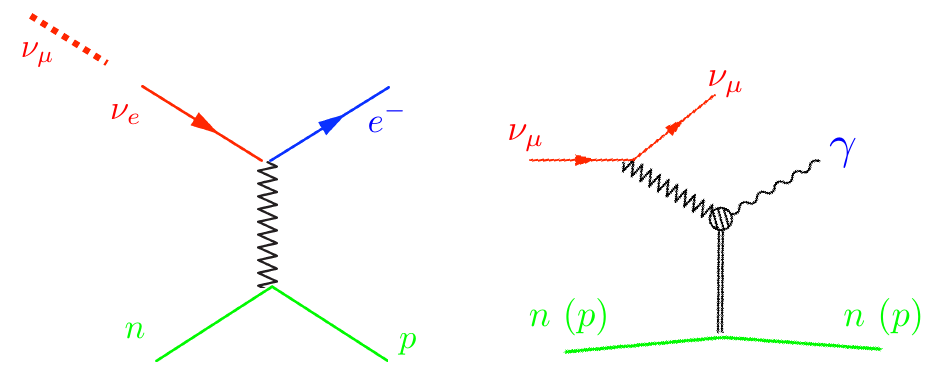}
\caption{\label{fig:osc_vs_NCphoton} Diagrams shown by Richard Hill illustrating the similarity of the neutrino NC $\gamma$ production to $\nue$ appearance oscillation.}
\end{figure}

Richard Hill summarized the issues in his theoretical overview talk on this subject.  The obvious importance of NC $\gamma$ events in interpreting neutrino oscillations results was discussed together with implications for proton decay and neutron star cooling.  
The anomalous term that couples the $\omega$, $Z$, and $\gamma$ can lead to diagrams that contribute to the energetic photon event rate; however, this contribution is challenging to calculate.  Since the original idea in Reference~\cite{Harvey:2007rd},
more complete calculations show that other effects such as $\Delta$ production with subsequent radiative decay are
the dominant processes and the anomalous $\omega Z \gamma$ term is small, albeit with large uncertainties.

Teppei Katori explained the experimental situation, starting with the approach used
by MiniBooNE to calculate the rate of NC $\gamma$ events \cite{MBnuedata:2012}.
MiniBooNE assumed that the origin of this background in their measurements arose from the radiative decay of $\Delta$ resonances ($\Delta^{+/0}\to (p/n)+\gamma$) produced in NC reactions. Rather than calculate the rate from a nuclear model, MiniBooNE used its own measurement of NC~$\pi^{0}$ production to infer the rate of NC $\Delta$ resonance production in carbon. Using the branching fraction of the radiative decay $\Gamma_\gamma/\Gamma_\Delta = 0.0056\pm0.0004$ for a free $\Delta$ at the resonance pole~\cite{Eidelman:2004wy} and accounting for the various interactions of the $\Delta$ and the pion from its decay in the nuclear medium,   $\left(\Gamma_\gamma/\Gamma_{\pi^0}\right)_{\Delta,^{12}\mbox{C}}$ the ratio of $\gamma$ to $\pi^0$ production in the decay of a $\Delta$ resonance in $^{12}\mbox{C}$, is established and scaled by the observed $\pi^0$ production. 
This model was incorporated into the {\texttt nuance}~\cite{Casper:2002sd} neutrino event generator to determine the rate and detailed topology of the NC $\gamma$ background contribution. The T2K experiment uses a similar approach for their calculation of this background in the {\texttt neut} generator.

There are very few measurements of neutrino-induced single photon production.
A direct search of NC $\gamma$ events has been performed by the NOMAD collaboration~\cite{Kullenberg:2011rd} with a $\sim 25$~GeV neutrino beam, providing a null result and an upper limit.  Currently running experiments,  T2K and MINERvA, as well as the near-future MicroBooNE, utilizing the capability of LAr-TPCs to discriminate electrons from photons, may have some sensitivity to measure NC $\gamma$ events. These will be challenging analyses due to the low rate of this process.  It is likely that such efforts are necessarily conducted jointly with $\nue$-appearance searches.

Given the importance to oscillation physics, it is highly desirable that additional and independent evaluation of this important background be carried out.  It was fortunate that two new evaluations of the NC $\gamma$ yield were reported at the workshop. The approaches were quite different, but both aimed at a more complete evaluation of  nuclear effects than was the case in References~\cite{Hill:2009ek,Hill:2010zy}. 

\begin{figure}
\centering
\includegraphics[width=10 cm]{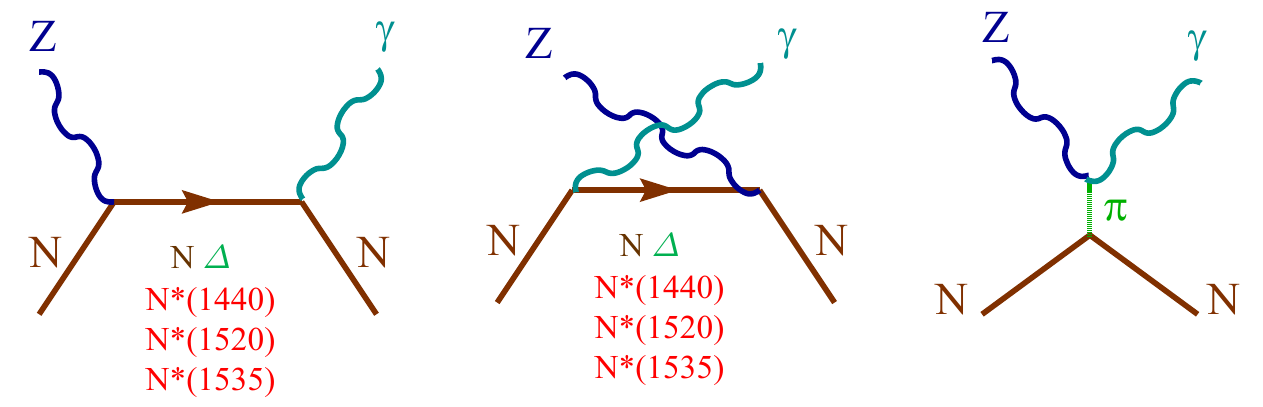}
\caption{\label{fig:photon_diagrams} Feynman diagrams of the hadronic weak couplings considered in the calculation of neutral current production of energetic photons \cite{Wang:2013wva}.}
\end{figure}

\begin{figure}
\centering
\includegraphics[width=10 cm]{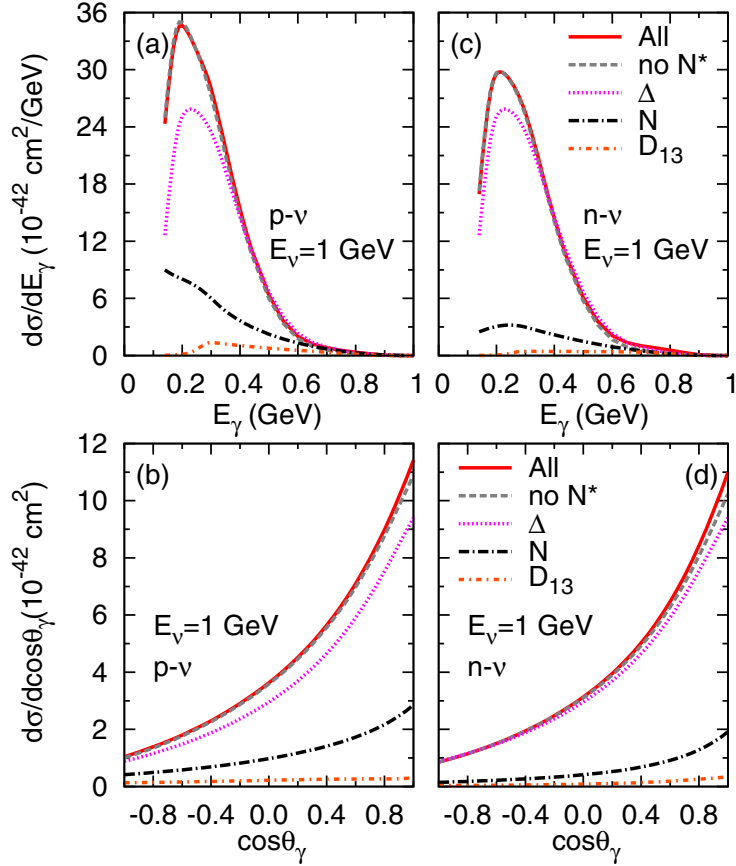}
\caption{\label{fig:nu_photon_xsections} Differential cross section for neutral current production of photons ($E_\gamma>120$~MeV) for $\nu-p$ (left) and $\nu-n$ (right) interactions at $E_\nu=1$~GeV as a function of photon energy (top) and photon emission angle relative to the incident neutrino (bottom). Partial contributions to the total cross sections are also indicated\cite{Wang:2013wva}.}
\end{figure}

\begin{figure}
\centering
\includegraphics[width=12 cm]{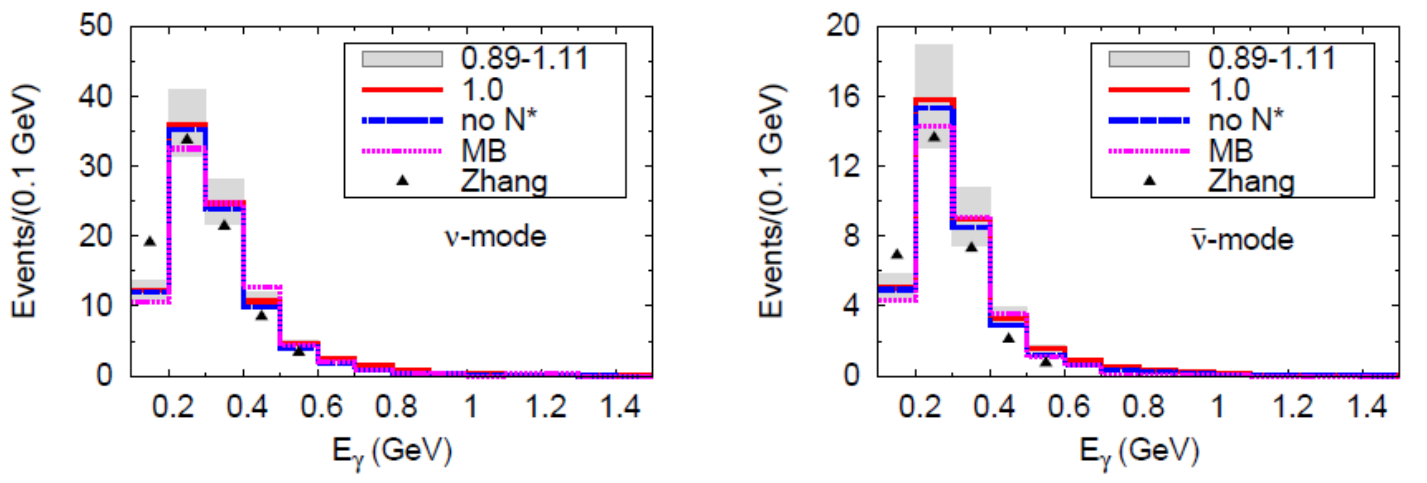}
\caption{\label{fig:nc_photon_evis} The predicted $\gamma$ spectrum of the Valencia group (red) assuming the MiniBooNE neutrino flux and detector response \cite{Wang:2014nat}.  Also shown are the results of Reference ~\cite{Zhang:2012xn} and the $\gamma$ spectrum calculated by MiniBooNE~\cite{MBnuedata:2012}.}
\end{figure}

The work of the Valencia group was reported  by Luis Alvarez-Ruso where the NC processes shown in Figure~\ref{fig:photon_diagrams}~\cite{Wang:2013wva, Wang:2014nat} were considered. 
The vector form factors for the baryons were taken from various electromagnetic measurements, with the axial vector form factors coming from the application of the partially-conserved axial current hypothesis. The axial mass in the dipole form factor for all baryons was assumed to be 1~GeV. Figure~\ref{fig:nu_photon_xsections} shows their results for the NC $\gamma$ process on free protons and neutrons, where it is clear that the dominant contribution comes from the $\Delta$ resonance.  Figure~\ref{fig:nc_photon_evis} shows their prediction for NC $\gamma$ production using the MiniBooNE neutrino flux. The gray band indicates the uncertainty resulting from the axial form factor for $\Delta$ production. It appears that good agreement exists between the model calculations and the background yield assumed by MiniBooNE.

Xilin Zhang presented the work carried out with Brian Serot on this subject. They largely focused on extension of their previous work~\cite{Serot:2012rd,Zhang:2012aka,Zhang:2012xi} to higher neutrino energies. This approach is very different from that employed by the Valencia group in that it has nucleons and $\Delta$ resonances moving in relativistic mean fields with strong vector and scalar potentials. In spite of the very different approach, there is very good agreement with the Valencia group for the cross section of the NC $\gamma $ process except at higher neutrino energies. 
For reconstructed neutrino energies relevant to MiniBooNE, there is also good agreement with the predictions used by MiniBooNE as shown in 
Table~\ref{tab:Zhang_event_table}~\cite{Zhang:2012xn}. 
Thus, it appears that the convergence of these predictions with  more sophisticated treatments of nuclear effects, and their agreement with the initial estimates used by MiniBooNE, dampen earlier speculations indicating that the MiniBooNE electron neutrino excess is due to mismodeling the NC $\gamma$ process.
 
 \begin{table}
\centering
\begin{tabular}{lccc}\hline\hline
$E_\nu^{QE}$ (GeV) & [0.2, 0.3] & [0.3, 0.475] & [0.475, 1.25] \\ \hline
Coherent   & 1.5 (2.9) & 6.0 (9.2) & 2.1 (8.0) \\
Incoherent & 12.0(14.1) & 25.5(31.1) & 12.6 (23.2) \\
Hydrogen   & 4.1(4.4)  & 10.6(11.6) & 4.6 (6.3) \\ \hline
Total      & 17.2 (21.4) & 42.1 (51.9) & 19.3 (37.5) \\ \hline
MiniBooNE  & 19.5 & 47.3  & 19.4 \\  \hline
Excess     & $42.6 \pm 25.3$  & $82.2 \pm 23.3$ & $21.5\pm 34.9$ \\ \hline
\end{tabular}
\caption{\label{tab:Zhang_event_table} Number of predicted photons detected, using the MiniBooNE $\nu$ energy distribution, in the indicated $E_\nu^{QE}$ interval from coherent and incoherent scattering oncarbon and hydrogen by Serot and Zhang \cite{Serot:2012rd}. The values in parentheses represent the calculated upper limits. The row labeled ``MiniBooNE'' are the event rates used in the MiniBooNE $\nu_e$ appearance analysis, and ``Excess''' refers to the observed excess of $\nu_e$ candidates following the subtraction of other background sources~\cite{MBnuedata:2012}.   
}
\end{table}

\begin{figure}
\centering
{\includegraphics[width=10cm]{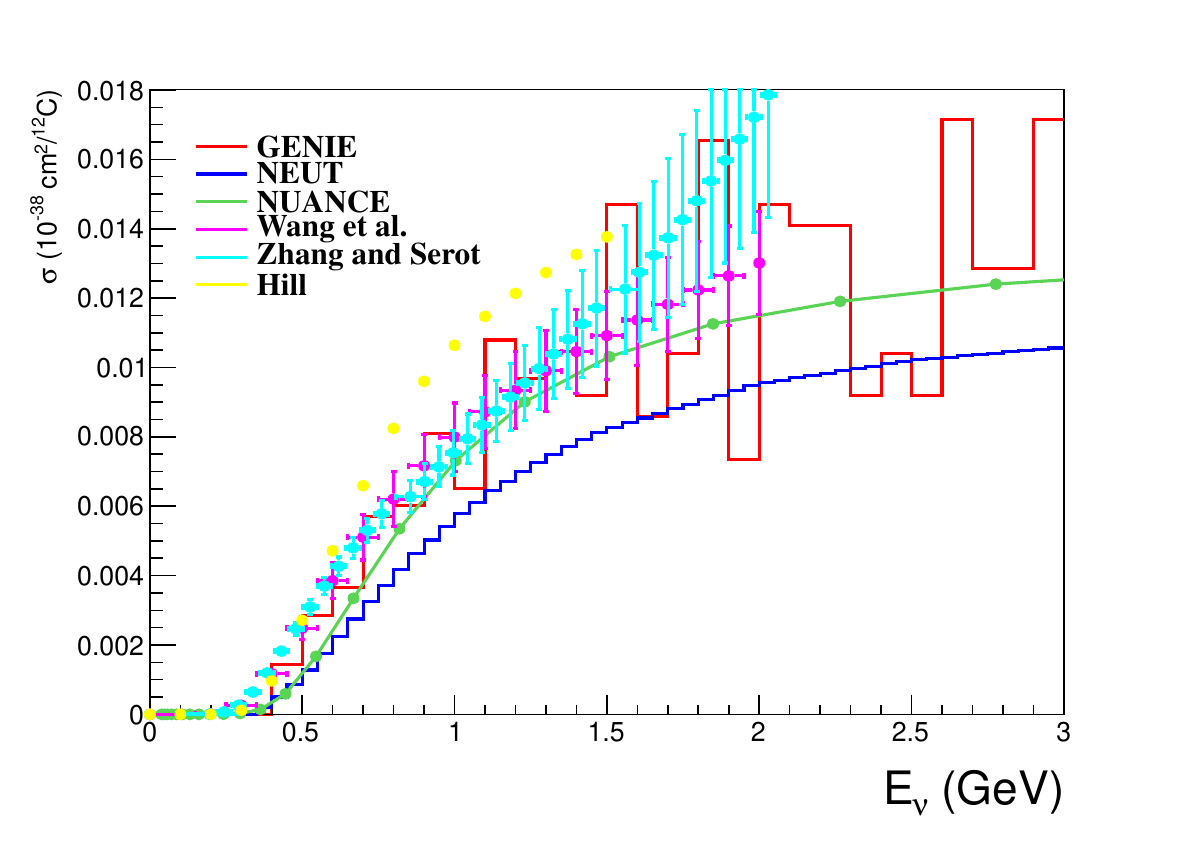}}
\caption{Comparison by T. Katori of neutrino-induced NC $\gamma$ production cross section on carbon as function of
neutrino energy as calculated
by Wang et al.~\cite{Wang:2013wva}, Zhang and Serot~\cite{Zhang:2012xn}, and Hill~\cite{Hill:2009ek}
and compared to three neutrino event generators.  The NUANCE curve corresponds 
to the MiniBooNE estimate for this process. Note that the most important region
in this plot is at the MiniBooNE flux peak of 800~MeV.  The Zhang and Serot model
is acknowledged to contain divergent terms above 1.5~GeV.} 
\label{fig:NCpho_model_comp}
\end{figure}

An important summary plot of the situation was 
presented by Teppei Katori (Figure ~\ref{fig:NCpho_model_comp}).  In this plot, the total cross section for the NC $\gamma$ process as predicted by three neutrino event generators is shown together with the predictions from the three aforementioned models. In the $E_\nu \sim 800$~MeV region, there is good agreement between the theoretical models and the {\texttt nuance} event generator employed by MiniBooNE.  This is consistent with the theory-experiment agreement shown in Table~\ref{tab:Zhang_event_table}.  The predictions  appear to diverge at higher energies where some terms are less constrained, but this region is less relevant for interpreting the MiniBooNE excess.
  
Overall, the situation for neutrino NC $\gamma$ production seems to be converging and under control.  The theoretical models agree with the experimentally determined predictions for this process as determined by MiniBooNE.  However, there are possible contributions, namely from the anomalous $\omega Z \gamma$ term, where theoretical uncertainties remain large and the calculation is not robust.  Because of the overall importance to $\nu_e$ appearance studies, the issue is not completely closed and we should keep a close eye on possible future theoretical or experimental advances.

\section{Conclusions: Settled matters, open questions, future needs}
While there were no formal conclusions drawn from the discussions at the workshop, there was accord on a number of issues, which we try to capture here.

As expected, the importance of understanding the CCQE interaction was reinforced by the workshop participants, as it remains the primary channel for many current and planned neutrino oscillation experiments. These experiments must assign a flavor and incoming neutrino energy to observed neutrino interactions based on the observed final state particles. The updated theoretical view of the inclusive CCQE process is much more complicated than past assumptions based primarily on a Relativistic Fermi Gas (RFG) model of the nucleus supplemented with the impulse approximation (IA). Additional processes associated with the multi-nucleon dynamics of the nucleus can result in ambiguities of nearly several hundred MeV in the assignment of neutrino energies based on the outgoing lepton kinematics. With detectors that can directly measure at least part of the energy in the hadronic final state, this uncertainty can be reduced. 

In addition to those from the theoretical community, there were detailed studies performed by MiniBooNE and T2K assessing the impact of multi-nucleon effects in CCQE interactions on their neutrino oscillation results. For T2K, it appears that the current level of statistical precision is such that the potential effect  do not significantly impact measured oscillation parameters. However, as more data is accumulated and other sources of uncertainty shrink, this will likely become an important effect requiring careful treatment. The various models of these processes also differ significantly in the magnitude of the multi-nucleon contribution; thus theoretical convergence is also needed to predict the impact of multi-nucleon processes in CCQE with sufficient accuracy. The situation is similar for MiniBooNE, where an initial study of multi-nucleon effects concluded that these do not significantly impact the existing interpretation of the observed $\nu_e$ and $\bar{\nu}_e$ candidates. However, as a more ambitious short baseline program takes shape at Fermilab, the demand on the theory will continue to grow. 

While  an enormous amount of progress has occurred since  RFG/IA-based models of CCQE events at MiniBooNE resulted in underpredicting the cross sections by nearly 40\%, it is also clear that much remains to be learned from the study of electron QE scattering, where well-characterized beams and spectrometry of the outgoing electrons allow much more precise and controlled measurements than are possible with any neutrino experiment. These measurements have already yielded a robust characterization of the nuclear spectral function, with approximately 20\% of the nucleons in high momentum correlated pairs. Any model of neutrino CCQE interactions must account for what is observed in the corresponding electron QE data. To this end, the impulse approximation must be partially abandoned, and known correlations created by nucleon-nucleon interactions  incorporated to give a successful prediction consistent with observed CCQE measurements using free nucleon form factors as a foundation.

From the various theoretical approaches presented at the workshop, there seems to be convergence on the characterization of the multi-nucleon  effects in the inclusive CCQE process, where the process is considered integrating over possible hadronic final states. However, the current and upcoming suite of neutrino experiments utilizing fine grained tracking detectors and LAr-TPCs will demand a more exclusive description of the CCQE process in order to interpret their data. A rigorous treatment  would require many more form factors to be determined, and is beyond the scope of the present theoretical tools.  This presents both a challenge and opportunity: the enormous difficulties in producing a robust treatment of the exclusive hadronic states were manifestly obvious to all the participants. However, reliable and robust measurements with unprecedented detail are becoming available. In the mean time, more phenomenological approaches can and should be examined and extensively tested with electron scattering data. 

The increasing sophistication and computational demands of the requisite theory also call for a paradigm shift in the way their predictions are encoded into the neutrino event generators used by the experiments. To date, most such generators recalculate the relevant physics (for example, the RFG model) from scratch, either on-the-fly or in a separate pre-computational step to produce numerical tables or other parametrization that (drastically) reduce computational time. However, it is unreasonable to even contemplate directly incorporating some of the methods presented at the workshop ({\em e.g.} a Green's Function Monte Carlo calculation) into an event generator, and thus the results of such calculations must be provided more directly from the theorists to the event generator. Some new methods were already in place at the workshop, where tables of differential cross sections from RPA calculations were directly inserted into an event generator, while new ideas, such as precomputing the hadronic tensor from such a calculation, emerged from the discussions. This is an interface which will have to be actively maintained by the experimental and theoretical community in order to facilitate the incorporation of new theoretical predictions into the event generators, a necessary step to enable  a robust confrontation of these predictions with experimental data.

The discussion on  single photon production in neutral current interactions (NC $\gamma$) concluded with a very satisfying convergence of the predicted rates of this process from two sophisticated theoretical frameworks. Furthermore, these calculations were in reasonable agreement with the estimates made by MiniBooNE in determining the background from this process in their $\nu_\mu\to\nu_e$ and $\bar{\nu}_\mu\to\bar{\nu}_e$ oscillation searches. Following a search for this process at NOMAD in a relatively high energy neutrino beam,  we can look forward to further experimental study of this process with LAr-TPCs such as MicroBooNE.  MINERvA and ND280 at T2K could also provide measurements of this process. Due to the inherently perilous nature of this process as a background to studying electron neutrinos, continued vigilance both from the theory side in identifying any possible mechanisms for NC $\gamma$ interactions, and from experiments in directly identifying and studying this process, will be important.

\vskip 0.5 cm
Over the course of the workshop, some general and advice and guidance emerged directed to the experimentalists and theorists which we summarize here: \\ \\
\noindent{\bf Experimentalists}
\begin{itemize}
\item Continued efforts to improve the neutrino flux estimates will certainly produce additional dividends. Effective use of hadron production measurements is essential, but likely insufficient, to achieve better than $5\%$ uncertainty on the predicted absolute flux, where effects of secondary and tertiary interactions, and uncertainties related to the primary proton beam optics, beam line alignment, etc. may become significant, possibly dominant, sources of systematic uncertainty once the primary production is measured precisely. To this end, the community may need to consider developing a program of necessary measurements to understand these additional hadronic processes and investing in the appropriate beam line monitoring to ensure that the properties of the beam line are sufficiently understood.
\item It is essential to reduce any barrier between the reported measurements and the quantities that can be predicted from theory. There has already been significant progress in moving from measurements of effective parameters to more model-independent quantities that can be directly calculated by theorists with neutrino flux predictions provided by experimentalists. It will be critical for all experimental efforts to engage and continuously develop this interface in order to provide theorists with the most effective input.
\item We have seen that precisely defining what is meant by ``CCQE'' in a measurement is critical for interpreting the results. Clarity in the event selection and observables will become even more important as engagement continues with theorists and the electron scattering community, where terms such as ``quasi-elastic scattering" and ``final state interactions" may mean something substantially different.
\end{itemize}

\noindent{\bf Theorists}
\begin{itemize}
\item There is consensus now on the important role of the enhancement in the transverse vector response observed in electron scattering in predicting the neutrino CCQE cross section. At the workshop, we saw various approaches as to whether there is a corresponding enhancement in the axial response, and  indications from {\em ab initio} methods that such an enhancement is indeed present. It is critical to better understand this situation fundamentally ({\em i.e.} what is the nature and magnitude of this enhancement) and to clarify within each theoretical framework (RPA, RMF, SuSA, etc.) what assumptions were made with regard to these enhancements in calculating the CCQE cross section.
\item While there are domains where a non-relativistic treatment may suffice, it is clear that a complete picture of the CCQE process will require a thorough accounting of relativistic effects. 
\item There were clear specific demands from the experimental community. First, new calculations must be provided in a form that can be incorporated into a neutrino event generator so that the results can be used by neutrino experiments. Experimentalists will  require guidance in estimating the uncertainties in the prediction in order to use it effectively in an analysis. Finally, experiments will need predictions on the exclusive final states of neutrino-induced CCQE interactions. The path forward on the latter point was particularly unclear, but there was consensus that theoretical guidance and input on the above is crucial in moving forward.
\end{itemize}

\section{Acknowledgments}
The authors are pleased to acknowledge the enormous support from the Institute for Nuclear Theory at the University of Washington, which kindly hosted the workshop, arranged travel support to its participants, and provided extensive logistical support before, during, and after the workshop. We also thank Los Alamos National Laboratory and Fermi National Laboratory for providing travel support for some of the organizers. The authors are also extremely grateful to all the participants of the workshop for their insightful contributions and discussions. This work was supported by the Fermi National Accelerator Laboratory under US Department of Energy contract
No. DE-AC02-07CH11359.

\appendix
\section{Definitions}
\label{sec:definitions}
Due to the various meanings applied to commonly used terms, Bill Donnelly kindly provided and led a discussion on defining several key terms, which we reproduce below:
\vskip 0.5 cm
\noindent{\bf Quasielastic}\\
From very early work in electron scattering from nuclei the inclusive quasielastic 
contribution is usually taken to be the peak seen at roughly $Q^2/2M$. The reason 
for the terminology comes from a simple (and somewhat naive) model: were the 
process to be simply electron scattering from a non-interacting nucleon at rest in 
a nucleus, a delta-function peak at the above energy loss would be the answer. 
The nucleons in the nucleus are in fact interacting and moving, and thus the 
delta function is smeared out (Fermi smearing). Actually things are more 
complicated than this and both initial- and final-state interactions are important; 
furthermore, the nucleons are not on-shell and thus their energies and momenta 
are not trivially related as some models suggest.

For many theorists the quasielastic contributions are distinguished by their 
being produced by one-body operators, in contrast to effects arising from 
two-body operators, such as meson exchange current contributions (see below). 
Note that this does not equate to one-, two-, three-, etc. nucleon knockout, 
however, as one type of current operator can give rise to different numbers 
of nucleons in the final state, dependent on what channels are open. 
In neutrino studies, in contrast to electron scattering, different signatures occur 
for contributions where a pion is detected versus where it is not; the latter is 
called quasielastic, but is in that usage really the net effect obtained using the 
full electroweak current with one- and two-body contributions. In fact, even then 
there is an issue that pions can be produced, but be absorbed and so not detected 
and accordingly these contributions are counted as ``quasielastic''. Naturally 
some model is typically invoked to account for these effects that corrupt the 
strict meaning of quasielastic, although this means some model dependence 
has been introduced.\\

\noindent{\bf Inelastic contributions}\\
Above pion production threshold one has inelastic contributions coming 
from various sources: non-resonant pion production, production in the 
region where the Delta dominates, or where other baryon resonances 
are expected to play significant roles, multi-pion production, kaon 
production, etc., eventually to deep inelastic scattering. In the inclusive 
cross section these are not really distinguishable, but all pieces of the 
total cross section. For instance, duality studies indicate that on the average 
effects from specific hadrons in the final state give rise to the same overall 
result as does DIS. Thus modeling in this region must be done with care to 
avoid double counting.\\

\noindent{\bf Meson exchange current contributions}\\ 
In models such as the relativistic Fermi gas (RFG) one can catalog the various 
contributions from one-body current operators, two-body contributions counted 
as MEC versus correlation effects, with both single-nucleon and two-nucleon ejection 
(the former interfere with the one-body single-nucleon amplitudes), and so on. This 
can be done because the many-body wave functions are especially simple, namely, 
on-shell (non-interacting) plane-wave states in Slater determinants. For models with 
interactions present one has a problem separating the MEC and correlation effects 
computed as matrix elements of two-body operators from effects already present in 
the wave functions themselves. In fact, the very concepts are not observables but are 
representation-dependent. A sophisticated interacting many-body description may 
already have some (but likely not all) correlation and MEC effects incorporated, in 
contrast to a simple model where they may not already be present. 

Another comment on MEC effects: These are not optional, but are required for any 
interacting system by gauge invariance. Any model with interactions must confront 
the requirement of having the corresponding two-body MEC contributions and 
many models (almost all) cannot do this consistently.\\

\noindent{\bf Correlations, both long- and short-range}\\
In naive models such as the RFG one can include long-range p-h correlation effects 
within the context of perturbation theory, and can make things gauge invariant to a 
given order. These arise typically from the longest-range part of the NN interaction, 
namely that arising from pion exchange. Short-range effects are sometimes also 
included, although there may be issues with their validity at high energies. 
Once one goes to more sophisticated models the meaning of short- and long-range 
correlations change, these are not observables and are representation dependent. 
For instance, in one approach a strong repulsive core might be included to allow 
saturation of nuclear matter and might thereby influence the electroweak cross 
sections, especially in promoting strength to high missing energy (this needs defining 
as well; it can be covered in discussions). Alternatively, in relativistic mean field 
approaches (\a la Walecka) many of the correlation effects are already present via 
strong scalar and vector meson exchanges and therefore one should not be adding 
them willy nilly or one will run the risk of double counting. Note that this approach 
also saturates nuclear matter.\\

\noindent{\bf Inclusive versus semi-inclusive and more exclusive reactions}\\
Inclusive electroweak cross sections are total hadronic cross sections: only 
the final-state lepton is presumed to be detected, but nothing from the nuclear 
side of the scattering diagram. As such, even very naïve models such as the RFG 
can give reasonable answers. The models tend to satisfy basic symmetries such 
as unitarity, Lorentz covariance (often, but not always which is serious), maybe 
gauge invariance, etc. This being so, one tends to get roughly the correct answers 
since sum rules are being enforced. The main issues with such simple modeling 
for inclusive reactions is that the strength is often not quite correctly distributed 
in energy-momentum. 
In contrast, when (say) a nucleon in the final state is detected in coincidence 
with the final-state lepton one has a very different problem. The details of how 
that nucleon interacts with the rest of the nucleons in the final state is a much 
more complicated problem. Typical modeling that may be adequate for inclusive 
cross sections can be very bad for semi-inclusive studies. 

\section{The Relativistic Mean Field Model}
This appendix provides background on the use of the relativistic mean field 
(RMF) model to characterize the nucleus, as many are no longer familiar with its use. The depth of the central potential that nucleons encounter in a nucleus is typically taken to be 50 MeV, allowing a non-relativistic description of their wave functions via the use of the Schr{\"o}dinger equation. In the RMF model, the 50 MeV depth is the result of a strong attractive scalar potential of $\sim -400 \mbox{MeV}$ and a strong repulsive vector potential of  $\sim +350 \mbox{MeV}$ as shown in Figure \ref{fig:svecpotential}a If nucleons are treated as Dirac particles, the following Dirac equation is obtained where the Coulomb potential has been omitted for simplicity:
\begin{equation}
\left[{\bf \alpha}\cdot{\bf p} + \beta\left( m+U_S(r) \right)  + U_{V0}({\bf r})\right]\psi({\bf r}) = E\psi({\bf r})
\end{equation}
where

\begin{equation}
{\bf \alpha}= \left(\begin{array}{cc}
0 & {\bf \sigma}  \\
{\bf \sigma} & 0
\end{array}\right)
~~~~~~
\beta= \left(\begin{array}{cc}
1 & 0 \\
0 & -1
\end{array}\right)
~~~~~~
\psi= \left(\begin{array}{c}
\psi_U \\
\psi_L
\end{array}\right)
\end{equation}
and $U_s$ is the scalar potential and $U_{V0}$ is the fourth component of a vector field. ${\bf \alpha}$ and $\beta$  are the usual Dirac matrices and $\psi$ the four component Dirac wave function with upper and lower components. By converting the above to a second order equation for the upper component, one obtains:
\begin{equation}
\label{eq:psiup}
\left[(|{\bf p}|+U_{V0})^2 + (m+U_S)^2 + iU_D {\bf r} \cdot {\bf p} \right]\psi_U = 0
\end{equation}
with $U_D$ termed the Darwin potential and
\begin{equation}
\label{eq:darwin}
U_{SO} = -\frac{1}{rA} \frac{\partial A}{\partial r} = -U_D~~~~~
A = E + m+ U_S -U_{V0}.
\end{equation}
Thus, $U_{V0}$ shifts the nucleon to higher energy and the cross term $2EU_{V0}$ provides the energy dependence visible in Figure \ref{fig:svecpotential}b . The scalar potential reduces the effective mass of the nucleon in the nuclear interior. The spin-orbit interaction is greatly enhanced as follows from Equation \ref{eq:darwin}.

The solutions of Equation \ref{eq:psiup} produce single particle nucleon bound states very similar to the familiar shell model orbits but with a large spin-orbit splitting as required by nuclear systematics. 

When the RMF model is carried over to treat the continuum, an optical model potential of the form 
\begin{equation}
\label{eq:opticalpotential}
\begin{array}{lll}
U_0 & = & (V_0+iW_0)f_0(r)\\
U_S & = & (V_S+iW_S)f_0(r) 
\end{array}
\end{equation}
is commonly used where the shape of the nuclear optical potential is consistent with the nuclear shape.  Employing a standard Woods-Saxon shape, this prescription has been extremely successful in describing proton-nucleus elastic scattering cross sections and analyzing powers  with a minimum number of parameters. When calculating  inclusive electron quasi-elastic scattering, the impulse approximation and the same real scalar and vector potentials are used to generate the initial bound and final continuum states of the struck nucleon.  All nuclear effects are in the interaction of the nucleon with the real mean field.  Using an optical potential appropriate to the final state energy of the struck nucleon does not work as well. This prescription, while not obvious, allows for current conservation, orthogonality of the nucleon wave functions and generates the asymmetry of the longitudinal response and enhancement in the transverse response as seen in Figure \ref{fig:psil_rmf}.

Apart from their phenomenological success, there is a body of theoretical work justifying the use of scalar and vector potentials. The scalar field is generated by a $\sigma$ meson, a very broad $T=0$, $J^{\pi}=0^+$ resonance formed by two pions, while the vector field is due to the $\omega$ meson with $T=0$, $J^{\pi}=1^-$.

\end{document}